\documentclass[iop,revtex4]{emulateapj}
\usepackage{lscape}
\usepackage{amsmath}
\usepackage{relsize}
\usepackage{epstopdf}
\usepackage{epsfig}
\usepackage{longtable}
\usepackage{graphicx}
\usepackage{afterpage}
\begin{document}

\shorttitle{Brown Dwarf Analogs to Exoplanets}
\shortauthors{Faherty et al.}

\title{Population Properties of Brown Dwarf Analogs to Exoplanets\footnote{This paper includes data gathered with the 6.5 meter Magellan Telescopes located at Las Campanas Observatory, Chile.}}

\author{Jacqueline K.\ Faherty\altaffilmark{1,2,9}, Adric R. Riedel\altaffilmark{2,3},  Kelle L. Cruz\altaffilmark{2,3,11}, Jonathan Gagne\altaffilmark{1, 10}, Joseph C. Filippazzo\altaffilmark{2,4,11}, Erini Lambrides\altaffilmark{2}, Haley Fica\altaffilmark{2}, Alycia Weinberger\altaffilmark{1}, John R. Thorstensen\altaffilmark{8},  C. G. Tinney\altaffilmark{7,12}, Vivienne Baldassare\altaffilmark{2,5}, Emily Lemonier\altaffilmark{2,6}, Emily L. Rice\altaffilmark{2,4,11}}

\altaffiltext{1}{Department of Terrestrial Magnetism, Carnegie Institution of Washington, Washington, DC 20015, USA; jfaherty@carnegiescience.edu }
\altaffiltext{2}{Department of Astrophysics, 
American Museum of Natural History, Central Park West at 79th Street, New York, NY 10034; jfaherty@amnh.org }
\altaffiltext{3}{Department of Physics \& Astronomy, Hunter College, 695 Park Avenue, New York, NY 10065, USA}
\altaffiltext{4}{Department of Engineering Science \& Physics, College of Staten Island, 2800 Victory Blvd., Staten Island, NY 10301 USA}
\altaffiltext{5}{Department of Astronomy, University of Michigan, 1085 S. University, Ann Arbor, MI 48109}
\altaffiltext{6}{Department of Physics \& Astronomy, Columbia University, Broadway and 116th St., New York, NY 10027 USA}
\altaffiltext{7}{School of Physics, UNSW Australia, 2052. Australia}
\altaffiltext{8}{Department of Physics and Astronomy, Dartmouth College, Hanover NH
03755, USA}
\altaffiltext{9}{Hubble Fellow}
\altaffiltext{10}{Sagan Fellow}
\altaffiltext{11}{Physics Program, The Graduate Center, City University of New York, New York, NY 10016}
\altaffiltext{12}{Australian Centre for Astrobiology, UNSW Australia, 2052. Australia}

\begin{abstract}
We present a kinematic analysis of 152 low surface gravity M7-L8 dwarfs by adding 18 new parallaxes (including 10 for comparative field objects), 38 new radial velocities,  and 19 new proper motions.  We also add low- or moderate-resolution near-infrared spectra for 43 sources confirming their low- surface gravity features.  Among the full sample, we find 39 objects to be high-likelihood or new bona fide members of nearby moving groups, 92 objects to be ambiguous members and 21 objects that are non-members.   Using this age calibrated sample, we investigate trends in gravity classification, photometric color, absolute magnitude, color-magnitude, luminosity and effective temperature. We find that gravity classification and photometric color clearly separate 5-130 Myr sources from $>$ 3 Gyr field objects, but they do not correlate one-to-one with the narrower 5 -130 Myr age range.  Sources with the same spectral subtype in the same group have systematically redder colors, but they are distributed between 1-4$\sigma$ from the field sequences and the most extreme outlier switches between intermediate and low-gravity sources either confirmed in a group or not.  The absolute magnitudes of low-gravity sources from $J$ band through $W3$ show a flux redistribution when compared to equivalent ly typed field brown dwarfs that is correlated with spectral subtype. Low-gravity late-type L dwarfs are fainter at $J$ than the field sequence but brighter by $W3$. Low-gravity M dwarfs are $>$ 1 mag brighter than field dwarfs in all bands from $J$ through $W3$.  Clouds, which are a far more dominant opacity source for L dwarfs, are the likely cause.   On color magnitude diagrams, the latest-type low-gravity L dwarfs drive the elbow of the L/T transition up to 1 magnitude redder and 1 magnitude fainter than field dwarfs at $M_{J}$ but are consistent with or brighter than the elbow at $M_{W1}$ and $M_{W2}$. We conclude that low-gravity dwarfs carry an extreme version of the cloud conditions of field objects to lower temperatures, which logically extends into the lowest mass directly imaged exoplanets. Furthermore, there is an indication on CMD's (such as $M_{J}$ versus ($J$-$W2$)) of increasingly redder sequences separated by gravity classification although it is not consistent across all CMD combinations.  Examining bolometric luminosities for planets and low-gravity objects, we confirm that (in general) young M dwarfs are overluminous while young L dwarfs are normal compared to the field. Using model extracted radii, this translates into normal to slightly warmer M dwarf temperatures compared to the field sequence while lower temperatures for L dwarfs with no obvious correlation with assigned moving group. 
\end{abstract}

\section{INTRODUCTION}

At masses $<$$\sim$ 75 M$_{Jup}$ -- the H-burning mass limit -- , the interior of a source changes significantly.  Below this mass limit, electron degeneracy pressure sufficiently slows contraction that the core of a given object is prevented from ever reaching the temperatures required for nuclear fusion (\citealt{Hayashi63}, \citealt{Kumar63}).   As a consequence, the evolution of substellar mass objects produces a temperature, age, and mass degeneracy that leads to an important, and at times completely indistinguishable, overlap in the physical properties of the lowest mass stars, brown dwarfs and planets. 

Objects with masses $<$$\sim$ 75 $M_{Jup}$ cool through their lives with spectral energy distributions evolving as their atmospheric chemistry changes with decreasing temperatures.    The spectral classification for sources in the range (3000 K $>$ T$_{eff}$ $>$ 250 K) corresponds to late-type M, L, T and Y with each class defined by the effects of changing molecular species available in the photosphere (\citealt{Kirkpatrick05}, \citealt{Burgasser02}, \citealt{Cushing11}). At the warmest temperatures, the atmosphere is too hot for the condensation of solids (\citealt{Allard95}; \citealt{Lodders99}).  But as the T$_{eff}$ falls below 2500 K, both liquid (e.g. Fe) and solid (e.g. CaTiO$_3$, VO) mineral and metal condensates settle into discrete cloud layers (\citealt{Ackerman01}; \citealt{Tsuji96,Tsuji96a}; \citealt{Woitke04}; \citealt{Allard01}).  

As temperatures cool further, cloud layers form at such deep levels in the photosphere that they have little or no impact on the emergent spectrum.  This transition between ``cloudy" to ``cloudless" objects occurs rapidly over a narrow temperature range (1200-1400 K, corresponding to the transition between L-type and T-type spectra) and drives extreme photometric, spectroscopic, and luminosity changes  (\citealt{Burgasser02a}, \citealt{Tinney03}, \citealt{Faherty12, Faherty14a}, \citealt{Dupuy12}, \citealt{Radigan12}, Artigau et al. 2009).   Violent storms (like those seen on Jupiter), along with magnetic-activity-inducing aurorae, have been noted as environmental conditions likely to be present at the L-T transition ( \citealt{Metchev15}, \citealt{Radigan12}, \citealt{Apai13}, \citealt{Hallinan15}, \citealt{Buenzli14},\citealt{Gillon13}, \citealt{Buenzli15}, \citealt{Faherty14a}, \citealt{Burgasser14}).

Confounding our understanding of cloud formation in low-temperature atmospheres is mounting evidence for a correlation between cloud properties and youth.  Low surface gravity brown dwarfs, thought to be young, have unusually red near to mid-infrared colors and a fainter absolute magnitude through $\sim$ 2.5 $\mu$m when compared to their older spectral counterparts with field surface gravities (\citealt{Faherty12}, \citealt{Faherty13}, \citealt{Filippazzo15}, \citealt{Liu13}, \citealt{Gagne15b}).   \citet{Metchev06} made the first connection between age and cloudiness in their study of the young companion HD 203030B, whose transition to the cloud-free T spectral class appears to be delayed by the presence of thick clouds.  In a detailed study of the prototypical isolated, young brown dwarf 2M0355+1133, \citet{Faherty13}  found that the deviant colors and fainter absolute magnitudes were best explained by enhanced dust, or thick photospheric clouds, shifting flux to longer wavelengths.  At the coldest temperatures, where clouds should all but have dispersed below the photosphere in field brown dwarfs, \citet{Burgasser10} studied the T8 dwarf Ross 458 C and found clouds must be considered as an important opacity source for young T dwarfs. 

Exoplanet studies have independently found similar trends with age and cloud properties.  The young planetary mass companions 2M1207 b ($<$ 10 M$_{Jup}$) and HR8799 b  ($<$ 10 M$_{Jup}$),  are exceedingly red in the near-infrared and up to 2 magnitudes fainter than field brown dwarfs of similar T$_{eff}$ (\citealt{Marois08,Marois10}, \citealt{Mohanty07}).  To reproduce their anomalous observables, theorists have developed ``enhanced" cloudy atmospheric models with non-equilibrium chemistry (\citealt{Marley12}, \citealt{Barman11,Barman11b}, \citealt{Madhusudhan11}) in which lower surface gravity alters the vertical mixing which then leads to high altitude clouds with differing physical composition (e.g. thicker or thinner aggregations).  

In general, M, L, T, and Y classifications identify brown dwarfs. If the source is older ($>$ 2 - 3 Gyr), late-type M and early L dwarfs are stars.  But if the source is young ($<$ 1 Gyr), even those warmer classifications will describe an object that is $<$ 75 $M_{Jup}$.  To date, all directly imaged giant exoplanets have observable properties which lead to their classification squarely in this well-studied regime.  Planetary mass companions such as 2M1207 b, 51 Eri b, $\beta$ Pictoris b, ROXs 42B b, and the giant planets orbiting HR8799, have observables that are similar to L or T brown dwarfs (\citealt{Chauvin04}, \citealt{Macintosh15}, \citealt{Lagrange10}, \citealt{Marois08,Marois10}, \citealt{Currie14}, \citealt{Kraus14a}).    Furthermore, there exists a population of ``classical'' brown dwarfs that overlap in effective temperature, age -- many in the same moving group -- , and mass with directly observed planetary mass companions (e.g. PSO 318, SDSS1110, 0047+6803, \citealt{Liu13}, \citealt{Gagne15}, \citealt{Gizis12, Gizis15}).  Studies of these two populations in concert may resolve questions of the formation of companions versus isolated equivalents as well as untangle atmosphere, temperature, age, and metallicity effects on the observables.  

In this work we examine this new population of suspected young, low surface gravity sources that are excellent exoplanet analogs.  In section 2 we explain the sample examined in this work and in section 3 we describe the imaging and spectral data acquired.  In section 4 we review new near-infrared spectral types designated in this work and in section 5 we discuss how we measured new radial velocities.  In section 6 we assess the likelihood of membership in nearby moving groups such as $\beta$ Pictoris, AB Doradus, Argus, Columba, TW Hydrae, and Tucana Horologium.  In section 7 we review the diversity of the whole sample in spectral features, infrared colors, absolute magnitudes, bolometric luminosities and effective temperatures.  In section 8 we place the young brown dwarf sample in context with directly imaged planetary mass companions.  Conclusions are presented in section 9.

\section{The Sample}
\label{section:sample}
Given the age-mass degeneracy of substellar mass objects and an age range of $\sim$5 - 130 Myr for groups such as TW Hydrae (5-15 Myr; \citealt{Weinberger13}), $\beta$ Pictoris (20-26 Myr; \citealt{Binks14}, \citealt{Malo14a}), and AB Doradus (110 - 130 Myr; \citealt{Barenfeld13}, \citealt{Zuckerman04a}) the target temperature for our sample was T$_{eff}$ $<$ 3000 K, or, equivalently, sources with spectral types of M7 or later.  This cut-off restricted us to $<$ 0.07 $M_{\sun}$ or the classic brown dwarf boundary  (\citealt{Kumar63}, \citealt{Hayashi63}). 

The suspicion of membership in a nearby moving group should be accompanied by observed signatures of youth as kinematics alone leave doubt about chance contamination from the field sample.  As such, for this work we focused on $>$ M7 objects with confirmed spectral signatures of low-gravity in either the optical or the infrared.  We note that while there are isolated T dwarfs thought to be young (e.g. SDSS 1110, \citealt{Gagne15a}, CFBDSIR 2149, \citealt{Delorme12}), their spectral peculiarities appear to be subtle making them more difficult to identify and investigate (see also \citealt{Best15}). 

For isolated late-type M and L dwarfs suspected to be young, there are strong spectral differences in the strength of the alkali lines and metal oxide absorption bands as well as the shape of the near-infrared $H$- band ($\sim$1.65 $\mu$m) compared to older field age counterparts (e.g. \citealt{Lucas01};  \citealt{Gorlova03}; \citealt{Luhman04}; \citealt{McGovern04}; \citealt{Allers07}; \citealt{Rice10,Rice11}, \citealt{Patience12}; \citealt{Faherty13}).    Physically this can be explained by a change in the balance between ionized and neutral atomic and molecular species, as a result of lower surface gravity and, consequently, lower gas densities in the photospheric layers (\citealt{Kirkpatrick06}). Furthermore, a lower surface gravity is linked to an increase in collision induced H$_2$ absorption (see e.g. \citealt{Canty13}, \citealt{Tokunaga99}). Changes in the amount of this absorption result in the $K$- band ($\sim$ 2.2 $\mu$m) being suppressed (or enhanced) and the shape of the $H$ band being modeified to cause the  ``peaky" H band relative to water opacity seen in young sources (see \citealt{Rice11}).

The collection of brown dwarfs with spectral signatures of a low-surface gravity is increasing\footnote{Our group maintains a listing of known isolated field objects noted as having gravity sensitive features on a web-based compendium$http://www.bdnyc.org/young\_bds$}. \citet{Gagne14d, Gagne15b, Gagne15} presented a bayesian analysis of the brown dwarf population looking for potential new moving group members and uncovered numerous low-gravity sources. \citet{Allers13} presented a near-infrared spectroscopic study of a large number of known sources.  Aside from those extensive studies, objects have been reported singly in paper (e.g.  \citealt{Kirkpatrick06}, \citealt{Liu13}, \citealt{Gagne14a,Gagne14b, Gagne15a}, \citealt{Faherty13}, \citealt{Rice10}, \citealt{Gizis12}, \citealt{Gauza15}) or included as a subset to a larger compilation of field objects (e.g. \citealt{Cruz07}, \citealt{Kirkpatrick10}, \citealt{Reid07}, \citealt{Thompson13}).   The objects within this paper were drawn from the literature as well as our ongoing search for new low-surface gravity objects. 

The 152\footnote{While this paper was awaiting acceptance, \citet{Aller16} reported a sample of AB Doradus late-type M and L dwarfs. Those objects are not included in this analysis but should be considered in future work.} low-surface gravity objects examined in this work are listed in Table~\ref{tab:photometry} with their coordinates, spectral types, gravity classifications (optical and infrared as applicable), 2MASS and WISE photometry.  There are 48 sources in this sample that are lacking an optical spectral type and 8 sources lacking an infrared spectral type.  Of the 96 objects with both optical and near-infrared data, there are  67 sources (70\%) that have a different optical spectral type than the infrared although the majority are within 1 subtype of each other.  

For assigning a gravity designation, there are two classification systems based on spectral features. The \citet{Cruz09} system uses optical spectra and assigns a low-surface gravity ($\gamma$), intermediate gravity ($\beta$), or field gravity (`---' in Table~\ref{tab:photometry} and throughout) based on the strength of metal oxide absorption bands and alkali lines.  In certain cases a classification of $\delta$ is also used for objects that look more extreme than $\gamma$ (see \citealt{Gagne15b}).  In this work we label objects as $\delta$ in tables but plot them and discuss them along with $\gamma$ sources. On the \citet{Cruz09} scheme, $\gamma$ and $\beta$ objects are thought to be younger than the Pleiades (age $<$ $\sim$ 120 Myr, \citealt{Stauffer98}). The \citet{Allers13} system uses near-infrared spectra and evaluates spectral indices to assign a very low-gravity (vl-g), intermediate gravity (int-g), or field gravity (fld-g) to a given source.  As discussed in \citet{Allers13}, the optical and near-infrared gravity systems are broadly consistent. However, to anchor either requires an age-calibrated sample to ground the gravity designations as age-indicators.

\begin{figure*}[!t]
\begin{center}$
\begin{array}{cc}
\includegraphics[width=3.5in]{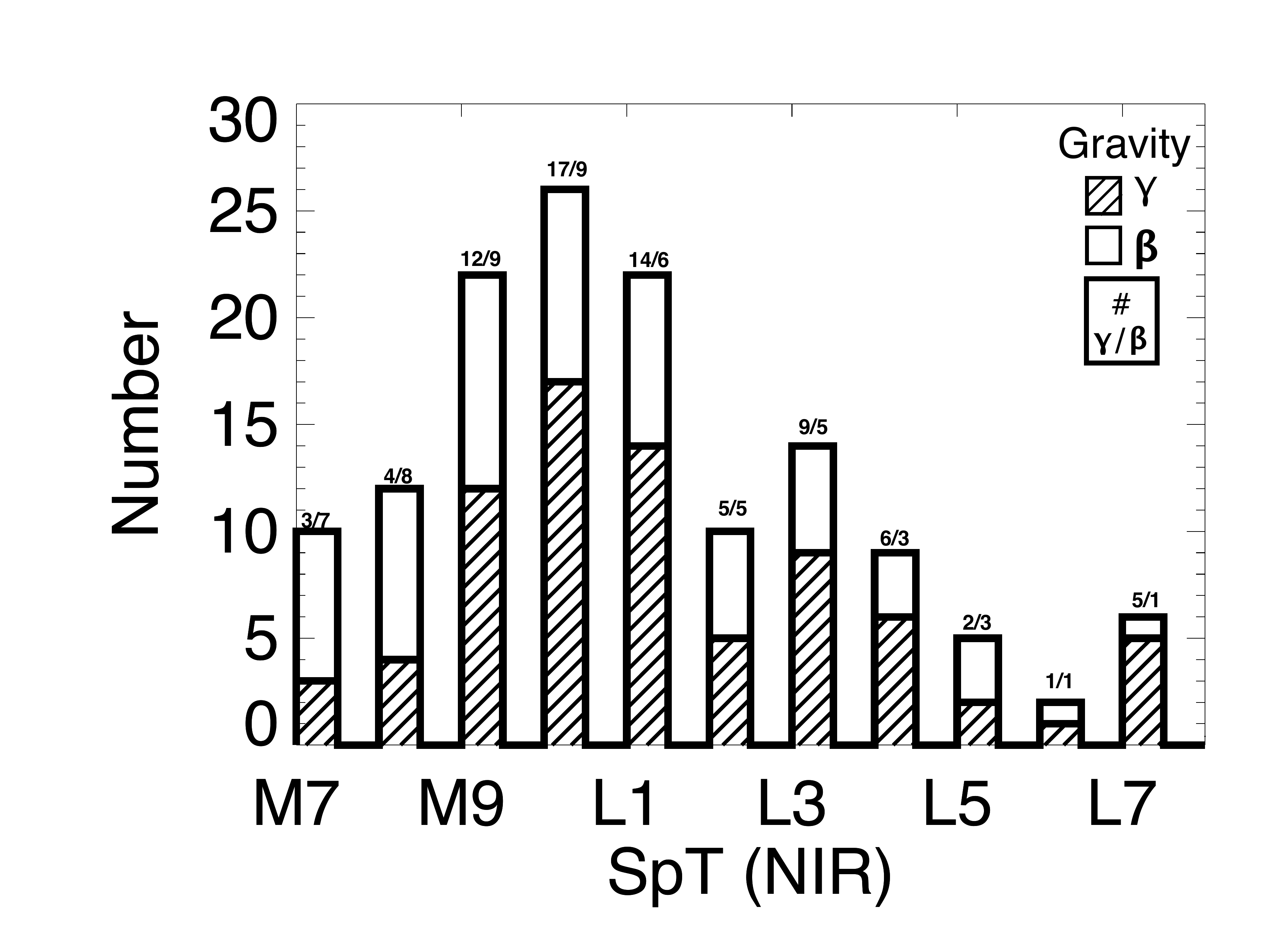}&
\includegraphics[width=3.5in]{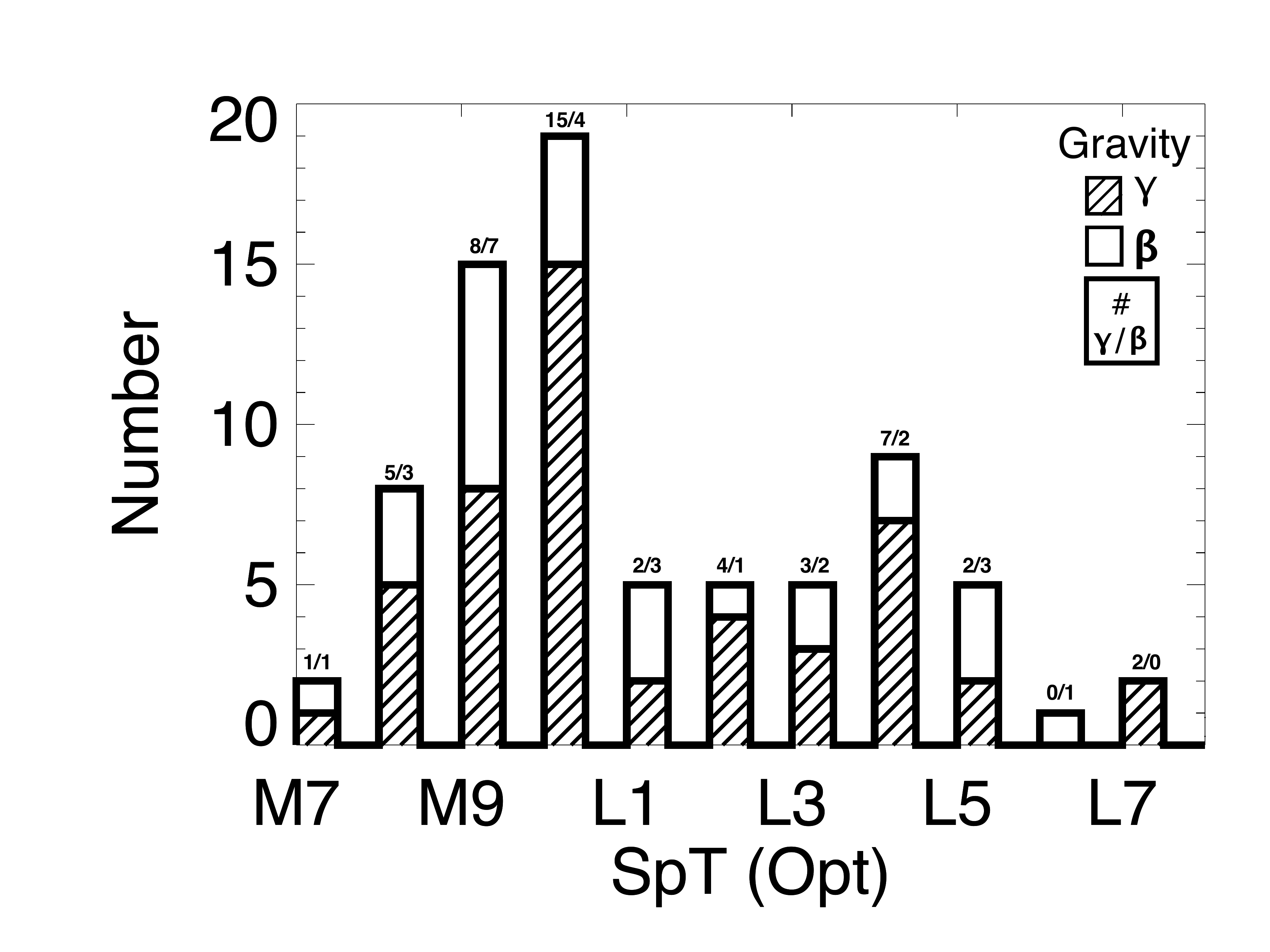}  \\
\end{array}$
\end{center}
\caption{The distribution of objects analyzed in this work organized by spectral subtype and gravity classification in either the near-infrared (left) or optical (right). For ease of labels in this work, we have chosen to label VL-G and INT-G objects classified using the \citet{Allers13} spectral indices as $\gamma$ and $\beta$ respectively. On this plot we also show the number of $\gamma$ and $\beta$ at each subtype as a ratio of \# $\gamma$ / \# $\beta$. }
\label{fig:Histogram}
\end{figure*}

Figure~\ref{fig:Histogram} shows a histogram distribution of the spectral subtypes in the optical and the infrared highlighting the gravity classification.  There are 51 objects classified optically as $\gamma$  and 80 with the equivalent infrared classification.  There are 27 objects classified optically as $\beta$ and 57 with the equivalent infrared classification. Of the objects which have both optical and infrared gravity designations, 16 sources (17\%) have different gravity classifications from the two methods and 23 objects (24\%)  have a low-gravity infrared classification but are not noted as peculiar in the optical.  For simplification of the text (and in large part because the two systems are generally consistent), we have adopted the convention that any object classified as vl-g or  int-g  in the infrared is referred to as $\gamma$ or $\beta$ (respectively) in the text, Tables, and Figures.

\section{DATA}
The sample of 152 M7-L8 ultracool dwarfs comprising our sample were placed on follow-up programs -- either imaging (parallax, proper motion), spectroscopy (radial velocity) or both -- to determine kinematic membership in a nearby moving group.  Below we describe the data collected for the suspected young brown dwarf sample.  

\subsection{Parallax and Proper Motion Imaging}
The astrometric images for this program were obtained using three different instruments and telescopes in the northern and southern hemispheres.  We report parallaxes for eight low-gravity and ten field dwarfs.  For 19 objects, we report proper motion alone as we lack enough epochs to decouple parallaxes.   An additional 13 objects have not yet been imaged by either the northern or southern instruments, and so we report proper motions using the time baseline between 2MASS and WISE.

\subsubsection{Northern Hemisphere Targets}
For Northern Hemisphere astrometry targets, we obtained $I$-band images with the MDM Observatory 2.4m Hiltner telescope on Kitt Peak, Arizona. Parallaxes are being measured for both low-surface gravity and field ultracool dwarfs and for this work we report 5 of the former and 10 of the latter (field dwarfs used for comparison in the analysis discussed in Section~\ref{sec:diversity}).  

For most observations at MDM, we used a thinned SITe CCD detector (named ``echelle") with 2048x2048 pixels and an image scale of $0''.275$ pixel$^{-1}$.  This suffered a hardware failure and was unavailable for some of the runs. As a substitute,  we began using ``Nellie", a thick, frontside-illuminated STIS CCD which gave $0''.240$ pixel$^{-1}$. The change in instrument made no discernible difference to the astrometry. Table \ref{tab:parallaxes} gives the pertinent astrometric information.  In addition to the parallax imaging, we took $V$-band images  and determined $V-I$ colors for the field stars for use in the parallax reduction and analysis. For a single target field (1552+2948) we used SDSS colors instead. The reduction and analysis was similar to that described in \citet{thorstensen03} and \citet{thorstensen08}, with some modifications.  

Parallax observations through a broadband filter are subject to differential color refraction (DCR). The effective wavelength of the light reaching the detector for each star will depend on its spectral energy distribution. Consequently, the target and reference stars can be observed to have different positions depending on how far from the zenith the target is observed at each epoch. In previous studies we approximated the $I$-band DCR correction as a simple linear trend with $V - I$ color, amounting to 5 mas per unit $V - I$ per unit $\tan z$ (where z is the zenith distance of each observation).  We checked this by explicitly computing the correction for stars of varying color using library spectra from \citet{pickles98}, a tabulation of the $I$ passband from \citet{bessell90}, and the atmospheric refraction as a function of wavelength appropriate to the observatory's elevation.  The synthesized corrections agreed very well with the empirically-derived linear correction. However, the library spectra did not extend to objects as red as the present sample and most are so faint in $V$ that we could not measure $V-I$ accurately.  We therefore computed DCR corrections using the spectral classifications of our targets and spectra of L and T-dwarfs assembled by Neill Reid\footnote{Available at {\tt http://www.stsci.edu/$\sim$inr/ultracool.html}}. The resulting corrections typically amounted to $\sim 25$ mas per unit $\tan z$, that is, the DCR expected on the basis 
of the linear relation for a star with $(V-I) \approx 5$. 

To minimize DCR effects, we restricted the parallax observations to hour angles within $\pm 2$ h of the meridian.  The effect of DCR on parallax is mainly along the east-west (or $X$) direction, and the $X$-component of refraction is proportional to $\tan z \sin p$, where $p$ is the parallactic angle. This quantity averaged 0.12 for our observations, reflecting a slight westward bias in hour angle, and its standard deviation was 0.15.  We are therefore confident that the DCR correction is not affecting our results unduly.

As in previous papers, we used our parallax observations to estimate distances using a Bayesian formalism that takes into account proper motion, parallax, and a plausible range of absolute magnitude \citep{thorstensen03}.  For these targets we assumed a large spread in absolute magnitude, so that it had essentially no effect on the distance, and used a velocity distribution characteristic of a disk population to formulate the proper-motion prior. For most of the targets, the parallax $\pi$ was precise enough that the Lutz-Kelker correction and other Bayesian priors had little effect on the estimated distance, which was therefore
close to $1/\pi$.

Table~\ref{tab:distances} lists our measured parallaxes and proper motions and Table~\ref{tab:parallaxes-duplicates} shows the comparison with literature values for four sources with previously reported values.  In the case of MDM measured proper motions, they are relative to the reference stars, and not absolute.  Although the formal errors of the proper motions are typically 1-2 mas yr$^{-1}$,  this precision is spurious in that the dispersion of the reference star proper motions is usually over 10 mas yr$^{-1}$, so the relative zero point is correspondingly uncertain.

\subsubsection{Southern Hemisphere Targets}
We observed 16 of the most southernly targets with the Carnegie Astrometric Planet Search Camera (CAPSCam) on the 100-inch du Pont telescope and five with the FourStar imaging camera (\citealt{Persson13}) on the Magellan Baade Telescope.  In the case of both programs, we are continually imaging objects for the purpose of measuring parallaxes.  However, for this work we report parallaxes (and proper motions) for only 3 objects with CAPScam.  The remaining 18 objects (13 -- CAPScam, 5 -- FourStar) need more epochs to  decouple parallax from proper motion and will be the subject of a future paper.

A description of the CAPSCam instrument and the basic data reduction techniques are described in \cite{Boss09} and \citet{Anglada-Escude12}. CAPSCam utilizes a Hawaii-2RG HyViSI detector filtered to a bandpass of 100 nm centered at 865 nm with 2048 x  2048 pixels, each subtending 0.196$\arcsec$ on a side.  CAPSCam was built to simultaneously image bright target stars in a 64 x 64 pixel guide window allowing short exposures while obtaining longer exposures of fainter reference stars in the full frame window (e.g. \citealt{Weinberger13}).  The brown dwarfs targeted in this program were generally fainter than the astrometric reference stars so we worked with only the full frame window.  Data were processed as described in \citet{Weinberger13}.  For exposure times,  we used 30s - 120s for our bright targets and 150s - 300s for our faint targets with no coadds in an 8 - 12 point dither pattern contained in a 15$\arcsec$ box. Pertinent astrometric information for parallax targets is given in Table~\ref{tab:parallaxes}.

FourStar is a near-infrared mosaic imager  (\citealt{Persson13}) with four 2048 x 2048 Teledyne HAWAII-2RG arrays that produce a 10.9$\arcmin$ x 10.9$\arcmin$ field of view at a plate scale of 0.159$\arcsec$ pixel$^{-1}$.  Each target was observed with the J3 (1.22--1.36\,$\mu$m) narrow band filter and centered in chip 2.  This procedure has proven successful in our astrometric program for late-T and Y dwarfs (e.g. \citealt{Tinney12,Tinney14}, \citealt{Faherty14}).  Exposure times of 15s with 2 coadds in an 11 point dither pattern contained in a 15$\arcsec$ box were used for each target. The images were processed as described in \citet{Tinney14}.  

In the case of the 18 proper motion only targets from either CAPSCam or FourStar, we combine our latest image with that of 2MASS ($\Delta$t listed in Table~\ref{tab:PMs}).  Proper motions were calculated using the astrometric strategy described in \citet{Faherty09}.  Results are listed in Table~\ref{tab:PMs}.  

For three CAPSCam targets there is sufficient data to solve for both proper motions and parallaxes.  For these sources, the astrometric pipeline described in  \citet{Weinberger13} was employed. Table~\ref{tab:distances} lists our measured parallaxes and proper motions and Table~\ref{tab:parallaxes-duplicates} shows the comparison with literature values for 0241-0326.

 \subsection{Low and Medium Resolution Spectroscopy}

\subsubsection{FIRE}
We used the 6.5m Baade Magellan telescope and the Folded-port InfraRed Echellette (FIRE; \citealt{Simcoe13}) spectrograph to obtain near-infrared spectra of 36 sources.  Observations were made over 7 runs between 2013 July and 2014 September.  For all observations, we used the echellette mode and the 0.6$\arcsec$ slit  (resolution $\lambda$/$\Delta \lambda \sim$ 6000) covering the full 0.8 - 2.5 $\micron$ band with a spatial resolution of 0.18$\arcsec$/pixel. Exposure times for each source and number of images acquired are listed in Table~\ref{tab:spectra}.  Immediately after each science image, we obtained an A star for telluric correction and obtained a ThAr lamp spectra.  At the start of the night we obtained dome flats and Xe flash lamps to construct a pixel-to-pixel response calibration.  Data were reduced using the FIREHOSE package which is based on the MASE and SpeX reduction tools (\citealt{Bochanski09}, \citealt{Cushing04}, \citealt{vacca03}). 

\subsubsection{IRTF}
We used the 3m NASA Infrared Telescope Facility (IRTF) to obtain low-resolution near-infrared spectroscopy for 10 targets.  We used either the 0$\farcs$5 slit or the 0$\farcs$8 slit depending on conditions. All observations were aligned to the parallactic angle to obtain $R~\equiv~\lambda$ / $\Delta\lambda~\approx$~120 spectral data over the wavelength range of 0.7 -- 2.5 $\mu$m. Exposure times for each source and number of images acquired are listed in Table~\ref{tab:spectra}.  Immediately after each science observation we observed an A0 star at a similar airmass for telluric corrections and flux calibration, as well as an exposure of an internal flat-field and Ar arc lamp. All data were reduced using the SpeXtool package version 3.4 using standard settings (\citealt{Cushing04}, \citealt{vacca03}).

\subsubsection{TSpec}
We used the Triple Spectrograph (TSpec) at the 5 m Hale Telescope at Palomar Observatory to obtain near-infrared spectra of two targets.  TSpec uses a 1024 x 2048 HAWAII-2 array to cover simultaneously the range from 1.0 to 2.45$\mu$m (Herter et al. 2008). With a 1.1 x 43$\arcsec$ slit, it achieves a resolution of  $\sim$2500.   Observations were acquired in an ABBA nod sequence with an exposure time per nod position of 300s (see Table~\ref{tab:spectra}) so as to mitigate problems with changing OH background levels.  Observations of A0 stars were taken near in time and near in airmass to the target objects and were used for telluric correction and flux calibration. Dome flats were taken to calibrate the pixel-to-pixel response. Data were reduced using a modified version of Spextool (see \citealt{Kirkpatrick11}).

\subsection{High Resolution Spectroscopy}
\subsubsection{Keck II NIRSPEC}
The Keck II near-infrared SPECtrograph (NIRSPEC) is a Nasmyth focus spectrograph designed to obtain spectra at wavelengths from 0.95 -- 5.5$\mu$m \citep{McLean98}.  It offers a choice of low-resolution and cross-dispersed high resolution spectrographic modes, with optional adaptive optics guidance.  In high-resolution mode, it can achieve resolving powers of up to R=25000 using a 3 pixel entrance slit, with two orders visible on the output spectrum (selectable by filter).

Multiple observations of 17 sources were taken in high resolution mode on Keck II on 14, 15, and 16 September 2008, using the NIRSPEC-5 filter to obtain H-band spectra in Order 49 (1.545 -- 1.570$\mu$m).  Observational data for each source are listed in Table~\ref{tab:spectroscopic}. The data were reduced using the IDL-based spectroscopy reduction package REDSPEC. Many of our observations had very low signal-to-noise, and so multiple exposures were co-added before extracting spectra.  We tested this procedure by using objects with sufficient signal-to-noise prior to co-adding, and by comparing individual against co-added spectra.  The resulting individual exposure spectra were almost identical to those obtained by co-adding prior to running REDSPEC reductions. Heliocentric radial velocity corrections were calculated with the IRAF task $\it{rvcorrect}$, and applied using custom python code.

\subsubsection{Gemini South Phoenix}
The Phoenix instrument (previously on Gemini South) is a long-slit, high resolution infrared echelle spectrograph, designed to obtain spectra between 1 -- 5 $\mu$m at resolutions between R=50000 and R=80000.  Spectra are not cross-dispersed, leaving only a narrow range of a single order selectable by order-sorting filters. 

Observations of 18 sources were taken during semester 2007B and 2009B using the H6420 filter to select H-band spectra in Order 36 (1.551 -- 1.558 $\mu$m).  Spectra were reduced using the supplied IDL Phoenix reduction codes. Observational data for each source are listed in Table~\ref{tab:spectroscopic}.

\subsubsection{Magellan Clay MIKE}
The Magellan Inamori Kyocera Echelle (MIKE) on the Magellan II (Clay) telescope is a cross-dispersed, high resolution optical spectrograph, designed to cover the entire optical spectrum range (divided into two channels, blue: 0.32 -- 0.48 $\mu$m, and red: 0.44 -- 1.00 $\mu$m) at a resolving power of R$\sim$28,000 (blue) and R$\sim$22,000 (red) using the 1.0'' slit. The output spectra contain a large range of overlapping echelle orders, each covering roughly 0.02$\mu$m of the red optical spectrum.

Red-side spectra were taken on 4 July 2006, 1 November 2006, and 2 November 2006.  The observations comprise 17 target and standard spectra, with additional B-type and white dwarf flux calibrators.  Observational data for each source are listed in Table~\ref{tab:spectroscopic}. Spectra were reduced using the IDL MIKE echelle pipeline\footnote{http://web.mit.edu/\~burles/www/MIKE/ checked 19 JUNE 2014}, and orders 38 -- 52 (0.92 -- 0.65$\mu$m) were extracted from every spectrum.  Many of those orders were unusable for our purposes, and were not used in the final solutions.  Telluric atmospheric features dominate the wavebands covered by orders 38 (telluric $O_2$), 45 (A band), and 50 (B band).  The orders higher (bluer) than 44 typically had insufficient signal due to the extreme faint and red colors of the target objects.

\section{New Near-Infrared Spectral Types}
We obtained spectra with FIRE, SpeX, and TSpec for 43 targets to investigate near-infrared signatures of youth.  Each object had either demonstrated optical low-surface gravity features, but were missing (or had poor) near infrared data, or had low signal to noise near infrared spectra.  Determining the spectral type and gravity classification for peculiar sources has its difficulties. Primarily because one wants to ground the peculiar object type by comparing to an equivalent field source.  However, as will be seen in section ~\ref{sec:diversity}, the low gravity sequence does not easily follow from the field sequence.  In the infrared, \citet{Allers13} have presented a method for determining gravity classification using indices.  Alternatively, the population of low gravity sources (especially earlier L types) has grown in number such that templates of peculiar sources can be made for comparison (e.g. \citealt{Gagne15b}, Cruz et al. submitted). For this work, we have defaulted to a visual match to templates or known sources -- grounded by their optical data -- as our primary spectral typing method.  However we also check each source against indices to ensure consistency.  

\begin{figure*}[t!]
\resizebox{\hsize}{!}{\includegraphics[clip=true]{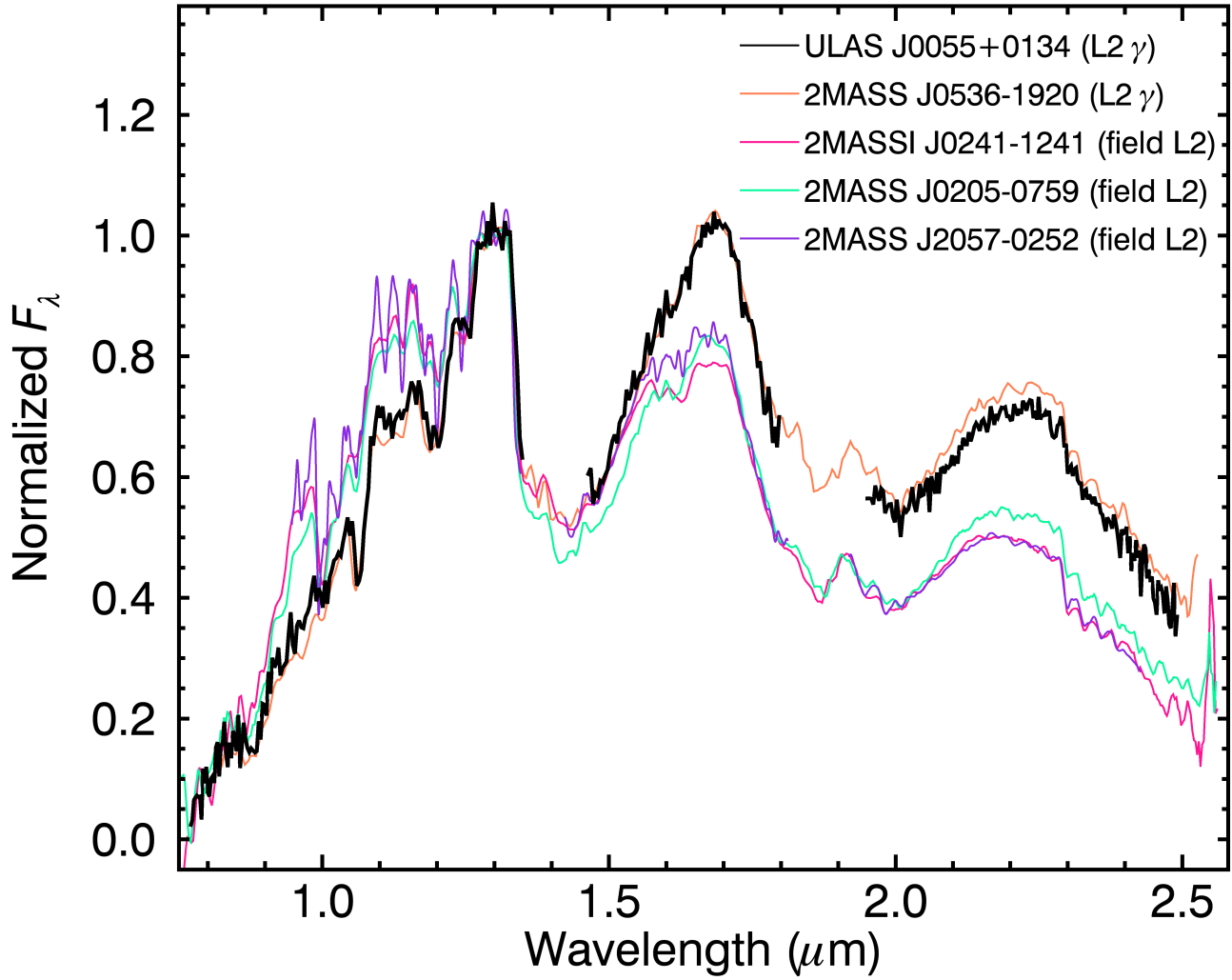}}
\caption{The SpeX prism spectrum of 0055+0134 over-plotted with a sample of field and very low-gravity subtype equivalents. 
\label{fig:spectra2}}
\end{figure*}

\begin{figure*}[t!]
\resizebox{\hsize}{!}{\includegraphics[clip=true]{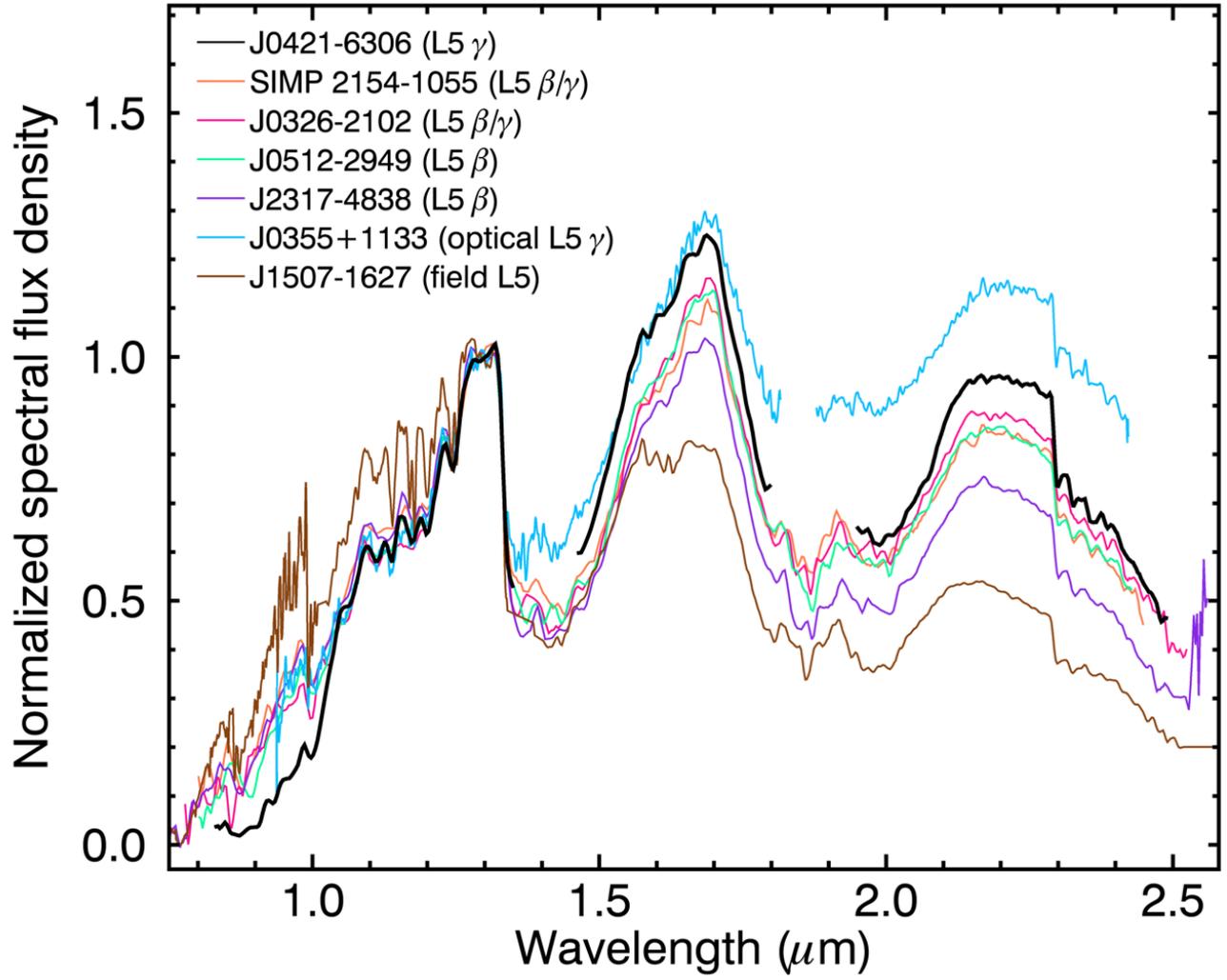}}
\caption{The FIRE spectrum of 0421-6306 binned to prism resolution and over-plotted with a sample of field and very low-gravity subtype equivalents. 
\label{fig:spectra1}}
\end{figure*}

In the case of each new spectrum, we visually compared to the library of spectra in the SpeX prism library\footnote{http://pono.ucsd.edu/$\sim$adam/browndwarfs/spexprism/} as well as the low gravity templates discussed in \citealt{Gagne15b}.  For the FIRE and TSpec echelle spectra, we first binned them to prism resolution ($\sim$ 120).  This visual check to known objects gave the match we found most reliable for this work in both type and gravity classification and it is listed in the ``SpT adopted" column in Table ~\ref{tab:spectra2}.  Figures ~\ref{fig:spectra2} - ~\ref{fig:spectra1} show two example spectra (prism and binned down FIRE data) compared visually to both field and very low gravity sources, as representations of our spectral typing method.

As a secondary check, we report the indices analysis of each object.  A near infrared spectral subtype is a required input for evaluating the gravity classification with the \citet{Allers13} system so we first applied the subtype indices (as described in \citealt{Allers13}) and list the results in the ``SpT Allers13" column in Table ~\ref{tab:spectra2}. Once we had determined the closest near infrared type, we evaluated the medium resolution (for FIRE and TSpec data) and/or the low resolution gravity indices (for Prism data).  We list the results of each in Table ~\ref{tab:spectra2}. In general we found that matching visually to low gravity templates or known objects versus using indices were consistent within 1 subtype.  
    
\section{Radial Velocities}
Radial velocities from NIRSPEC, Phoenix, and MIKE were calculated using a custom python routine, which uses cross-correlation with standard brown dwarfs to achieve 1 km s$^{-1}$ radial velocity precision.  Useable spectra have resolving power of R=20,000 or higher, and generally signal-to-noise of at least 20.   All data were corrected to heliocentric radial velocity by shifting the wavelength grid.  Brown dwarf standards -- sourced primarily from \citet{Blake10} -- were observed and corrected with the same settings as the targets, and paired with objects of matching spectral type.  Given that our radial velocity sample is bimodal with peaks at L0 and L4, our spectra are fit against relatively few standards.

The python code calculates radial velocities on the fly, and operates on optical and infrared data without modification.  The radial velocity inputs were the wavelength, flux, and uncertainty data as three one-dimensional arrays, for both the target and the standard.  The target and standard spectra were first cropped to only the portion where they overlap, and then interpolated onto a log-normal wavelength grid.  

From there, 5000 trials were conducted with different randomized Gaussian noise added to the wavelength grids of the target and standard, according to the per-element uncertainties.  This was done to account for the uncertainties on the fluxes and to provide a method of quantifying the uncertainty on the output radial velocity.  For each of the 5000 trials, the two spectra were cross-correlated to produce a wavelength shift between them.  A small 400-element region around the peak of the cross-correlation function was fit with a Gaussian and a linear term, to locate the exact peak of the cross-correlation on a sub-element basis.  The widths (and therefore per-measurement errors) of the cross-correlation peak were discarded, assumed to be accounted for in the actual spread of the resulting set of peaks.

The results of the 5000 trials formed (in well-determined cases) a Gaussian histogram centered on the radial velocity shift between the two systems. The width of that Gaussian was taken to represent the uncertainty in the measured radial velocity.  This pixel shift was converted into km $s^{-1}$ radial velocity, and corrected for the known velocity of the standard.  Semi-independent verification of the radial velocities was accomplished by measuring the velocity relative to two different standards, or where available, using multiple orders from the same spectrum.

The process was sensitive to virtually all unwanted processes that produce features in the brown dwarf spectra. Chief among these were cosmic rays and detector hot pixels, which were pixel-scale events removed from the spectra prior to interpolation onto the grid and provide very little signal in the RV correlation.  The most important effect was telluric lines, which appeared like normal spectral features but did not track the radial velocity of the star.  These were dealt with by identifying orders whose contamination was severe enough to produce discordant radial velocities and avoiding them in the analysis.

Table~\ref{tab:spectroscopic} lists the final radial velocity values for each source.  Table~\ref{tab:spectroscopic2} shows a comparison of sources for which there was a literature value.

\subsection{Instrument-specific Differences}
MIKE data has multiple echelle orders, and all stars were measured against two L2 dwarf standards: LHS~2924 and BRI~1222-1221 (\citealt{Mohanty03}).  All orders were examined by eye, and determinations were made as to whether they contained sufficient signal for a believable radial velocity. This was corroborated by cross-checking the result against other orders from the same star and radial velocity.  Some orders had broader cross-correlation functions, and the Gaussian was fit to a 200 pixel region around the peak rather than the standard 100 pixels.  In other orders, a small a-physical secondary peak in the final results appeared, and was removed from the Gaussian fit for the final radial velocity for that order.

After visual inspection, the most consistent orders -- both within the match between the two stars, and between the two standards -- were combined into a weighted mean and weighted standard deviation.  For all stars, the results of orders 39-44, collectively covering 0.77$\mu$m -- 0.89$\mu$m -- were deemed sufficiently reliable (despite the presence of telluric water features) to be used in the final result.

\section{Computing Kinematic Probabilities in Nearby Young Moving Groups}
Among the 152 brown dwarfs investigated for kinematic membership in this work, we report 37 new radial velocities, 8 new parallaxes, and 33 new proper motions (13 of which are reported as new from 2MASS to WISE proper motion measurements).  In total 27 targets have full kinematics (a parallax, proper motion and radial velocity), and the remaining 123 have only partial kinematics -- 16 have a parallax and proper motion but no radial velocity,  26 have radial velocity and proper motion but no parallax, and  81 have proper motions but no parallax or radial velocity measurement.  All astrometric information for this sample is listed in Table~\ref{tab:kinematics}.

As discussed in Section~\ref{section:sample}, there are 51 objects classified optically as $\gamma$  and 80 with the equivalent infrared classification.  There are also 27 objects classified optically as $\beta$ and 57 with the equivalent infrared classification. Given the gravity indications, we regard each object as a potentially young source and investigate membership in a group within 100 pc of the Sun.  To assess the likelihood of membership, we employed four different tools to examine the available kinematic data:
 \begin{itemize}
\item BANYAN-I Bayesian statistical calculator \citep{Malo13} and its successor, 
\item BANYAN-II \citep{Gagne14},
\item LACEwING \citep{Riedel15}, and
\item Convergent point method of \citet{Rodriguez13}.
 \end{itemize} 
  The plurality of measurements in combination with a visual inspection of an objects kinematics against bona fide members listed in \citep{Malo13} along with the individual kinematic boxes of \citet{Zuckerman04} drove our decision on group membership. 
    
The four different methods test for membership in different sets of groups - LACEwING considers 14 distinct groups, BANYAN I and II consider 7 groups (or 8 including the ``Old'' object classification), and the convergence method tests for membership in 6 groups. Each method has its benefits and flaws.  For instance Banyan I is a fast bayesian formalism that uses flat priors but assumes (probably unrealistically) that radial velocity, and proper motion in a given direction are Gaussian.  Banyan II deals better with transforming measurements to probabilities based on the distribution of known members (does not assume a Gaussian distribution) however it likely has incomplete/imperfect lists of bonafide members.  LACEwING is similar to Banyan I in its assumption that radial velocity, and proper motion in a given direction are Gaussian but it requires fitting a model to a (arguably much cleaner list of) bonafide members of multiple groups not covered by the other methods.  The convergent point method is a simple yet different approach that estimates the probability of membership in a known group by measuring the proper motions in directions parallel and perpendicular to the location of a given groups convergent point.  Unfortunately, this method does not take into account measured radial velocities or distances.  Given the benefits and flaws of each method, we chose to take the output of each into consideration as we decided on membership for each target.  For an adequate comparison, we only considered membership in six groups: TW Hydrae, $\beta$ Pictoris, Tucana Horologium, Columba, Argus (which is not tested by the Convergence code), and AB Doradus. All other groups could not be consistently checked.  Therefore they may be mentioned (e.g. Chamaeleon near, Octans, Hyades) but are only considered tentative until further kinematic investigation.  

Furthermore, the output of each code should be viewed slightly differently.  In \citet{Malo13}, the authors adopt a membership probability threshold of 90\% to recover bona fide members.  Banyan II supplements with a contamination probability and finds this number should be $<$ a few percent with a high membership probability (we impose $>$ 90\% based on Banyan I) in order to recover bona fide members (\citealt{Gagne14}).  LACEwING (as described in \citealt{Riedel15}) finds $<$ 20\% - 60\% is low probability and $>$ 60\% is high probability for group membership.  Convergent point reports distinct probabilities for each group between 0 - 100\% (hence objects can have $>$ 90\% probability of membership in more than one group).  As with Banyan I, we impose $>$ 90\% as a high probability threshold for membership on convergent point as well.

In assessing the membership probability, we found four different categories for describing an object: 
\begin{itemize}
\item {\em Non-member, \bf{NM}}: An object that is kinematically eliminated from falling into a nearby group regardless of future astrometric measurements 
\item {\em Ambiguous member, \bf{AM}}: An object that requires updated astrometric precision because it could either belong to more than one group or it can not be differentiated from the field 
\item {\em High-likelihood member, \bf{HLM}}: An object that does not have full kinematics but is regarded as high confidence ($>$ 90\% in Banayan I, $>$ 90\% in Banyan II with $<$ 5\% contamination, $>$ 90\% in convergent point, $>$ 60\% in LACEwING) in three of the four codes, and
\item{\em Bona fide member, \bf{BM}}: An object regarded as a high-likelihood member with full kinematics (parallax, proper motion, radial velocity) demonstrating that it is in line with known higher mass bona fide members of nearby groups.
 \end{itemize}
 
 \subsection{Full Kinematic Sample\label{sec:fullkinematics}}
\begin{figure*}[!ht]
\begin{center}
\epsscale{1.2}
\includegraphics[width=.55\hsize]{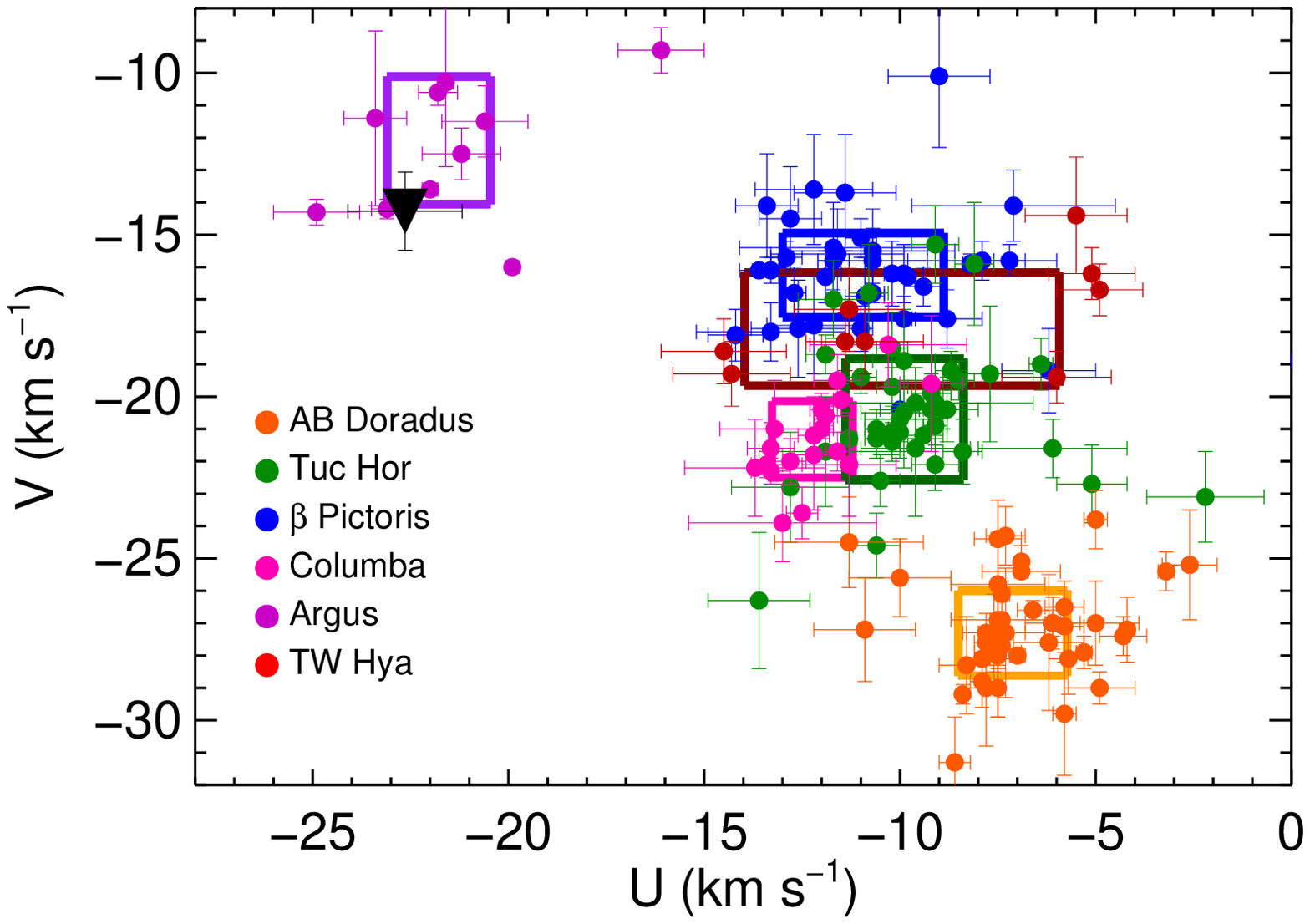}
\plottwo{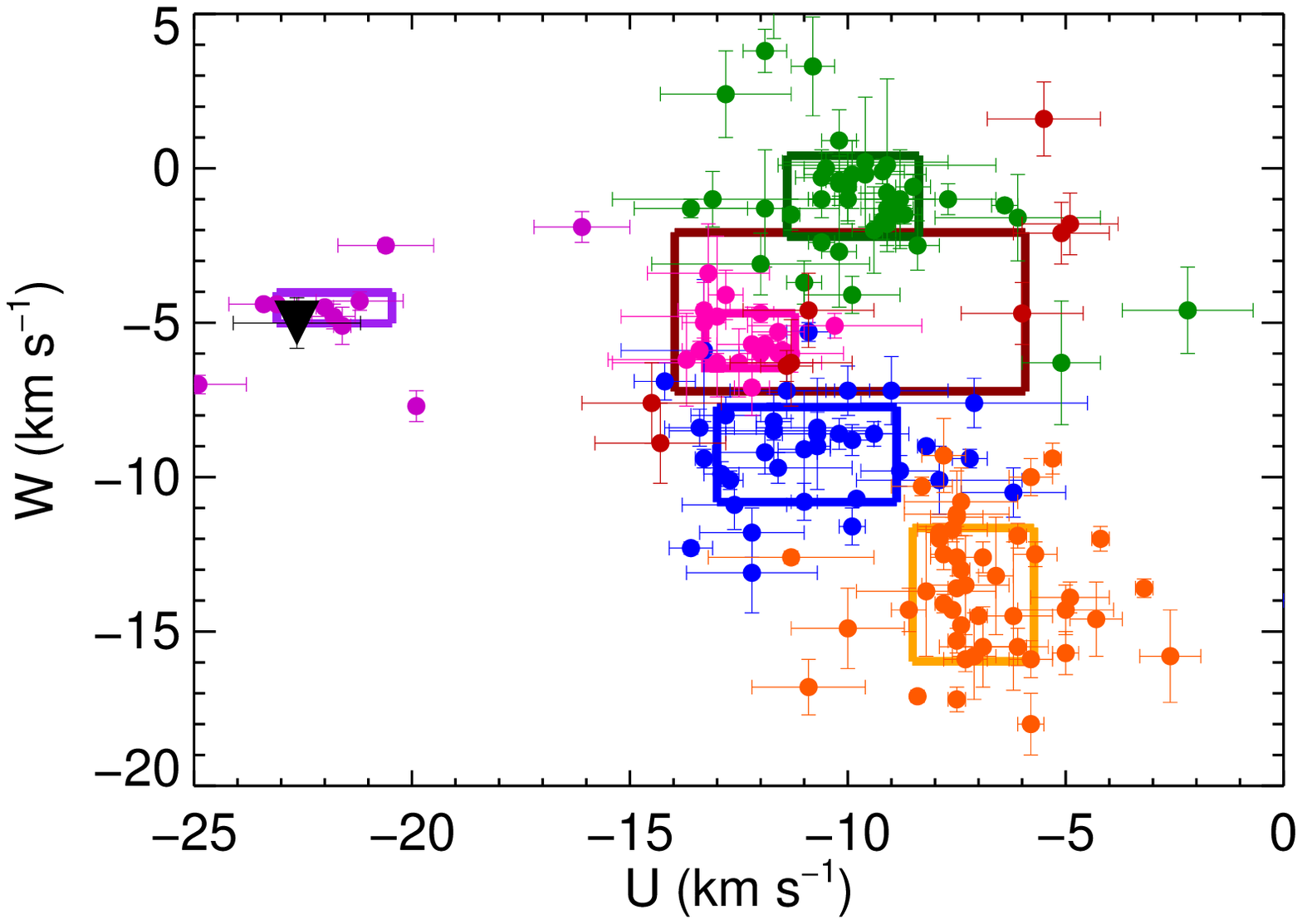}{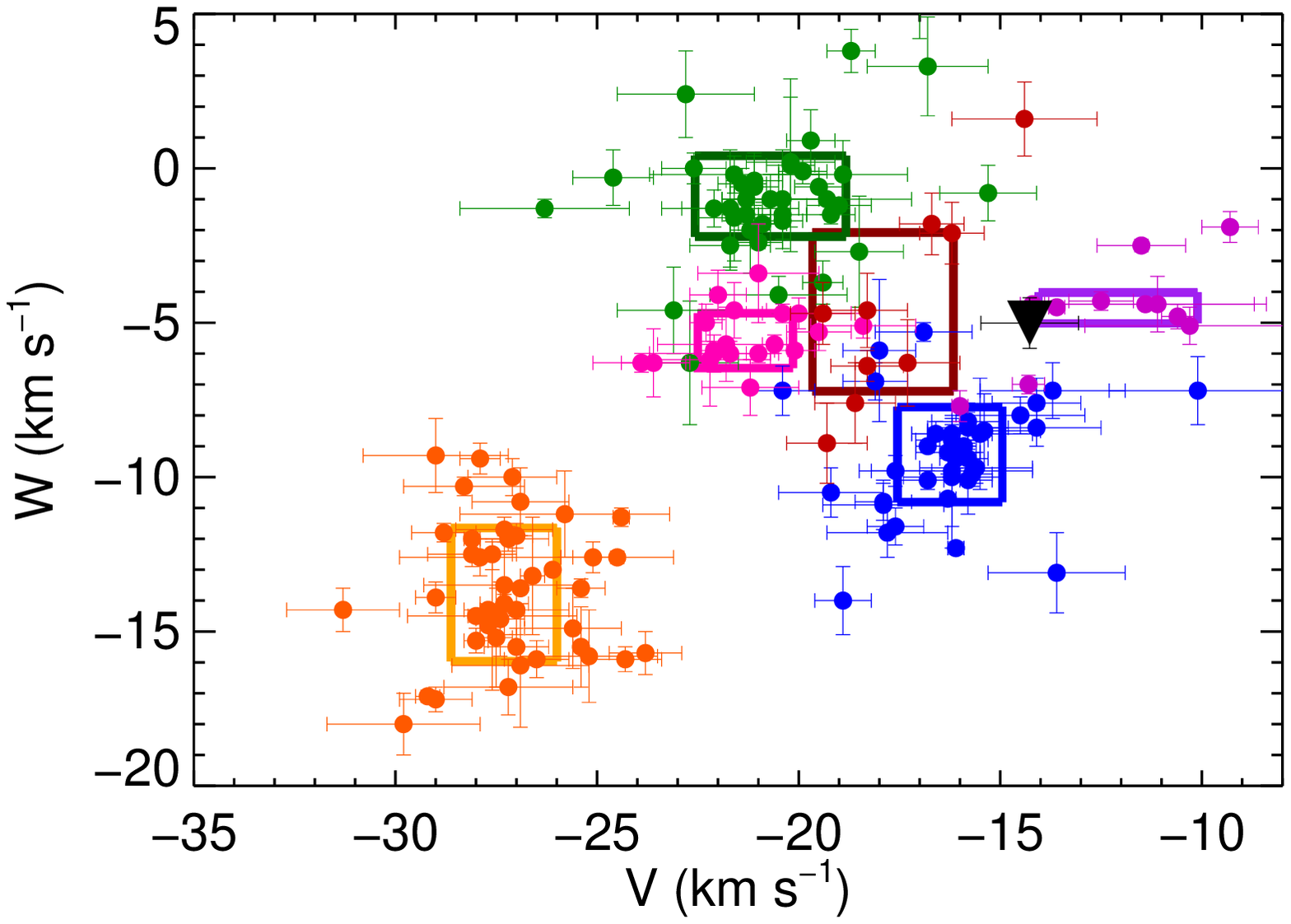}
\end{center}
\caption{The UVW properties of 0045+1634 (black filled triangle) compared to those of the members of the nearby young groups. Solid rectangles surround the furthest extent of highly probable members from \citet{Torres08} but their distribution does not necessarily fill the entire rectangle.   \label{fig:2M0045}} 
\end{figure*}

For the 28 targets with full kinematics, we compute the XYZ spatial positions and UVW velocities following the formalism of  \citet{Johnson87}, which employs U/X in the direction of the Galactic center, providing a right-handed coordinate system.  In general, the resulting values are limited by the parallax precision.   For these 28 objects, visual inspection against the positions and velocities of the bona fide members in each group (as listed by \citealt{Malo13}) gave an obvious and strong indication of membership. We used the four other methods listed above as confirmation for the visual inspection.  The XYZ spatial positions and UVW velocities for systems with full kinematic information are given in Table \ref{tab:uvw}. A visual example of the phase-space motions of 0045+1634, a new bona fide member of Argus, is shown in Figure~\ref{fig:2M0045}.   Among the full kinematic sample, we found 11 objects stand out as bona fide members, 9 objects are classified as ambiguous, and 8 are classified as non-members.  The outcome of assessing the likelihood of membership from each kinematic method is listed in Table~\ref{tab:memberships}.

 \subsection{Partial Kinematic Sample}
Having only partial kinematics for 124 objects limits our ability to definitively place these targets in a nearby group.  As stated above, the BANYAN I/II, Convergent Point, and LACEwING methods use varying techniques to yield membership probabilities.  We list the outcomes of assessing the likelihood of membership for each source in Table~\ref{tab:memberships}.  As can be seen from this tabulation, the results varied across methods.  In the case of an object like 2322-6151, all methods yield a probability of membership in the Tucana Horologium moving group with three of the 4 yielding $>$ 90\% membership.  The most difficult cases were objects like 1154-3400, where each method yielded a moderate to high probability in a different group (Banyan I and Banyan II both predict Argus, LACEwING predicts TW Hydrae, and Convergent point predicts Chameleon-near).  Our approach to the analysis was to be conservative with group membership to eliminate assigning objects to groups that were uncertain.  In all we concluded that there were 28 objects to be regarded as high likelihood members (HLM) of a known group, 83 objects that were ambiguous (AM), and 13 objects that were non-members (NM).   Adding the full kinematic sample the final tally is 28 high likelihood members (HLM), 11 bona fide group members (BM), 92 ambiguous (AM), and 21 non-members (NM).

\subsection{Comparison with Previous Works}
\label{section:comparison}
Among the 152 brown dwarfs examined in this work, 11 are newly identified as low-gravity and 141 have been previously discussed in the literature for membership in a nearby moving group (e.g. \citealt{Gagne15b, Gagne15}).   Several of the objects -- 2M0355, PSO318, 0047+6803, 1741-4642, 2154-1055, 0608-2753 -- have been the subject of single-object papers (\citealt{Faherty13}, \citealt{Liu13}, \citealt{Schneider14}, \citealt{Gagne14a, Gagne14b}, \citealt{Rice10}). 
The remaining 129 objects were examined for membership in a nearby group primarily by \citet{Gagne14} -- hereafter G14 -- , and \citealt{Gagne15b} -- hereafter G15 -- using BANYAN II.  

There are 69 objects from our sample included in G14, 73 objects in G15 and 6 in both.  In G14, there is a hierarchical probability structure that categorizes potential members as: (1) Bona fide, (2) High probability, (3) Moderate probability  and, (4) Low Probability\footnote{See the G14 paper for a detailed description of how to interpret each probability category'}.   That structure is not used in G15, but replaced by noting the probability of membership in a group (multiple groups if deemed necessary) along with its contamination potential.  

Of the 69 objects from our sample examined in G14, three objects -- 2M0355, 2M0123, and TWA 26 -- were declared bona fide members of AB Doradus, Tucana Horologium, and TW Hydrae respectively.   A further 29 objects were deemed high probability members of Argus (2), AB Doradus (5), $\beta$ Pictoris (4), Columba (4), and Tucana Horologium (14). Ten objects were deemed modest probability members of AB Doradus (3), $\beta$ Pictoris (3), Columba (1), Argus (1), and Tucana Horologium (2). Eight objects were deemed low probability members  of Argus (1), AB Doradus (1), $\beta$ Pictoris (4), and TW Hydrae (2).    There were also 16 objects designated as young field sources (aka ``no group membership possible'') and 3 objects designated as peripheral members or contaminants in a group.  

In Table~\ref{tab:memberships}, we use our new kinematics and show the predictions from BANYAN I, BANYAN II, LACEwING, the convergent point and our plurality decision based on reviewing the results from all four methods with updated astrometric measurements for many of the sources.  For the G14 overlap, we agree with 2M0355, 2M0123, and TWA 26 being considered bona fide members.  Among the high-likelihood sample from G14 we add a new radial velocity, parallax, and/or proper motion to 19 of the 29 objects and confirm 11 objects as high-likelihood members and demote 5 objects to ambiguous or non-member.  Our re-evaluation of the kinematics also leads us to demote 3 objects to ambiguous members rather than considering them high-likelihood sources in a given group. Among the moderate and low probability objects in G14,  we add new kinematics to 8 objects and find 5 remain ambiguous and 3 are demoted to non-members.  Our re-evaluation of the kinematics finds that 15 of the low or moderate probability sources are ambiguous therefore we can not say anything about membership.  The remaining 19 sources that were young field, periphery, or contaminants in G14 are ambiguous or non-members in our analysis.  

For the 67 objects in G15, we add three new kinematic points.  One object is deemed high-likelihood while the other 2 are ambiguous.  Otherwise, our re-evaluation of the kinematics leads us to classify 13 objects as high-likelihood members of groups while the remaining 54 are ambiguous (52) or non-members (2).  

In all, we find that when the Bayesian II analysis predictions find group members have a high-likelihood of membership in a single group ($>$ 99\%) while yielding a contamination probability of $<$1\%, the various kinematic methods are consistent in their predictions and we take this to mean that the source is a reliable member.

\section{Diversity of Young Brown Dwarfs}
\label{sec:diversity}
 Each one of these moving group members is a possible benchmark for examining the evolutionary properties of the brown dwarf and directly imaged exoplanet populations.  In this section we evaluate the homogeneity and diversity of the sample as a whole as well as the subsamples from each moving group.  

\subsection{Do Gravity Classifications Correspond With Age?}

In total there are 51 optically classified $\gamma$  objects  (80 infrared classified equivalents) as well as 27 optically classified $\beta$ objects ( 57 infrared equivalents). We confirm 20 (28) of the $\gamma$ objects  and 5 (7) of the $\beta$ objects respectively as high confidence or bona fide members of moving groups. There are an additional 19 (44) $\gamma$ objects and 17 (41) $\beta$ objects regarded as ambiguous members to a known group, and 12 (10) $\gamma$ objects and 5 (8) $\beta$ objects found to be non members. As stated in Section~\ref{section:sample}, $\gamma$ classified sources have spectral features indicating that they are a lower surface gravity than the $\beta$ classified objects.  Furthermore, $\beta$ classified sources are subtly but distinctively different from the field sample, indicating -- as noted in \citealt{Allers13} and \citealt{Cruz09}-- that they are also younger but not to the extent of the $\gamma$ objects.  The age calibrated sample allows us to test how well gravity features trace the age of an object.  

In Table~\ref{tab:Members}, we list the new members of each group as well as their optical and/or near-infrared spectral and gravity classification.  As stated in section~\ref{sec:fullkinematics}, nine bona fide objects have full kinematics.  For the 28 sources missing  a radial velocity or parallax but regarded as high confidence members to a group, we list the kinematically predicted radial velocity and/or parallax from BANYAN II -- checked to be consistent with LACEwING predictions -- in parentheses in Table~\ref{tab:Members}. 

In TW Hydrae and $\beta$ Pictoris, which are the two youngest groups at $\sim$ 10  and $\sim$ 20 Myr respectively, there are nine M7 or later objects, all of which have a gravity classification of $\gamma$ in both the infrared and optical.  In AB Doradus, the oldest association at $\sim$110 - 130 Myr, there are eight sources with 4 optical $\gamma$ (6 infrared) and 1 optical $\beta$ (2 infrared) objects.  Similarly, in Tucana Horologium, where we have the most number of bona fide or high-likelihood members, 20 M7 or later, there are 9 optical $\gamma$ (10 infrared) and 3 optical $\beta$ (5 infrared) objects.  

Splitting the sample of BM/HLM objects into $<$ 25 Myr ($\beta$ Pictoris, TW Hydrae), $\sim$ 40 Myr (Tucana Horologium, Columba, Argus), and $>$ 100 Myr (AB Doradus) categories, and using the default spectral type and gravity classes used in plots within this text, we find there are (9 $\gamma$, 0 $\beta$) in $<$ 25 Myr, (14 $\gamma$, 8 $\beta$) in $\sim$ 40 Myr, and (6$\gamma$, 2 $\beta$) in $>$ 100 Myr associations.  While these are still small numbers, the lack of a correlation of numbers of ($\gamma$, $\beta$) objects as a function of bin, indicates that spectral features do not correspond one to one with age.   We note that there are 25 objects that have both an infrared and optical spectral type and while 6 have differing gravity classifications, 19 are consistent with each other affirming the diversity.   Clearly, what have been assigned as gravity sensitive features are influenced by secondary parameters (see also discussions in \citealt{Allers13}, \citealt{Liu13}, G15).  

\subsection{Photometric Properties: What Do the Colors Tell Us?\label{sec:photometry}}
The majority of flux for a brown dwarf emerges in the infrared.  Interestingly,  an enormous amount of diversity among the population can be found by examining infrared colors alone (e.g. \citealt{Kirkpatrick08}, \citealt{Faherty09, Faherty13}, \citealt{Schmidt10}). 

\begin{figure*}[t!]
\center
\includegraphics[angle=0,width=1.0\textwidth]{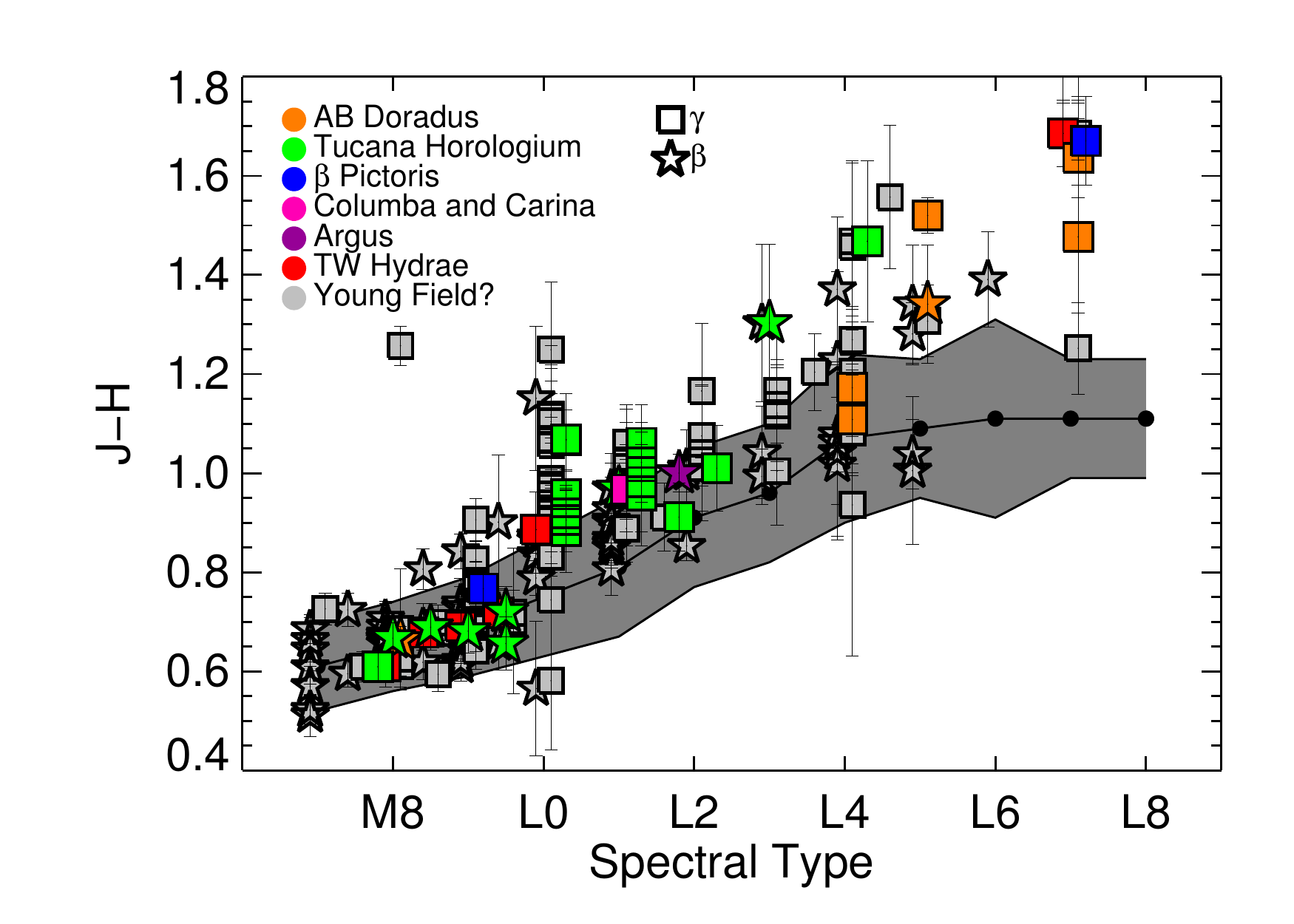}
\caption{The distribution of J-H color as a function of spectral type.  The black filled circle at each subtype is the mean value and the grey filled area marks the standard deviation spread. The isolated sources that compose this ``normal'' sample have no spectral peculiarities (e.g. subdwarfs, low-gravity, unresolved binarity), and were only included if they had a photometric uncertainty in each band $<$ 0.2 mag.  The full list was gathered from the dwarfarchives compendium and supplemented with the large ultra cool dwarf surveys from \citet{Schmidt10}, \citet{Mace13}, \citet{Kirkpatrick11}, and \citet{West08} (for M dwarfs).  Individual filled squares or five-point stars are $\gamma$ or $\beta$ (respectively) classified objects. Spectral types, as well as gravity classifications, are from optical data unless only infrared was available.   We note that most sources plotted have spectral type uncertainties of 0.5.  Objects are color coded by group assignments (or lack-thereof) discussed in this work. 2MASS photometry is used for $JHK_{s}$ bands.}
\label{fig:JmH}
\end{figure*}

\begin{figure*}[t!]
\center
\includegraphics[angle=0,width=1.0\textwidth]{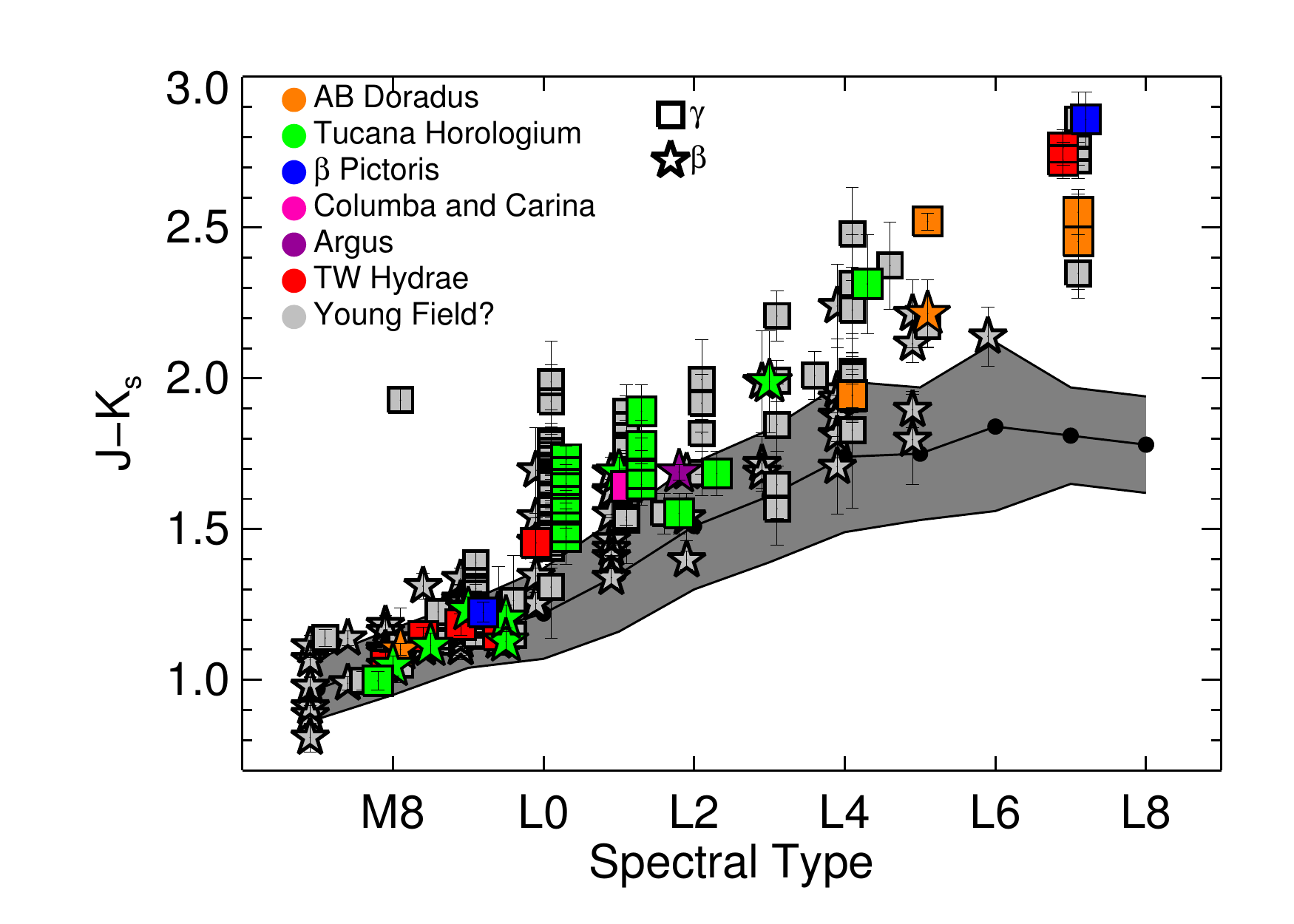}
\caption{The distribution of J-$K_{s}$ color as a function of spectral type.  Symbols are as described in Figure~\ref{fig:JmH}. }
\label{fig:JmK}
\end{figure*}

\begin{figure*}[t!]
\center
\includegraphics[angle=0,width=1.0\textwidth]{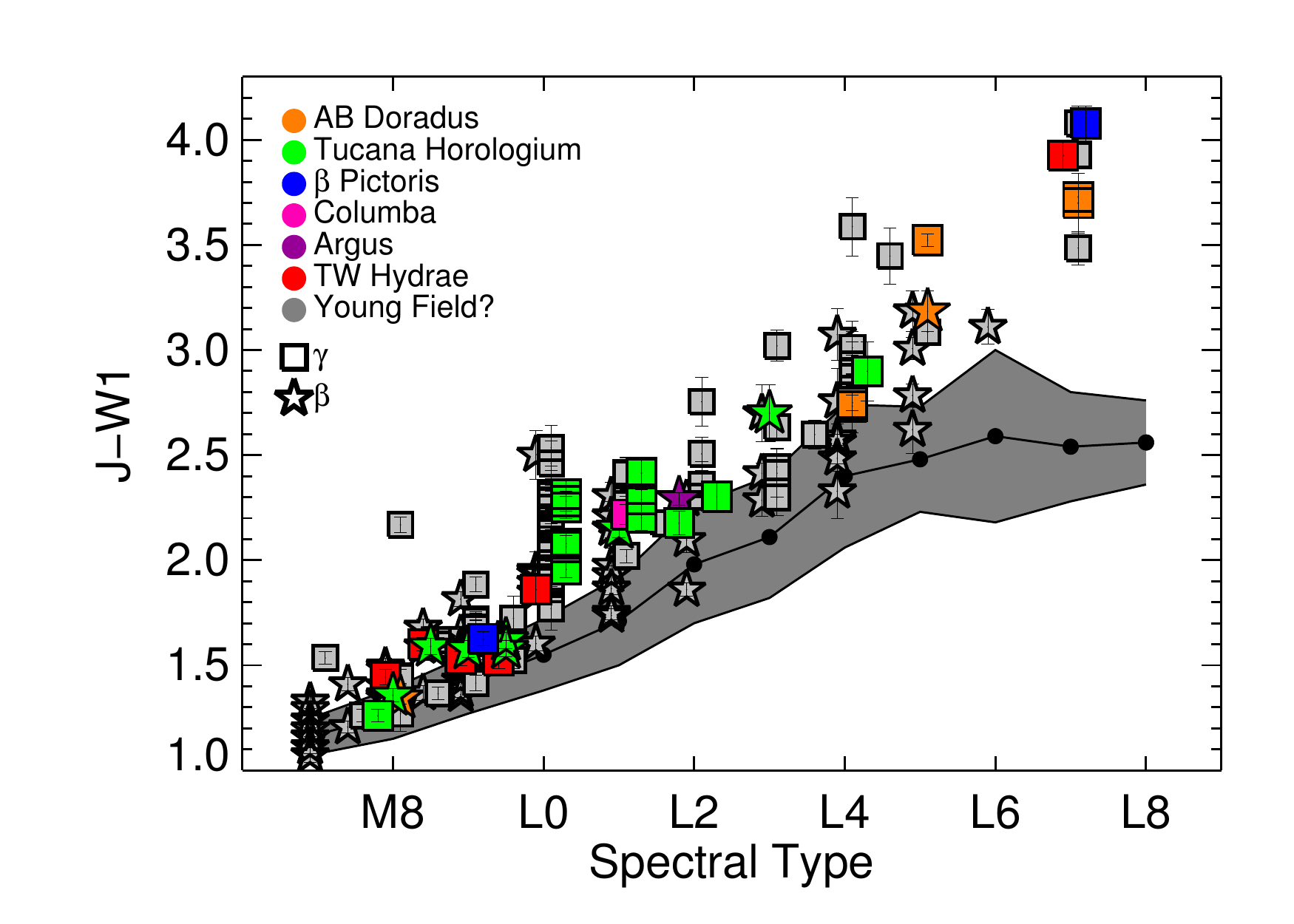}
\caption{The distribution of J-W1 color as a function of spectral type.  Symbols are as described in Figure~\ref{fig:JmH}.}
\label{fig:JmW1}
\end{figure*}

\begin{figure*}[t!]
\center
\includegraphics[angle=0,width=1.0\textwidth]{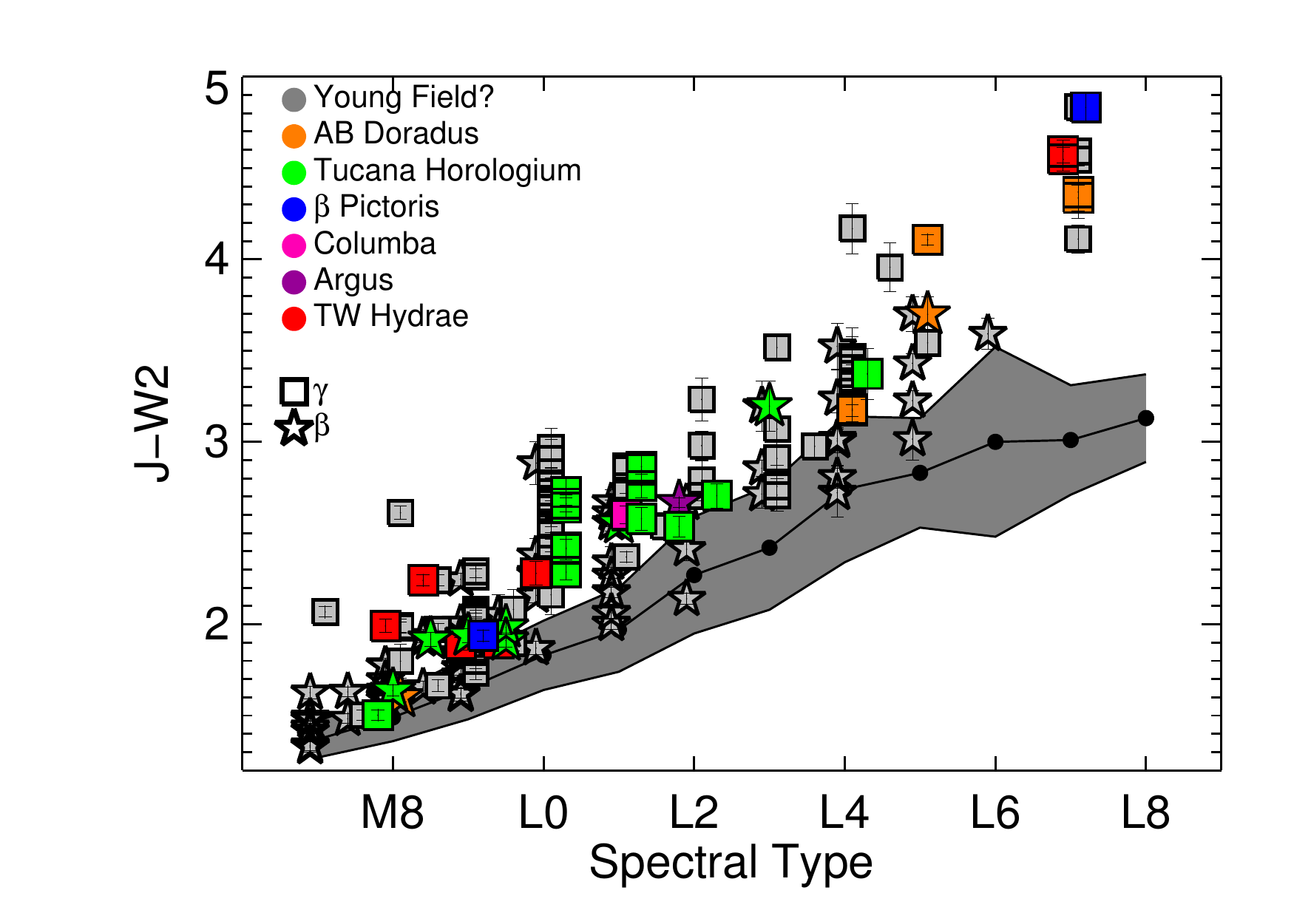}
\caption{The distribution of J-W2 color as a function of spectral type.  Symbols are as described in Figure~\ref{fig:JmH}.}
\label{fig:JmW2}
\end{figure*}

\begin{figure*}[t!]
\center
\includegraphics[angle=0,width=1.0\textwidth]{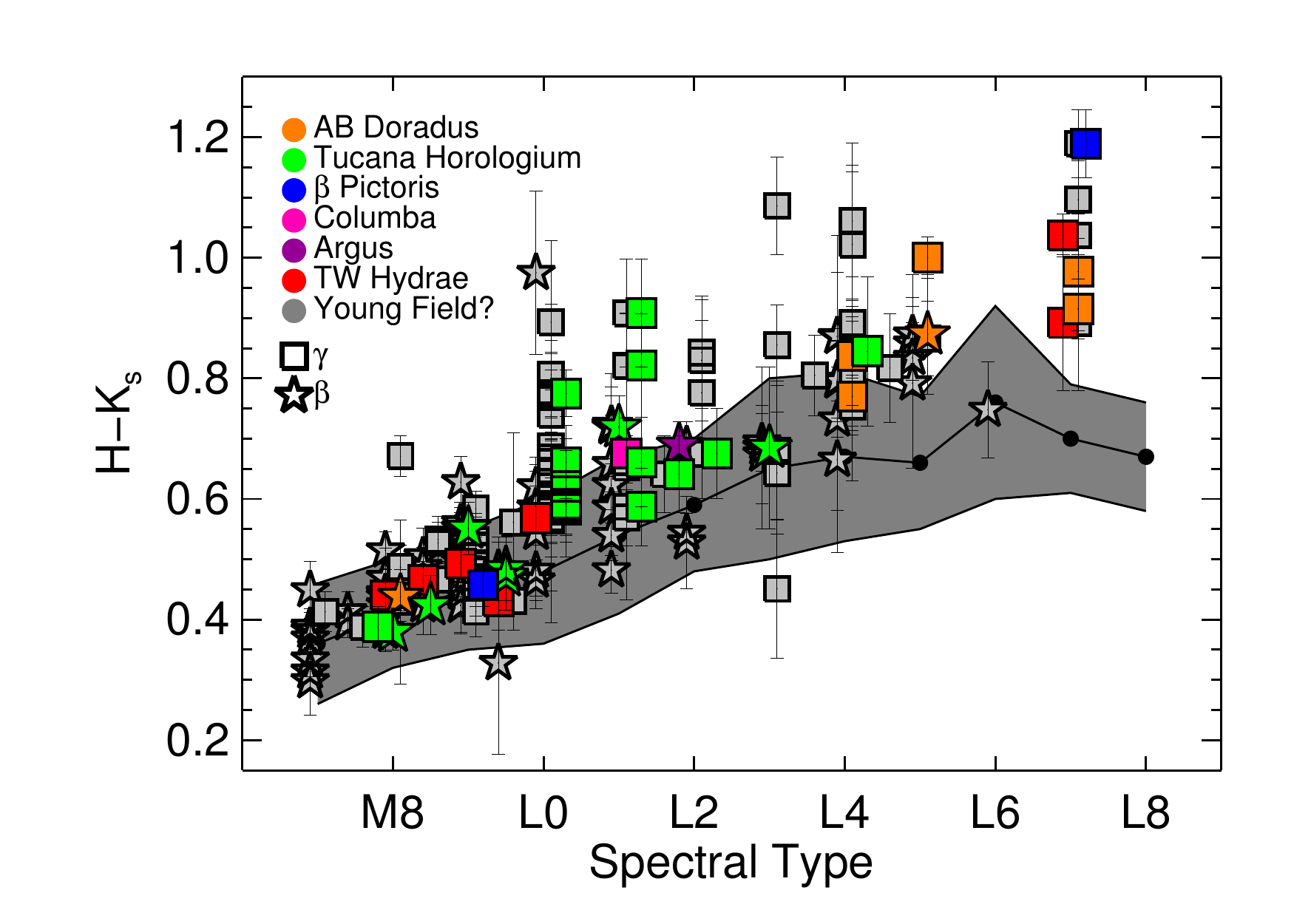}
\caption{The distribution of H-$K_{s}$ color as a function of spectral type.  Symbols are as described in Figure~\ref{fig:JmH}.}
\label{fig:HmK}
\end{figure*}

\begin{figure*}[t!]
\center
\includegraphics[angle=0,width=1.0\textwidth]{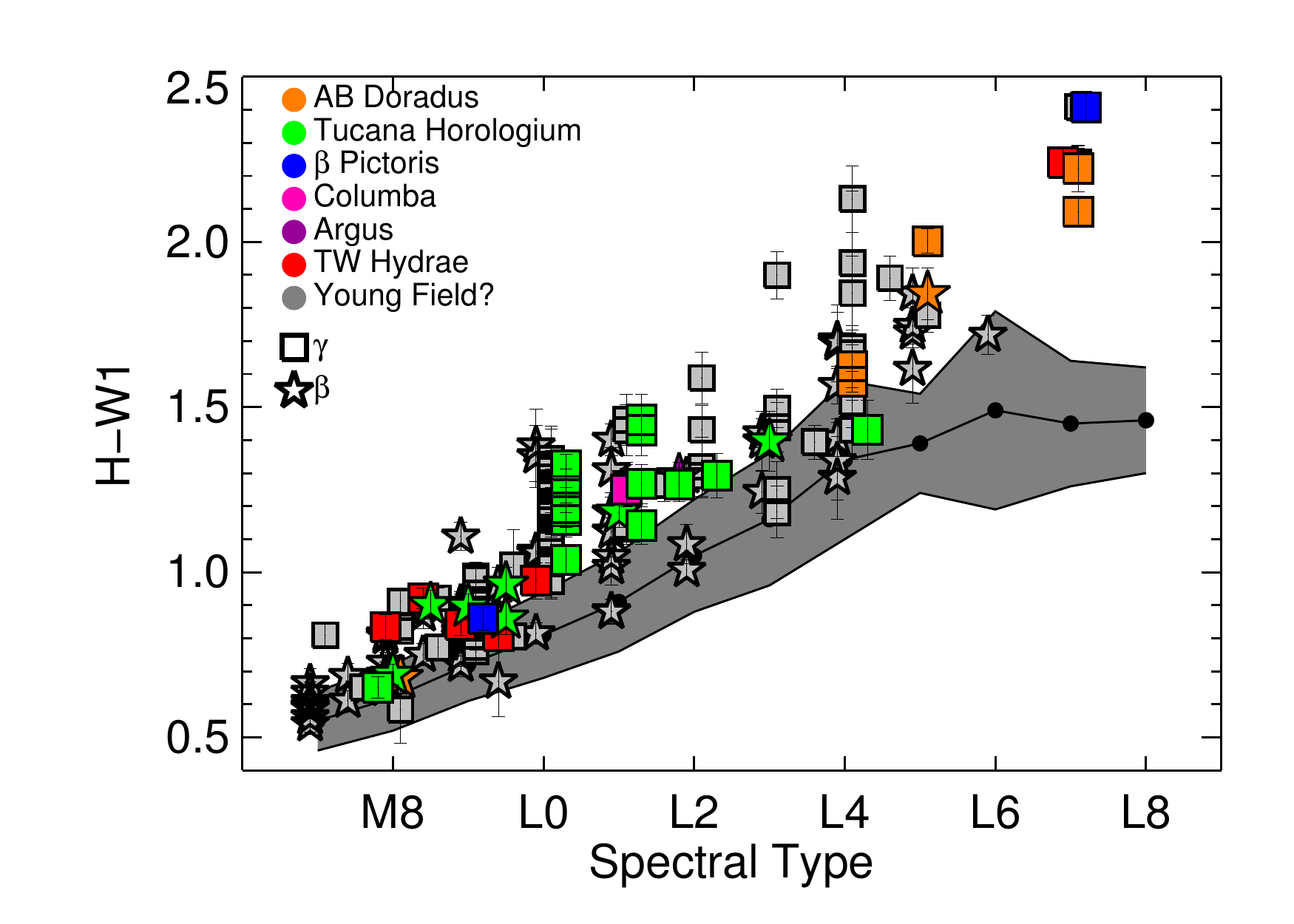}
\caption{The distribution of H-W1 color as a function of spectral type.  Symbols are as described in Figure~\ref{fig:JmH}.}
\label{fig:HmW1}
\end{figure*}

\begin{figure*}[t!]
\center
\includegraphics[angle=0,width=1.0\textwidth]{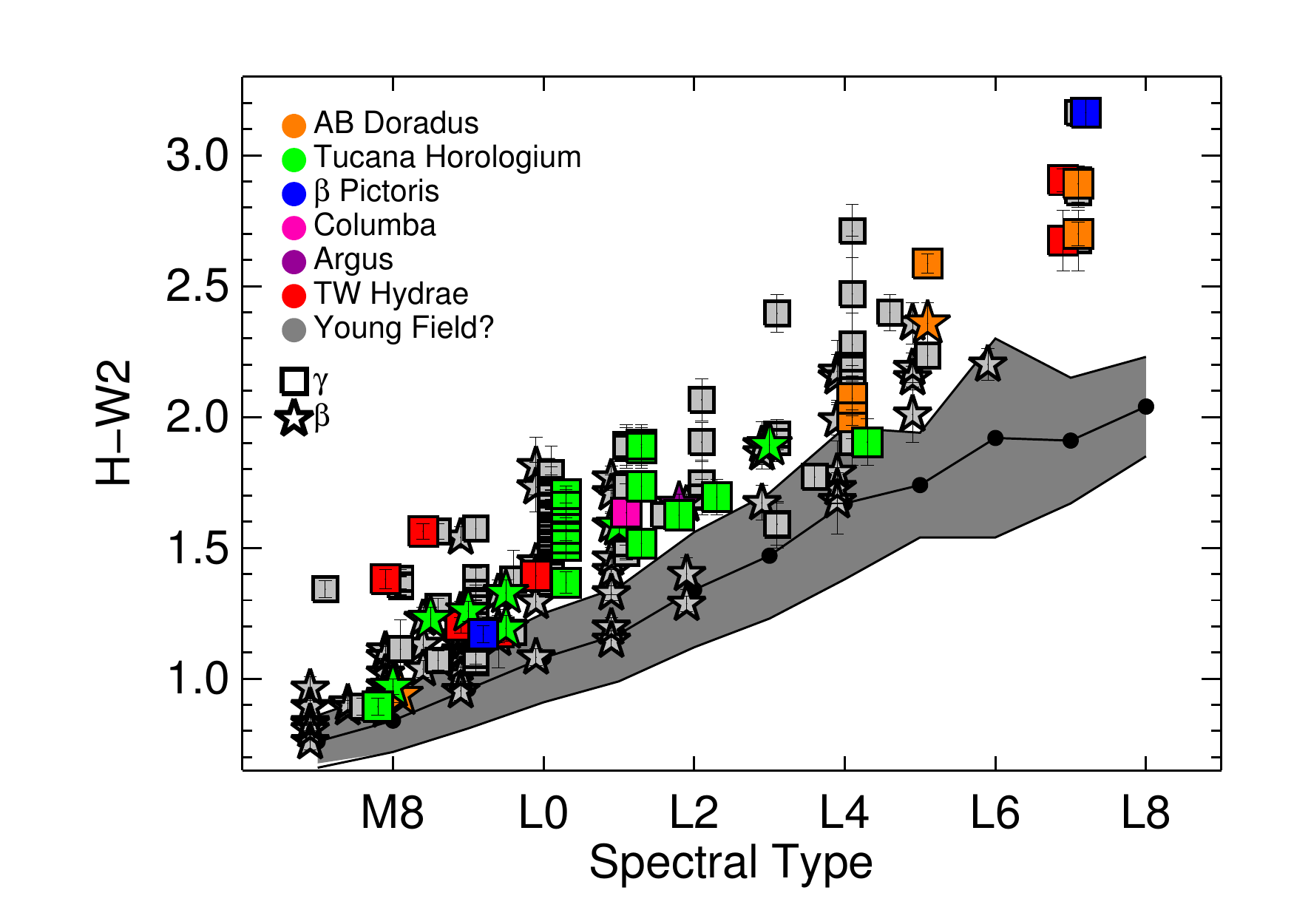}
\caption{The distribution of H-W2 color as a function of spectral type.  Symbols are as described in Figure~\ref{fig:JmH}.}
\label{fig:HmW2}
\end{figure*}

\begin{figure*}[t!]
\center
\includegraphics[angle=0,width=1.0\textwidth]{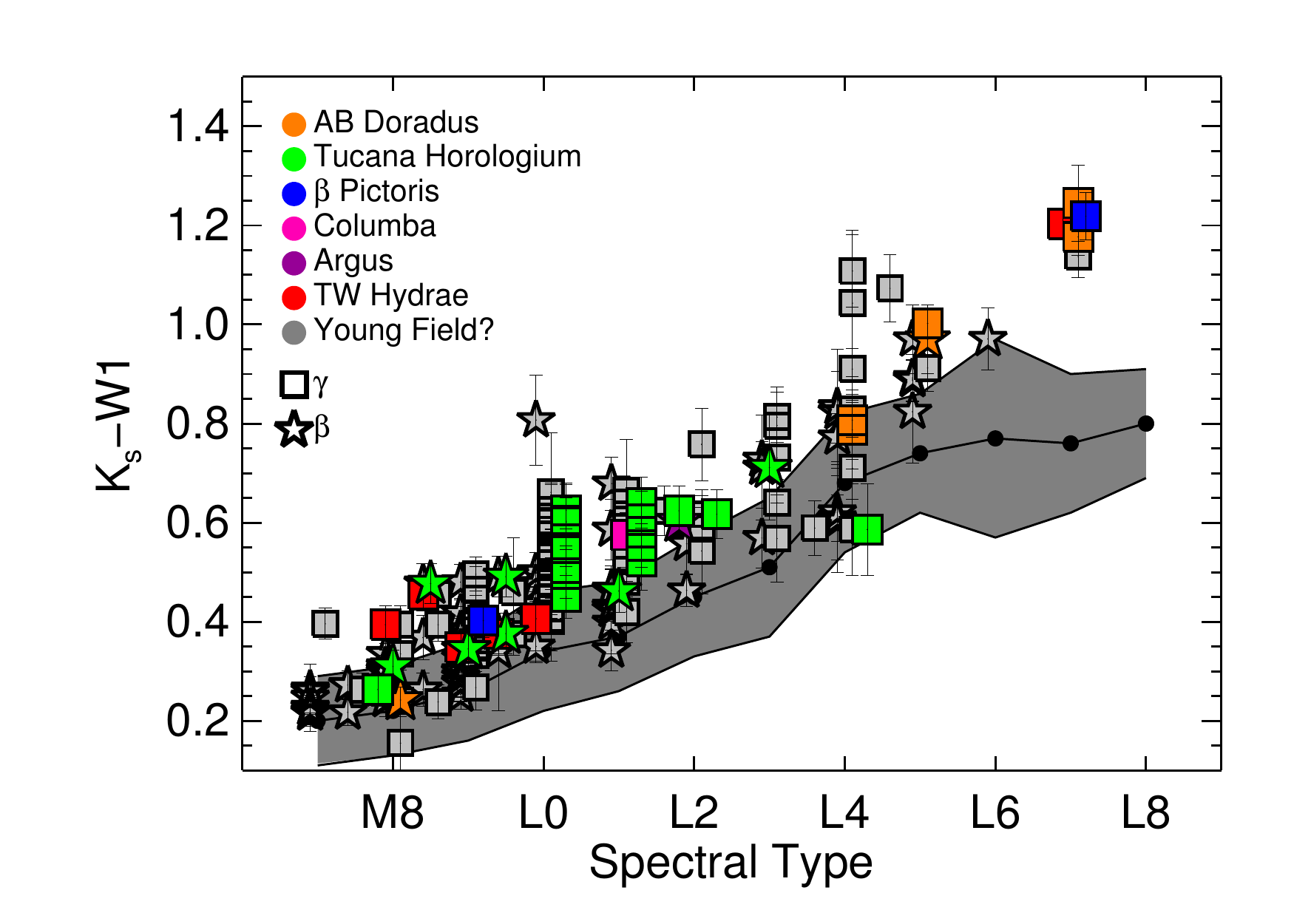}
\caption{The distribution of $K_{s}$-W1 color as a function of spectral type.  Symbols are as described in Figure~\ref{fig:JmH}.}
\label{fig:KmW1}
\end{figure*}

\begin{figure*}[t!]
\center
\includegraphics[angle=0,width=1.0\textwidth]{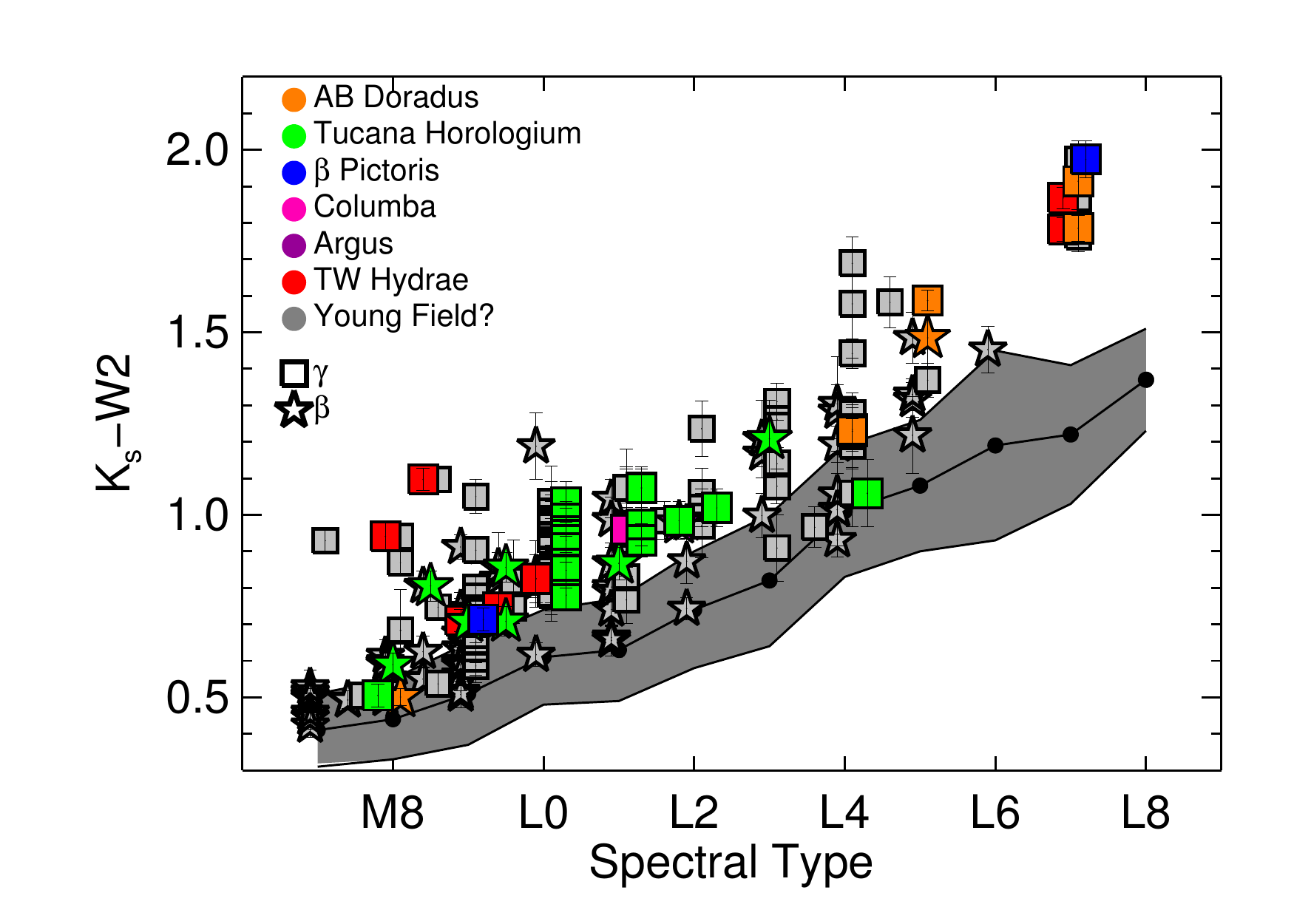}
\caption{The distribution of $K_{s}$-W2 color as a function of spectral type.  Symbols are as described in Figure~\ref{fig:JmH}.}
\label{fig:KmW2}
\end{figure*}

\begin{figure*}[t!]
\center
\includegraphics[angle=0,width=1.0\textwidth]{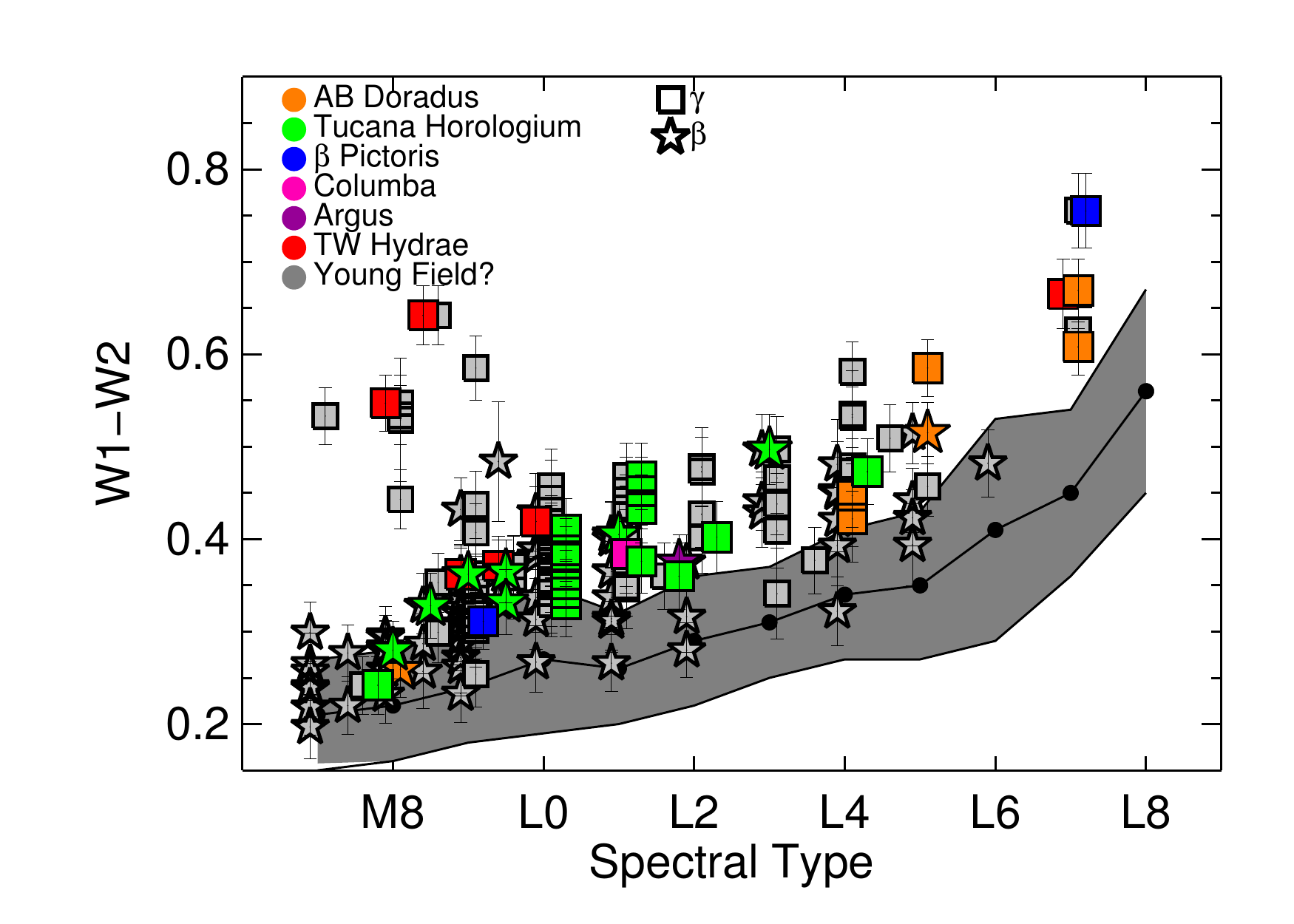}
\caption{The distribution of W1-W2 color as a function of spectral type.  Symbols are as described in Figure~\ref{fig:JmH}.}
\label{fig:W1mW2}
\end{figure*}

  As quantified in Tables~\ref{tab:meancolors1} -~\ref{tab:meancolors2} and visualized in Figures~\ref{fig:JmH} -~\ref{fig:W1mW2}, the scatter for ``normal'' sources is pronounced, especially among the mid- to late- L dwarfs. Past works have attributed this to variations in effective temperature, metallicity, age, or atmosphere conditions (\citealt{Faherty13}, \citealt{Kirkpatrick08}, \citealt{Patten06}, \citealt{Knapp04}). 

In Figures~\ref{fig:JmH} -~\ref{fig:W1mW2} we plot infrared color combinations for the population.  These can be used to examine trends or lack thereof among the new age-calibrated sample.  The spread for ``normal'' objects (grey shaded area) in Figures~\ref{fig:JmH} -~\ref{fig:W1mW2} was created by isolating sources without peculiar spectral features (e.g. subdwarfs, low-gravity objects, and unresolved binaries were eliminated),  and only keeping sources with photometric uncertainties in the color shown $<$ 0.2 mag.  The list was gathered from the dwarfarchives\footnote{http://www.dwarfarchives.org} compendium and supplemented with the large ultra cool dwarf surveys from \citet{Schmidt10}, \citet{Mace13}, \citet{Kirkpatrick11}, and \citet{West08} (for M dwarfs).  

For the low gravity sample, spectral types, as well as gravity classifications, are from optical data unless only infrared was available.   We note that most sources plotted have spectral type uncertainties of 0.5.  However, as can be seen in Table~\ref{tab:photometry}, low-gravity sources can have up to a 2 type difference in subtype, as well as differing gravity indications between the optical and the infrared.  We investigated whether isolating the smaller samples of optical only or infrared only yielded different trends than this mixed sample, but found similar results in all cases. Hence we default to optical classifications and designations where available, as this is the wavelength range where the original spectral typing schemes for this expected temperature range were created.  

Overplotted on the grey field sequences in Figures \ref{fig:JmH} -~\ref{fig:W1mW2} are the individual $\gamma$ or $\beta$ low-gravity objects.  We have color coded each source by the group it has been assigned or labeled it as ``young field?" for those with ambiguous or non conforming group kinematics.  We have also given the $\beta$ and $\gamma$ objects different symbols so their trends could be highlighted.  Tables~\ref{tab:meancolors3} -~\ref{tab:meancolors4} give the infrared color of each source as well as its deviation from the mean (see Tables~\ref{tab:meancolors1} -~\ref{tab:meancolors2}) of normal objects in its spectral subtype.  

In general, the $\gamma$ and $\beta$ sources are systematically redder than the mean for their subtype with the deviation from ``normal'' increasing with later spectral subtypes.  Objects are most deviant from normal in the ($J$-$W2$) color (Figure~\ref{fig:JmW2}) where they are an average of 2$\sigma$ (or up to 1.85 mag) redder than the subtype mean value across all types, and least deviant in the ($J$-$H$) color (Figure~\ref{fig:JmH}) where they are an average of $\sim$ 0.7$\sigma$ (or up to 0.6 mag) redder than the subtype mean value across all types.  Typically, the $\gamma$ sources mark the extreme red photometric outliers for each subtype bin, although there are a handful of extreme $\beta$ sources (e.g. 0153-6744, an infrared L3 $\beta$).  

The difference in age between the oldest moving group investigated and the average age for field sources is more than $\sim$ 3 Gyr (field age references \citealt{Burgasser15}, \citealt{Seifahrt10}, \citealt{Faherty09}).  As shown in Figures \ref{fig:JmH} -~\ref{fig:W1mW2}, the difference in photometric properties across this large age gap is distinct.  We also investigated the more subtle age difference changes to the photometric properties across the 5 - 130 Myr sample.  Isolating the sources that are confidently associated with a moving group, we conclude that there is no obvious correlation between the extreme color of an object and the age of the group. For example, 5-15 Myr TW Hydrae late-type M and L dwarfs, 25 Myr $\beta$ Pictoris spectral equivalents and 30 -50 Myr Tucana Horologium spectral equivalents have similar photometric colors.  The exceptions are TWA27A and TWA28, which are 2-5$\sigma$ redder than similar objects in several colors.  We note that \citet{Schneider12,Schneider12a} postulate that these sources may have disks hence their surroundings may be contributing to the colors.  

Comparing internally within the same moving group (assumed to have the same age and metallicity) we find that objects of the same spectral subtype show a large diversity in their photometric colors. The best example is at L0 where there are 5 Tucana Horologium members.  The objects have systematically redder colors,  but they are distributed between 1-4$\sigma$ from the field mean indicating that since this is a coeval group, diversity must be driven by yet another parameter.  We note that depending on the exact formation mechanism, metallicity variations can not be completely outruled as also contributing to the diversity among objects in the same group.

Plotted as grey symbols throughout Figures \ref{fig:JmH} - \ref{fig:W1mW2} are objects that are kinematically ambiguous or unassociated with any known group.  Many of these sources, (such as the L4$\gamma$ 1615+4953 and the very oddly reddened M8$\gamma$ 0435-1441 -- see discussion in \citealt{Allers13} and \citealt{Cruz03}), are among the reddest objects for their spectral bins and rival the associated kinematic members in their spectral and photometric peculiarities.  Conversely, there are several $\beta$ sources, such as the  L2$\beta$ 0510-1843, that fall within the normal range in each color and are only subtly spectrally different than field sources. Unfortunately, with no group association, we can not comment on the likelihood of age differences driving the diversity.   It is likely that within this sample, there are objects in moving groups not yet recognized, as well as sources that are not young but mimic low surface gravity features due to secondary parameters (e.g. atmosphere or metallicity variations).

\subsection{Flux Redistribution to Longer Wavelengths for Younger Objects\label{sec:absmags}}
\begin{figure*}[!ht]
\begin{center}
\epsscale{1.0}
\plotone{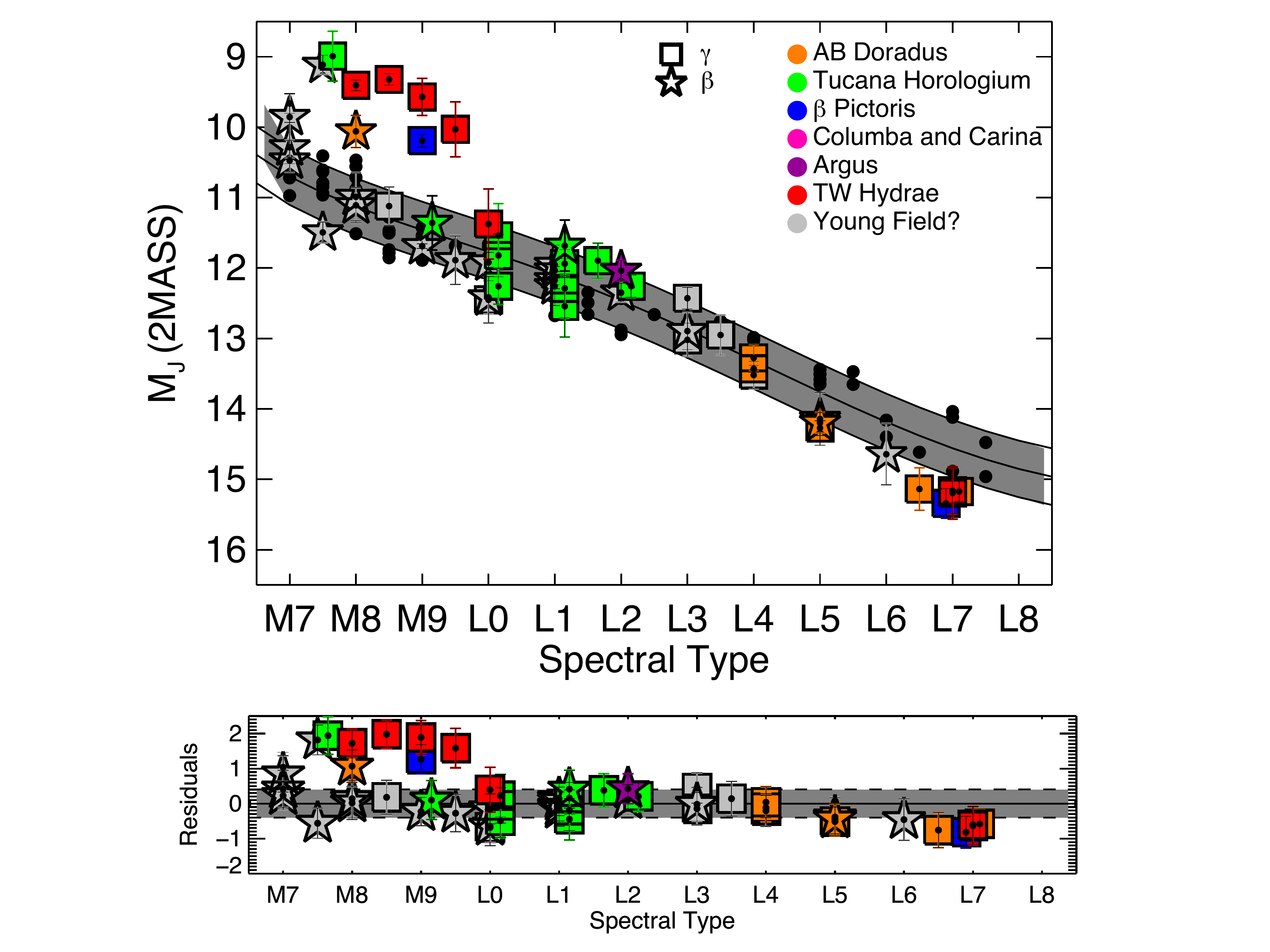}
\end{center}
\caption{The spectral type versus $M_{J}$ plot.  The field polynomial listed in Table~\ref{tab:polynomials} is represented by the grey area.  All $JHK$ photometry is from 2MASS.  Over-plotted are objects in this work with measured parallaxes or estimated kinematic distances from high confidence group membership.  Symbols distinguish very low ($\gamma$) from intermediate ($\beta$) gravity sources.  Objects are color coded by group membership.  For demonstration on the $M_{J}$ plot only, we also overplot individual field objects (with $M_{Jer}$ $<$ 0.5) as black filled circles.  Residuals of individual $\gamma$ and $\beta$ objects against the field polynomial are shown in the lower panel.  } 
\label{fig:MJ}
\end{figure*}

\begin{figure*}[!ht]
\begin{center}
\epsscale{1.0}
\plotone{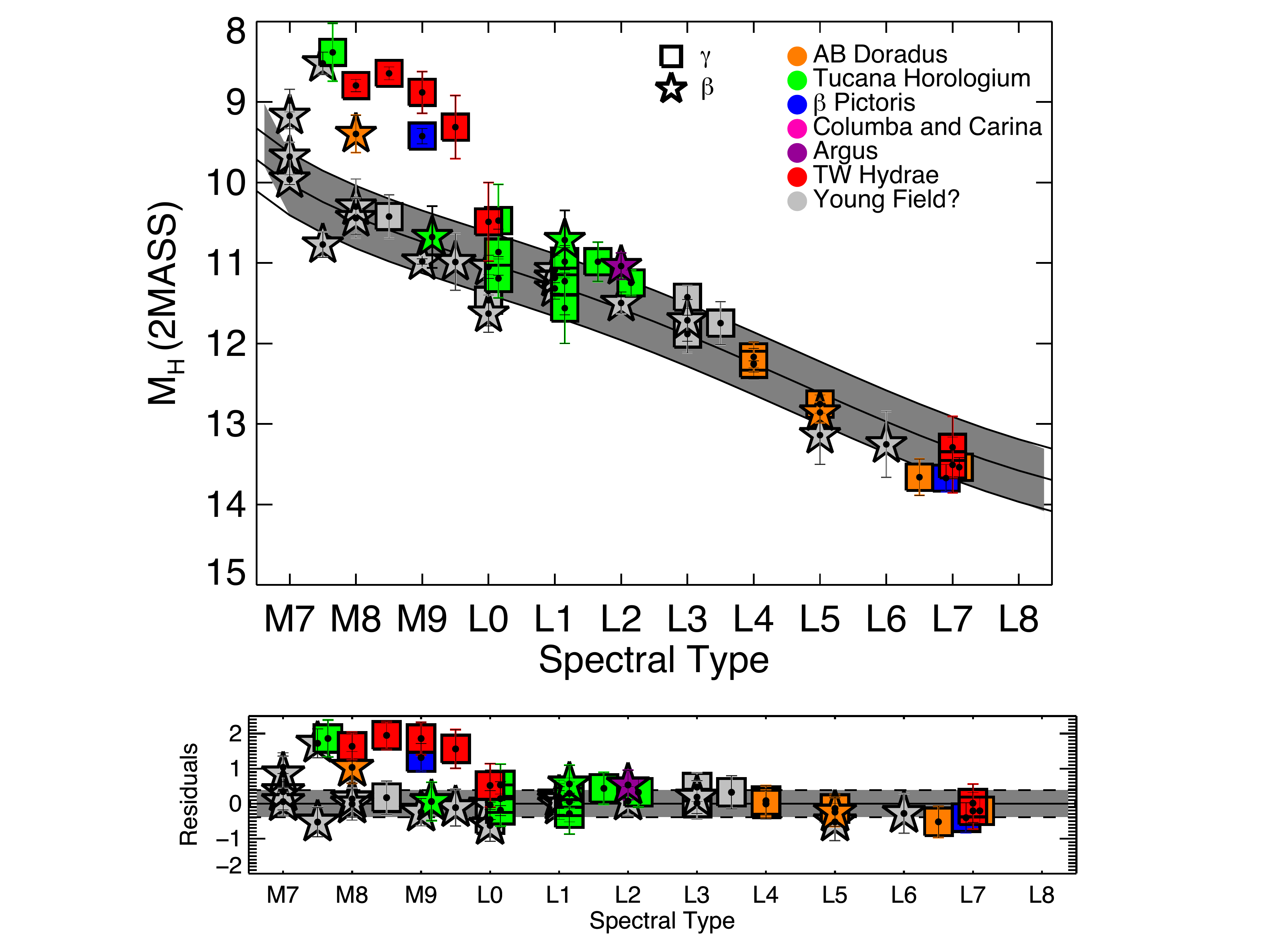}
\end{center}
\caption{The spectral type versus $M_{H}$ plot with residuals against polynomial relations (lower panel). Symbols are as described in Figure~\ref{fig:MJ}.} 
\label{fig:MH}
\end{figure*}

\begin{figure*}[!ht]
\begin{center}
\epsscale{1.0}
\plotone{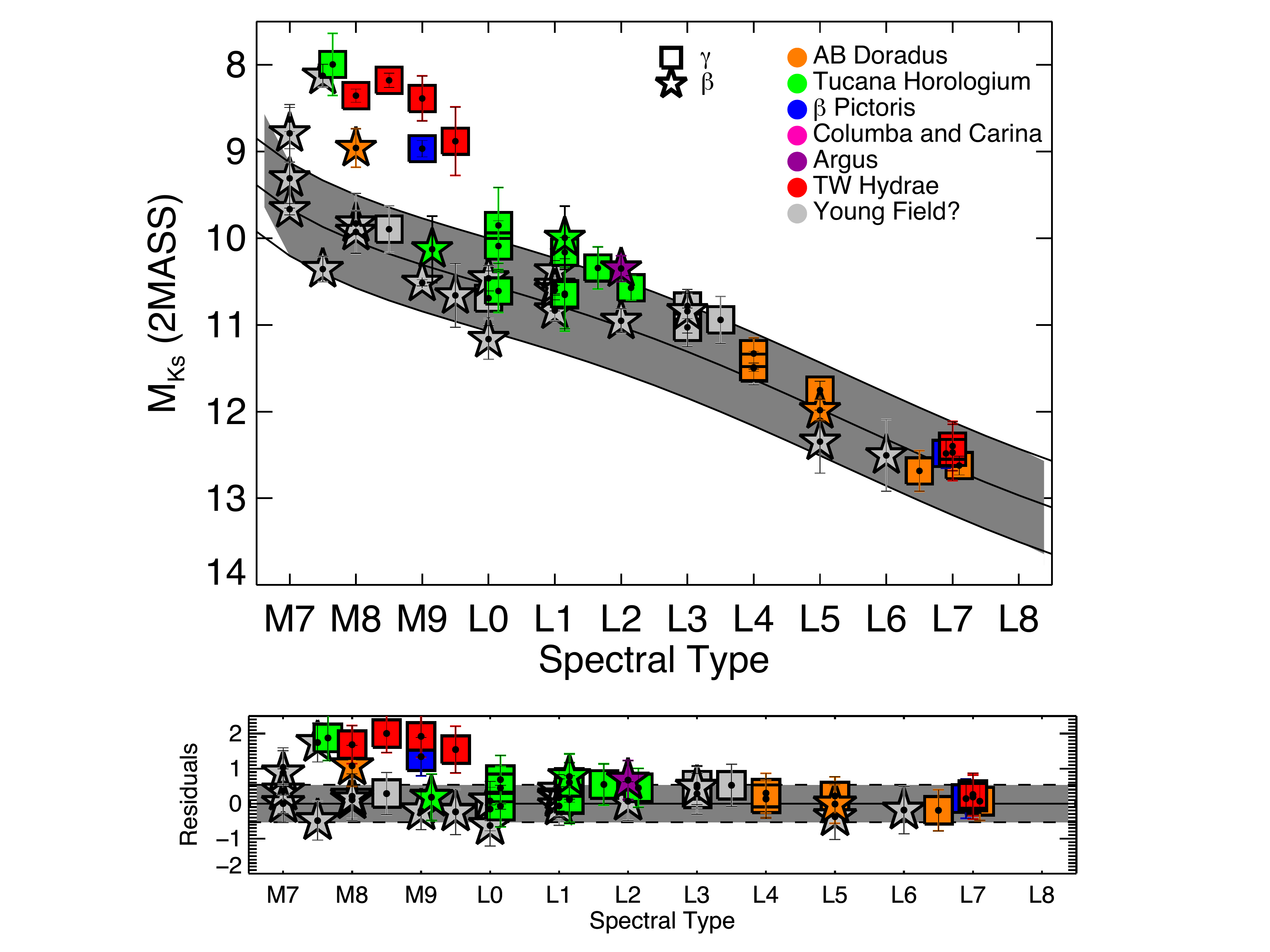}
\end{center}
\caption{The spectral type versus $M_{K_{s}}$ plot with residuals against polynomial relations (lower panel). Symbols are as described in Figure~\ref{fig:MJ}.  } 
\label{fig:MK}
\end{figure*}

\begin{figure*}[!ht]
\begin{center}
\epsscale{1.0}
\plotone{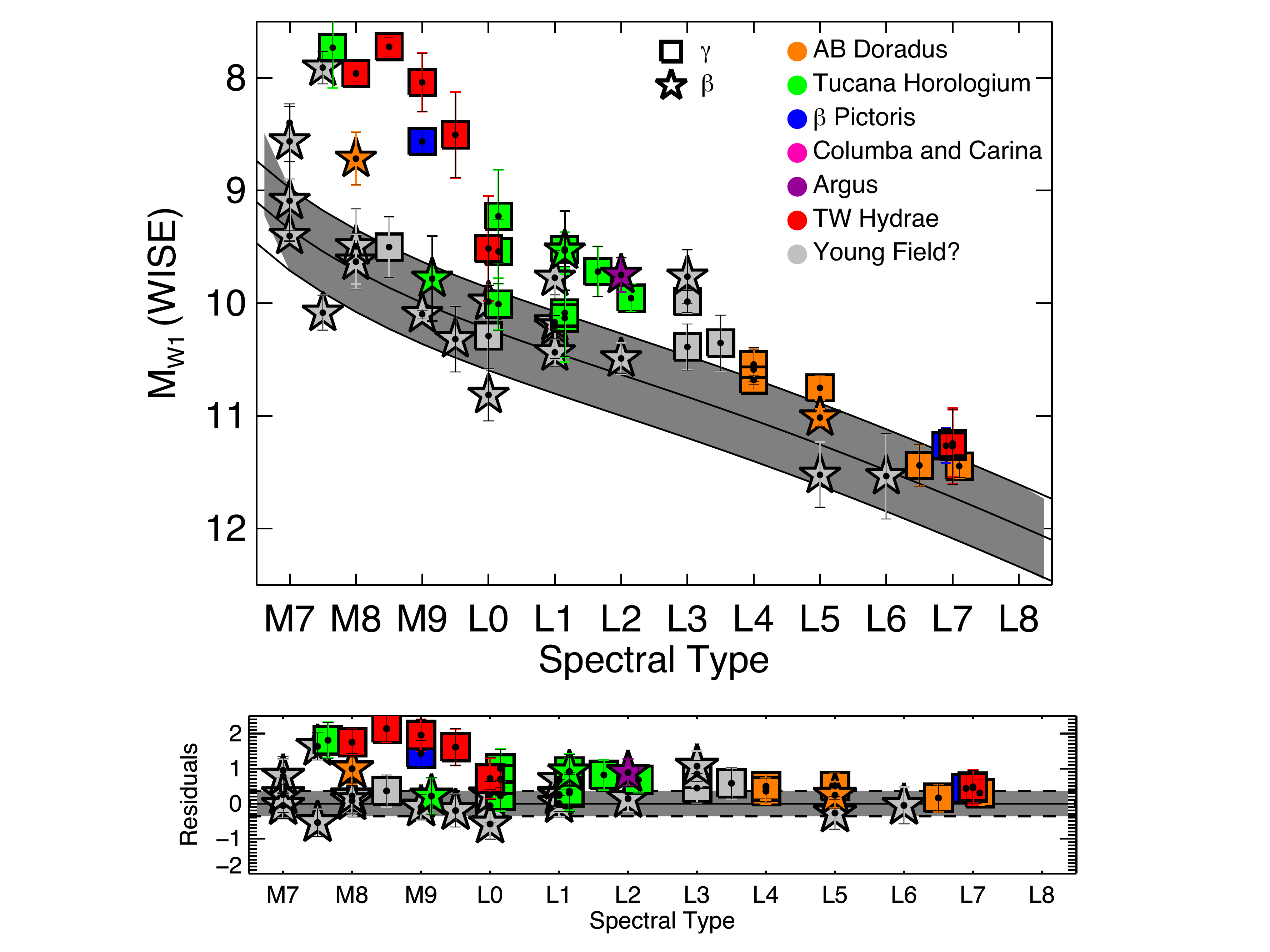}
\end{center}
\caption{The spectral type versus $M_{W1}$ plot with residuals against polynomial relations (lower panel). Symbols are as described in Figure~\ref{fig:MJ}.  } 
\label{fig:MW1}
\end{figure*}

\begin{figure*}[!ht]
\begin{center}
\epsscale{1.0}
\plotone{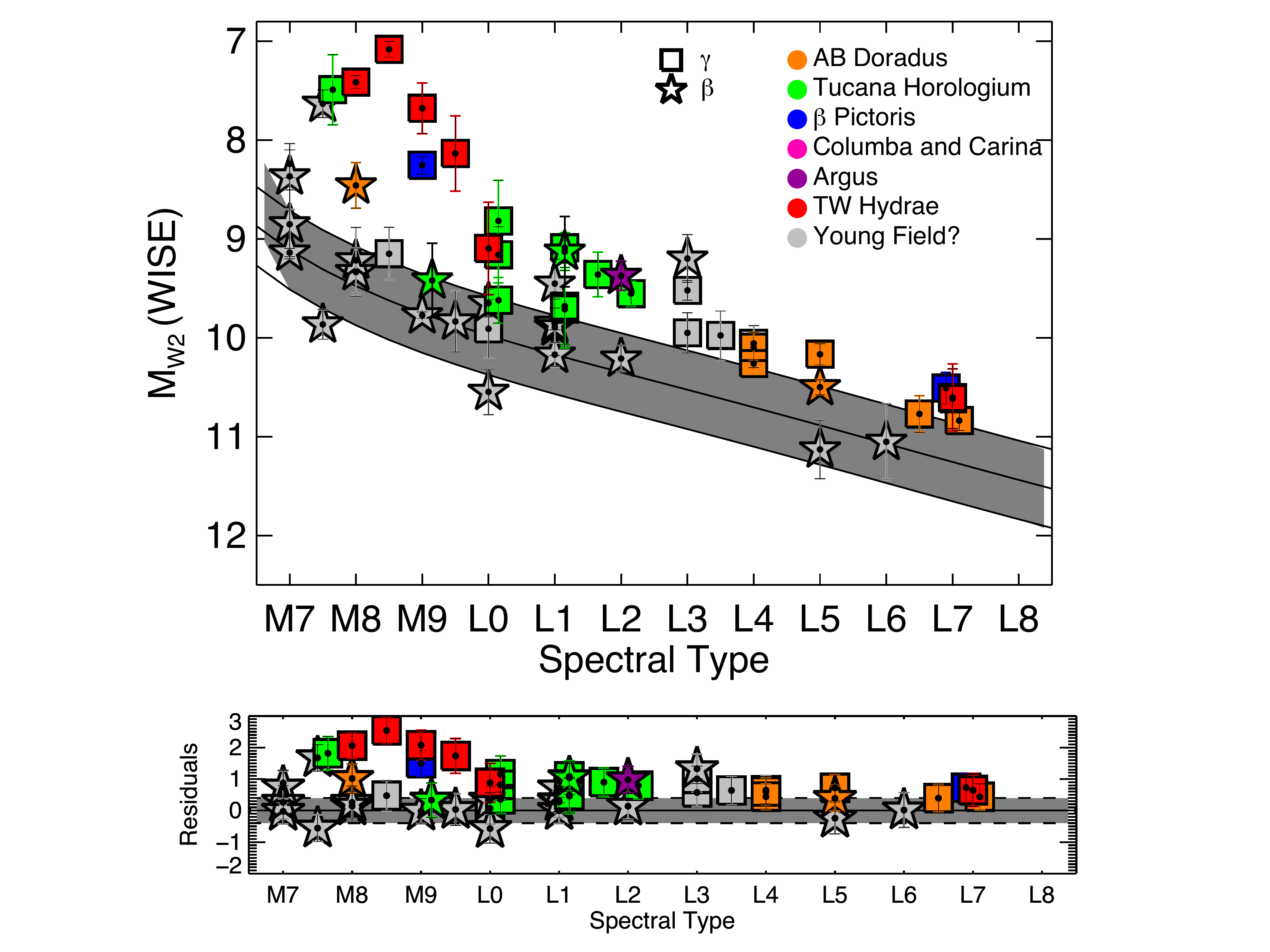}
\end{center}
\caption{The spectral type versus $M_{W2}$ plot with residuals against polynomial relations (lower panel). Symbols are as described in Figure~\ref{fig:MJ}.  } 
\label{fig:MW2}
\end{figure*}

\begin{figure*}[!ht]
\begin{center}
\epsscale{1.0}
\plotone{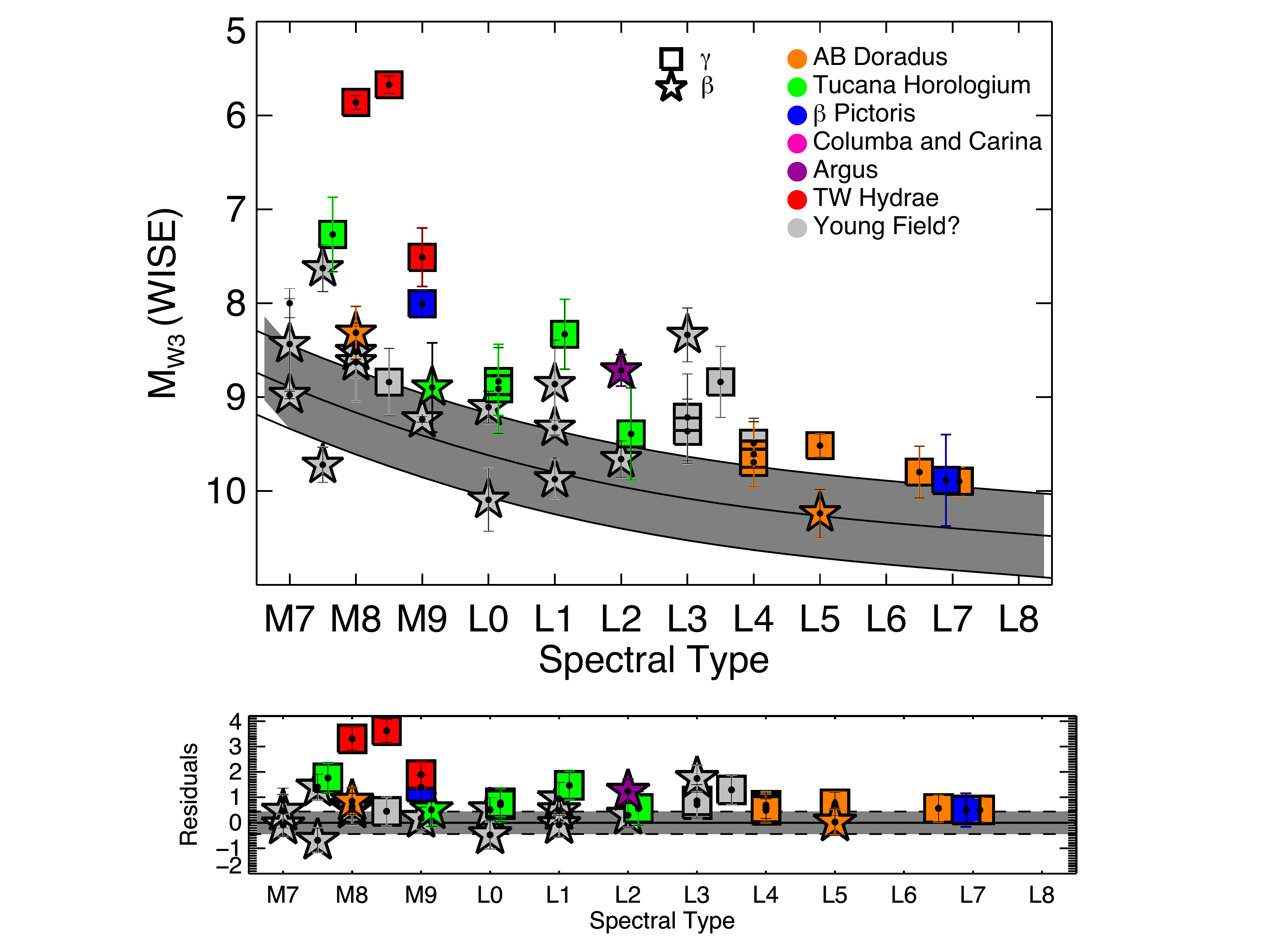}
\end{center}
\caption{The spectral type versus $M_{W3}$ plot with residuals against polynomial relations (lower panel). Symbols are as described in Figure~\ref{fig:MJ}.} 
\label{fig:MW3}
\end{figure*}

As discussed in section~\ref{sec:photometry} above, low surface gravity brown dwarfs are systematically redder for their given spectral types. Hence, one might expect that they would not logically follow the absolute magnitude sequence of field equivalents.  Figures~\ref{fig:MJ} -~\ref{fig:MW3} show $M_{J}$ through $M_{W3}$ versus spectral type for all low surface gravity sources with parallaxes or, in a select few cases, with kinematic distances.  In Table~\ref{tab:Members} we mark the 25 sources for which we lack a parallax but have assumed the kinematic distance to a moving group since the object was assessed as a high-likelihood member.  The grey area throughout each figure is the polynomial relation for the field population at each band recalculated with all known brown dwarfs with parallaxes. We list all new relations for both field and low gravity objects used in or calculated for this work in Table~\ref{tab:polynomials}.  Throughout Figures ~\ref{fig:MJ} -~\ref{fig:MW3}, individual low surface gravity objects are over plotted and color coded by group membership and given a symbol representing their gravity designation. The lower portion of each figure shows the deviation of each low-gravity source from the mean absolute magnitude value of the spectral subtype. 

\subsubsection{Trends with Spectral Type}
Focusing on the objects associated with known groups, there is a distinct difference between the behavior of low-gravity late-type M dwarfs and L dwarfs.  In Figure~\ref{fig:MJ},  which shows the $M_{J}$ band trend, the TW Hydrae and Tucana Horologium late-type M's are $\sim$ 2 magnitudes brighter than the field relations whereas the L dwarfs are normal to $\sim$1 mag fainter than the field relations.  Moving to longer wavelengths, the flux shifts. By $M_{W3}$, nearly all sources regardless of spectral type have brighter absolute magnitudes than the field polynomial.  One plausible explanation for this redistribution of flux is dust grains in the photosphere that absorb and reradiate at cooler temperatures (hence longer wavelengths).  Equally likely is the possibility that there exists thicker clouds or that there are higher lying clouds in the atmospheres of these sources (e.g. \citealt{Marocco13}, Hiranaka et al. 2015, \citealt{Faherty13}).

One main consequence of the young sources deviating from the field in some bands and not in others, is that the polynomial relations that use spectral type and photometry to obtain distances, are inappropriate for low-surface gravity objects.  At bluer near-infrared bands, they would over estimate distances, whereas at redder near-infrared bands they would under-estimate.  In Table~\ref{tab:polynomials} we have taken this into account and present new spectrophotometric polynomials for suspected young sources at J through W3 bands. As discussed in \citet{Filippazzo15}, the flux redistribution hinges around $K$ band.  As a result, we recommend this spectral distance polynomial relation for suspected young sources.

\subsubsection{Trends with Age}
Overall, we find that there is a clear difference in the behavior of the absolute magnitudes as a function of spectral type for the $>$ $\sim$ 3 Gyr field trends in each band compared to the behavior of the low surface gravity sources.  Narrowing in on the 5 - 130 Myr range and comparing equivalent spectral type sources in differing groups (such as the L7 sources in AB Doradus and $\beta$ Pictoris or the Tucana Horologium late- M and L dwarfs) we find that there is no obvious correlation with age and absolute magnitude trend.  In the case of PSO318 ($\sim$ 25 Myr), 1119-1137, and 1147-2040 ($\sim$ 5 -15 Myr) versus 0047+6803 ($\sim$ 110-130 Myr) or TWA 27A ($\sim$ 5-15 Myr) versus 0123-6921 ($\sim$30-50 Myr), the sources switch in brightness depending on the band, but stay within 1$\sigma$ of each other from $J$ ($\sim$ 1.25$\mu$ m) through W2 ($\sim$ 4.6$\mu$ m).  By $M_{W3}$ ($\sim$ 11.56$\mu$ m), TWA 27 A is over 1 magnitude brighter than 0123-6921 although this might be due to a disk and not due to the source (\citealt{Schneider12,Schneider12a}).  Regardless, it does not appear that the younger sources show an extreme version of the overall trend indicating that whatever causes the flux redistribution compared to the field ($>$ 3 Gyr sample) has a near equal impact from $\sim$5-15 Myr through $\sim$110-130 Myr.  

\subsubsection{Trends with Non-Group Members and Expanded Explanations for Diversity\label{sec:nonmembers}}
The sources with non-conforming group kinematics (grey points) do not all trace the behavior of the bona fide/high-likelihood members.  For instance, all but 2 of the $\gamma$ or $\beta$ ``Young Field?'' M7-L1 objects stay within the polynomial for each band.   Furthermore, all but one of those are classified as $\beta$, which is the more subtly altered gravity type. Conversely, the L3 and L4 $\gamma$ and $\beta$ sources move from  within the field polynomial band to being 1-2$\sigma$ brighter than equivalent sources between $M_{J}$ and $M_{W3}$. 

There are several explanations for why a spectroscopically classified low gravity object looks normal in other parameters.   Photometric variability may contribute slightly to their position on spectrophotometric diagrams (e.g. \citealt{Allers16}) but it is unlikely to contribute in a significant way.  As shown in works such as \citet{Radigan14}, and \citet{Metchev15}, large amplitude photometric variations ($>$ 5\%) among brown dwarfs are rare.  Alternatively, rotational velocity could contribute in a substantial way because it influences global circulation on a given source which causes or disrupts cloud patterns.  In this same vein, the distribution of clouds by latitude on a given object may not be homogeneous in structure or grain size. Consequently, (as first proposed by \citealt{Kirkpatrick10}) our viewing angle (e.g. pole-on or equatorial on) would impact the spectral and photometric appearance.  Unfortunately, there is little information on the rotational velocity distribution of the young brown dwarf sample so testing this parameter will require additional data.    

The simplest explanation is that sources falling within ``normal'' absolute magnitude and luminosity plots with non-conforming kinematics to any known group may not be young.  \citet{Aganze16} showed this to be the case for the d/sdM7 object GJ 660.1B that had a peculiar near infrared spectrum which hinted that it was young.  However this object was co-moving with a higher mass, low-metallicity star refuting that suggestion.  In the case of GJ 660.1 B, a low-metallicity likely helped to mimic certain spectral features of youth. In lieu of the fact that there may be some older contaminants in the sample, we present all new relations in Table~\ref{tab:polynomials} to be inclusive of all objects in this work with parallaxes as well as only objects that are considered bonafide or high likely members of groups.  

\subsection{Color Magnitude Trends for Young Brown Dwarfs\label{sec:CMDs}}
\begin{figure*}[!ht]
\begin{center}
\epsscale{1.0}
\plottwo{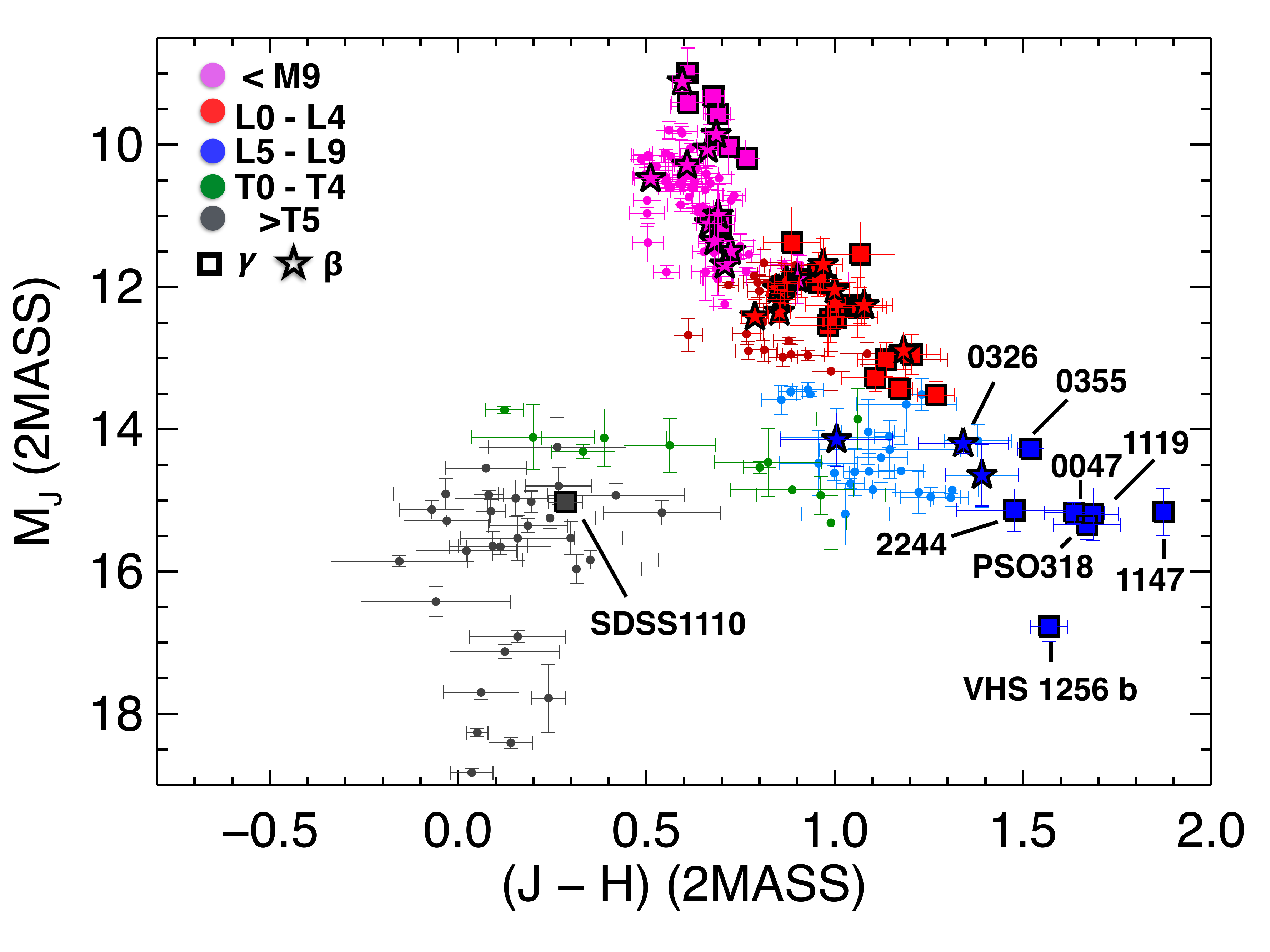}
{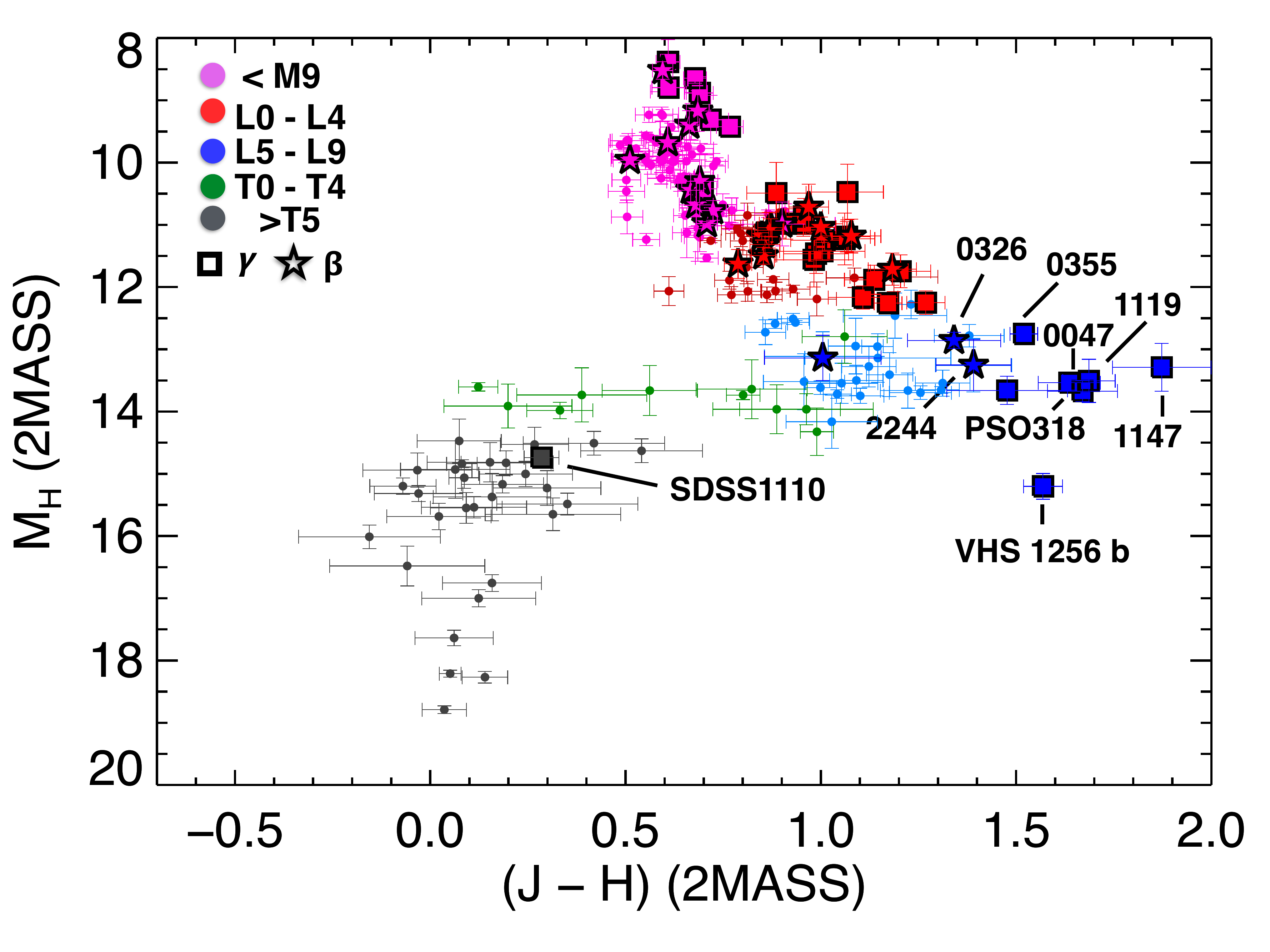}
\end{center}
\caption{The ($J$-$H$) versus $M_{J}$ (left) and $M_{H}$ (right) color magnitude diagram for late-type M through T dwarfs (Y dwarfs where photometry is available).  All $JHK$ photometry is on the MKO system.  Objects have been color coded by spectral subtype. Binaries, subdwarfs, spectrally peculiar sources and those with absolute magnitude uncertainties $>$ 0.5 have been omitted. Low-gravity objects are highlighted as bold filled circles throughout.  Objects of interest discussed in detail within the text have been labeled. \label{fig:JvJmH}} 
\end{figure*}

\begin{figure*}[!ht]
\begin{center}
\epsscale{1.0}
\plottwo{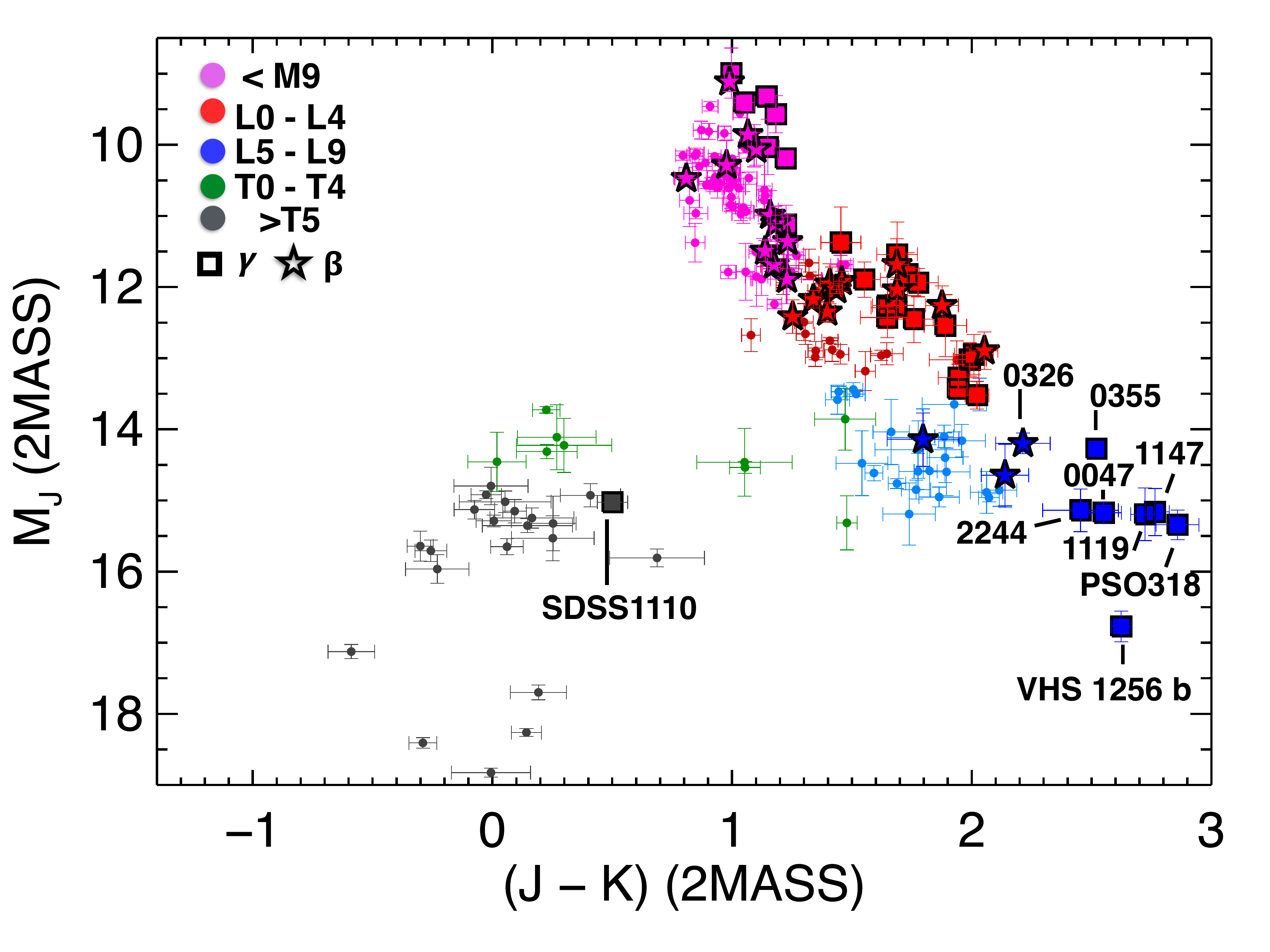}
{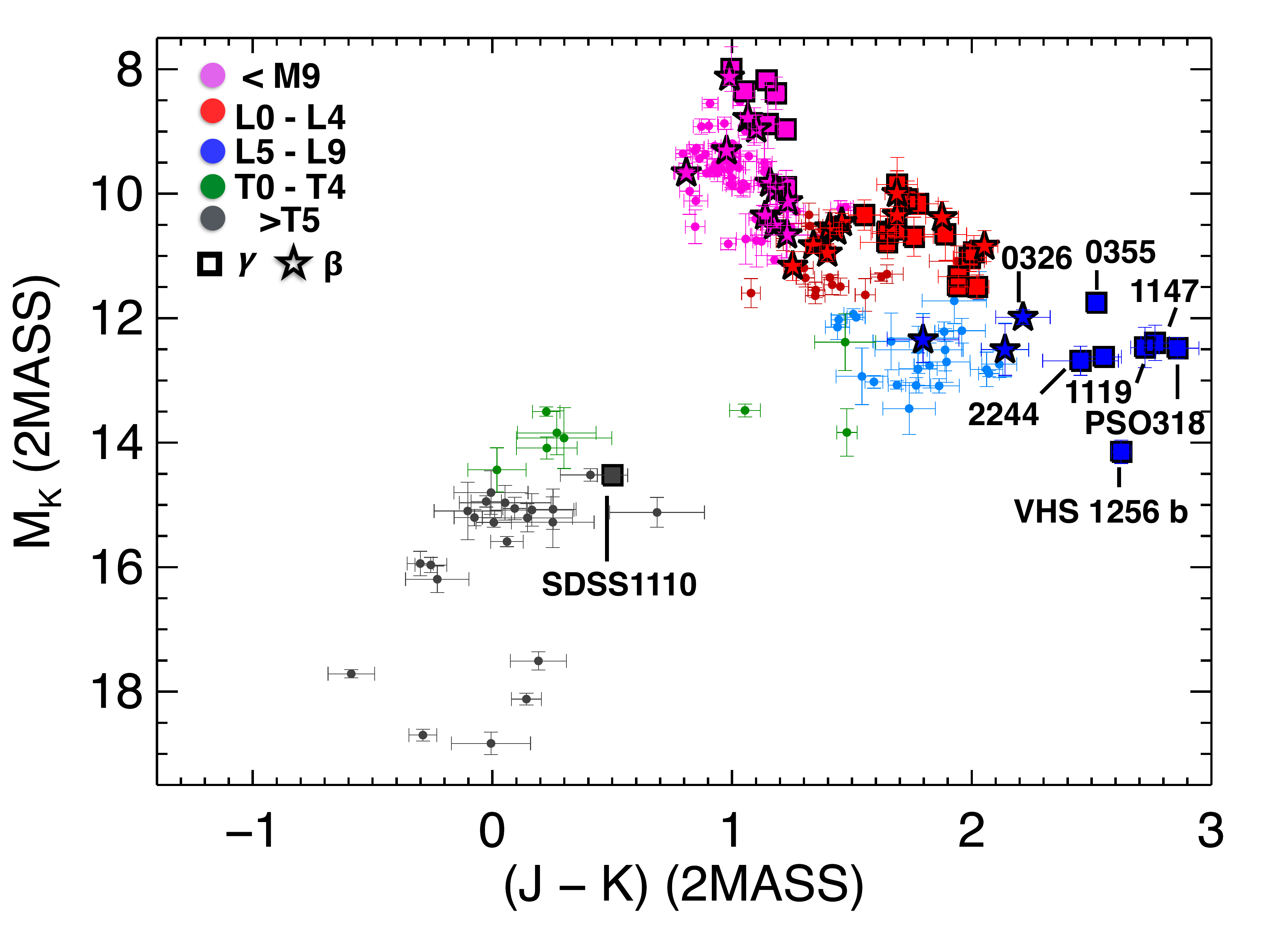}
\end{center}
\caption{The ($J$-$K$) versus $M_{J}$ (left) and $M_{K}$ (right) color magnitude diagram. Symbols are as described in Figure~\ref{fig:JvJmH}.
\label{fig:JvJmK}} 
\end{figure*}

\begin{figure*}[!ht]
\begin{center}
\epsscale{1.0}
\plottwo{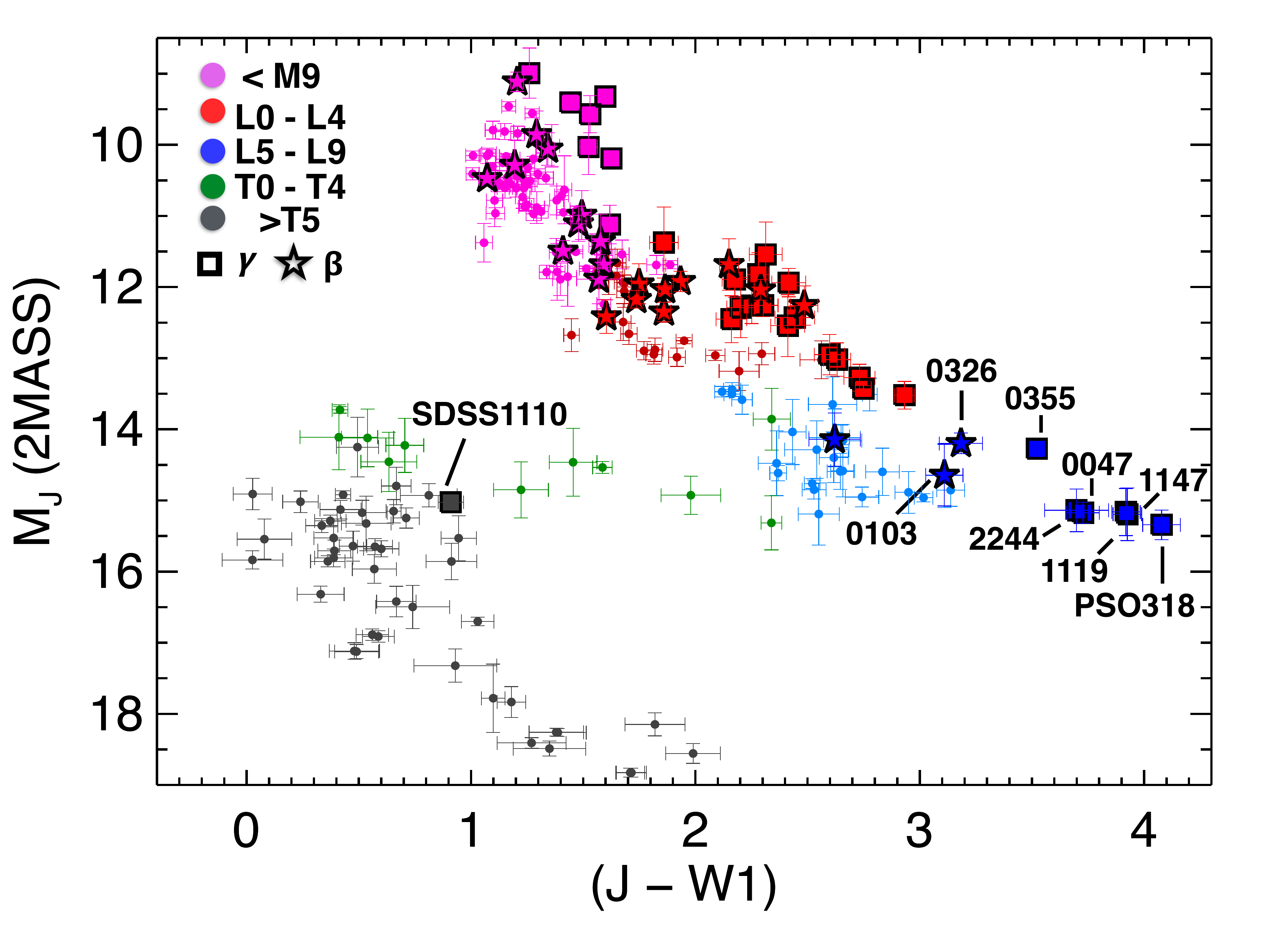}
{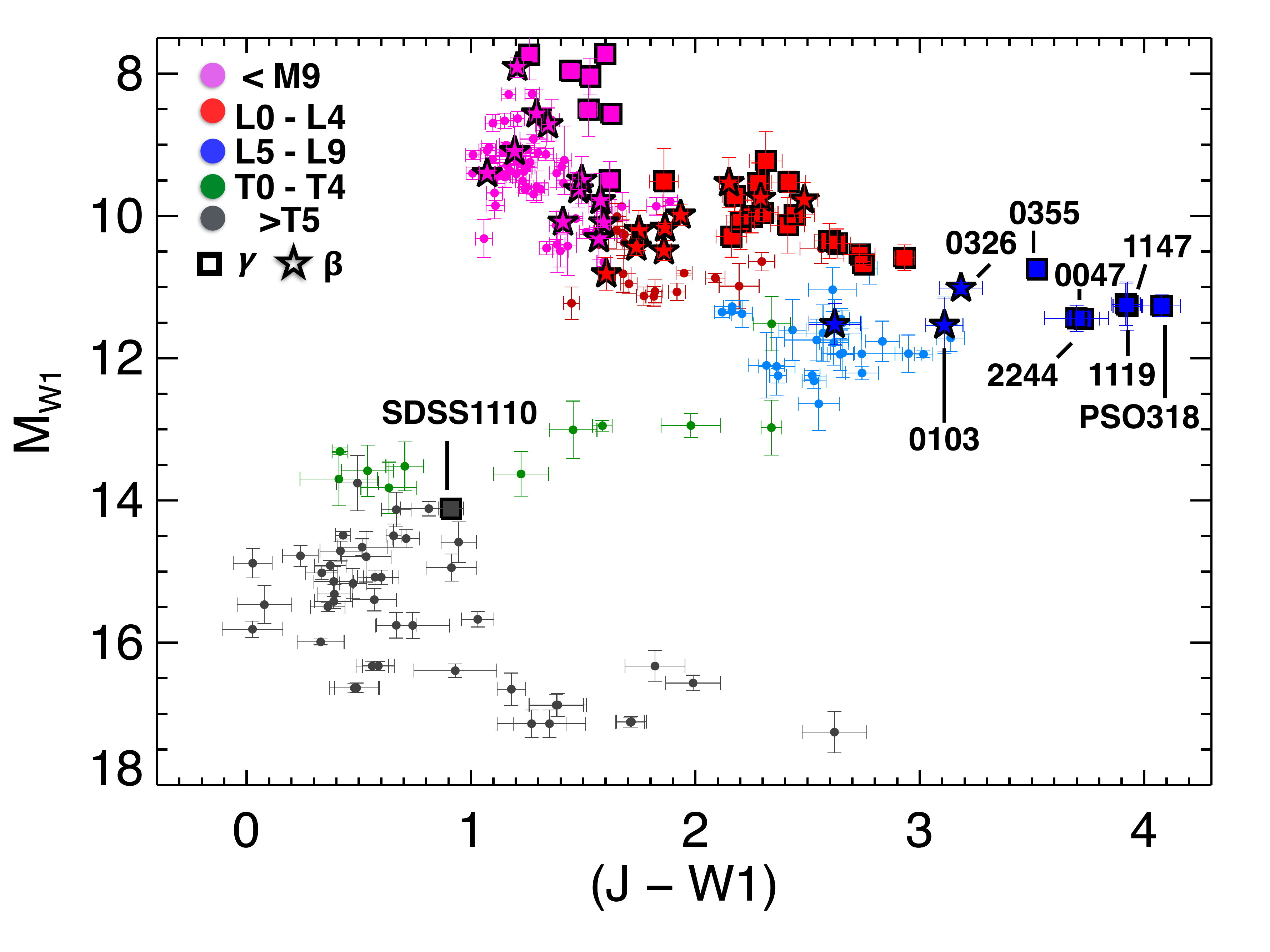}
\end{center}
\caption{The ($J$-$W1$) versus $M_{J}$ (left) and $M_{W1}$ (right) color magnitude diagram. Symbols are as described in Figure~\ref{fig:JvJmH}.  
\label{fig:JvJmW1}} 
\end{figure*}

\begin{figure*}[!ht]
\begin{center}
\epsscale{1.0}
\plotone{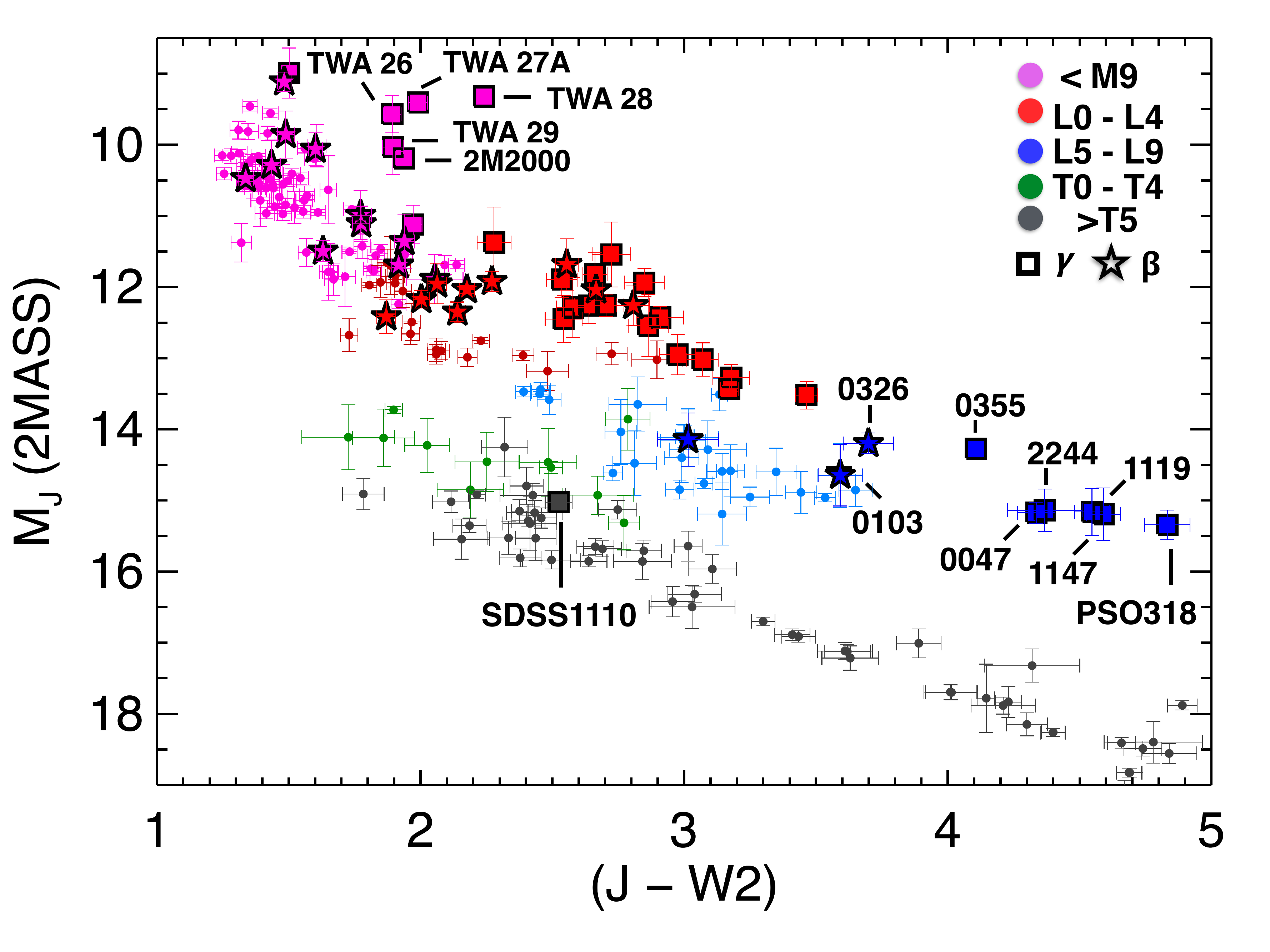}
\end{center}
\caption{The ($J$-$W2$) versus $M_{J}$ color magnitude diagram. Symbols are as described in Figure~\ref{fig:JvJmH}.   
\label{fig:JvJmW2A}} 
\end{figure*}

\begin{figure*}[!ht]
\begin{center}
\epsscale{1.0}
\plotone{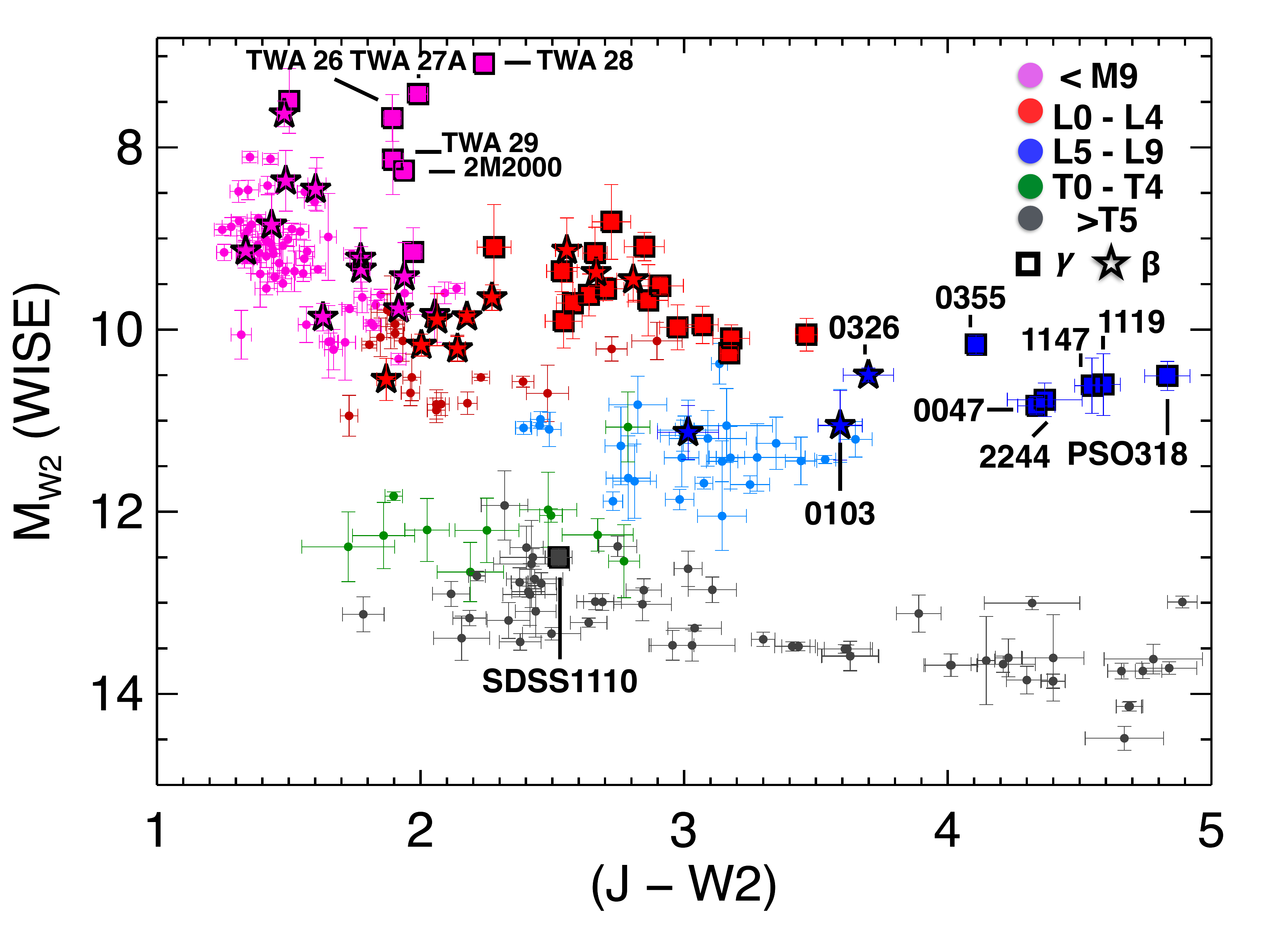}
\end{center}
\caption{The ($J$-$W2$) versus $M_{W2}$ (right) color magnitude diagram. Symbols are as described in Figure~\ref{fig:JvJmH}.   
\label{fig:JvJmW2B}} 
\end{figure*}

\begin{figure*}[!ht]
\begin{center}
\epsscale{1.0}
\plottwo{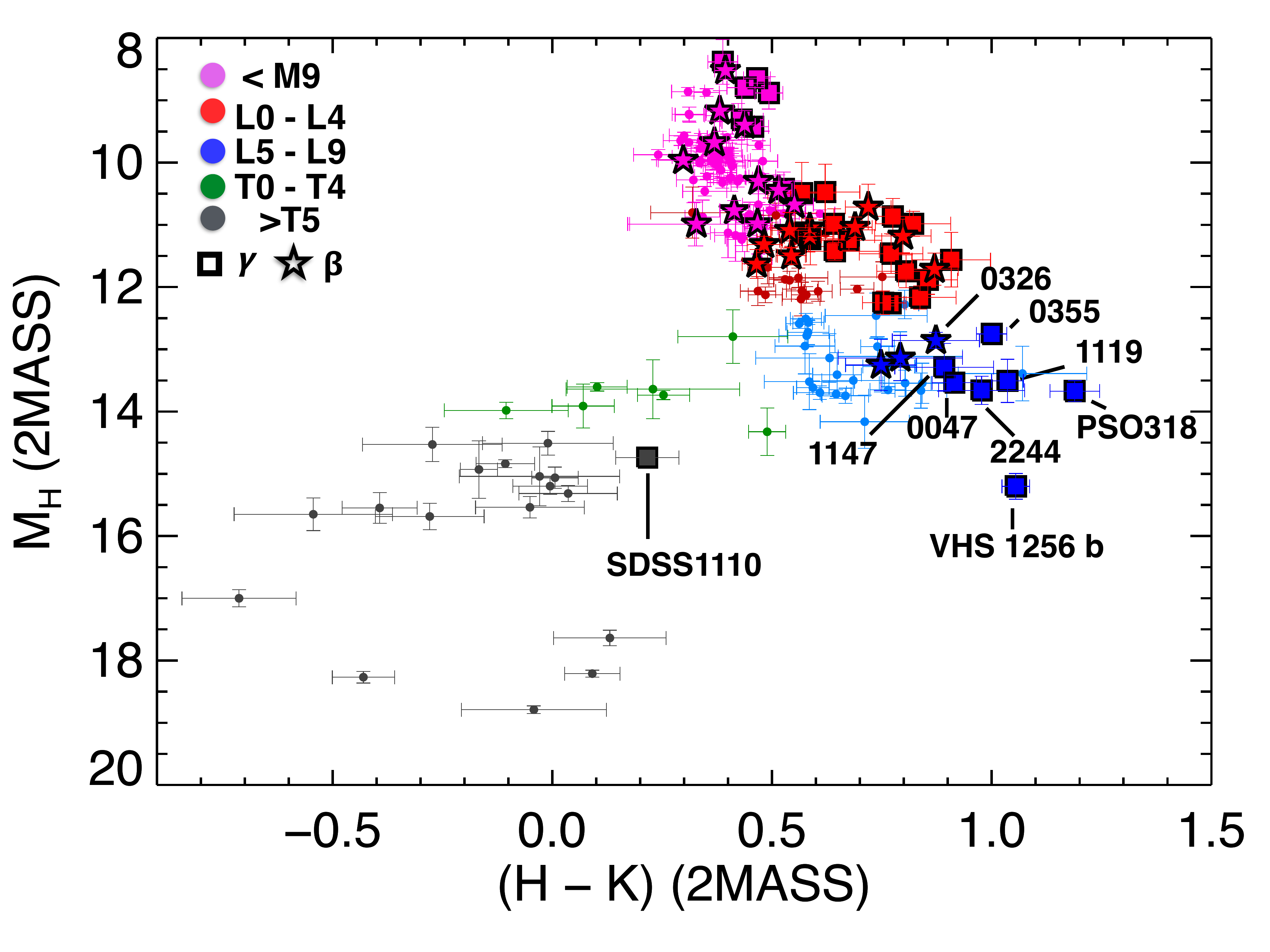}
{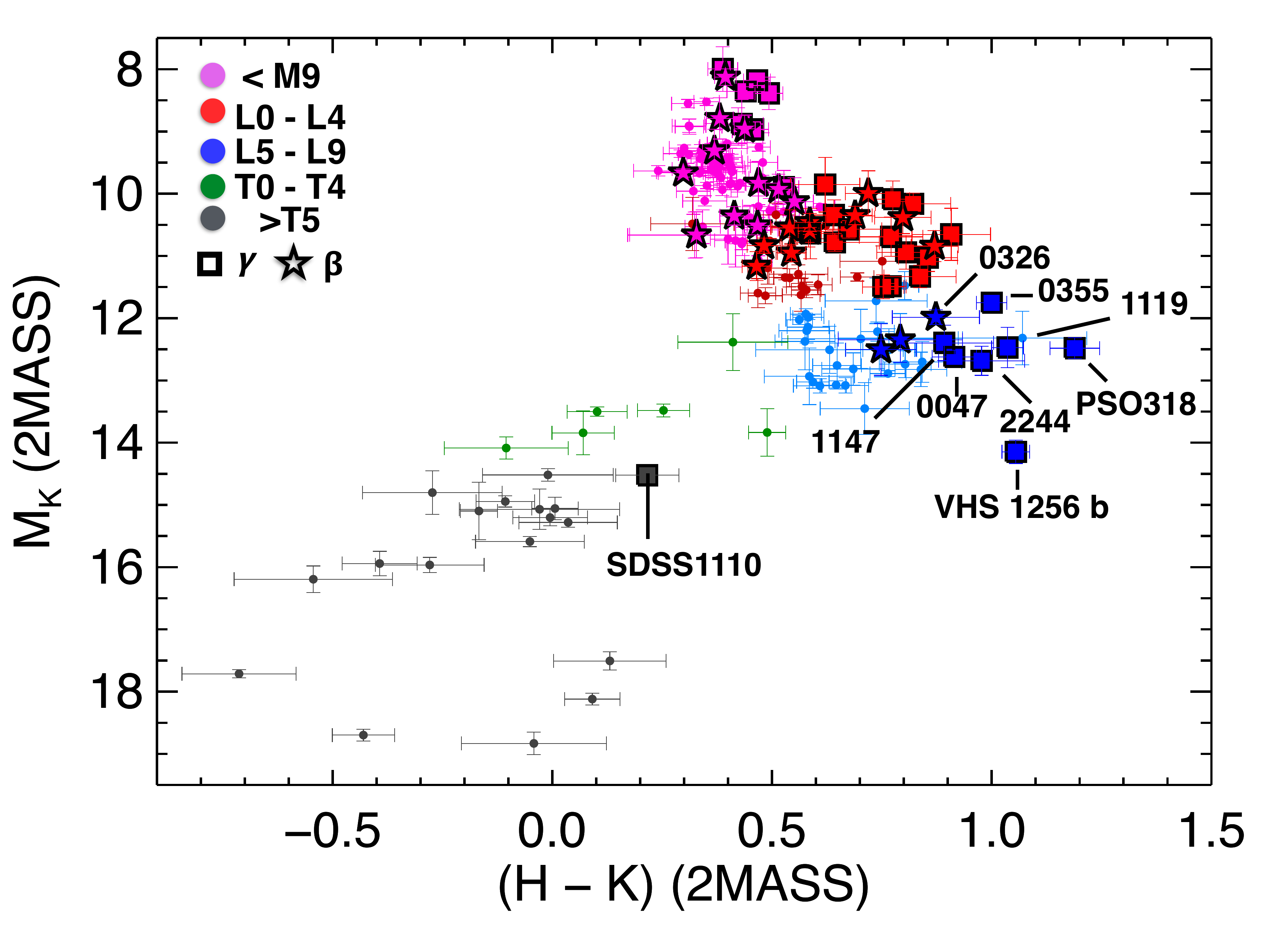}
\end{center}
\caption{The ($H$-$K$) versus $M_{H}$ (left) and $M_{K}$ (right) color magnitude diagram. Symbols are as described in Figure~\ref{fig:JvJmH}.   
\label{fig:HvHmK}} 
\end{figure*}

\begin{figure*}[!ht]
\begin{center}
\epsscale{1.0}
\plottwo{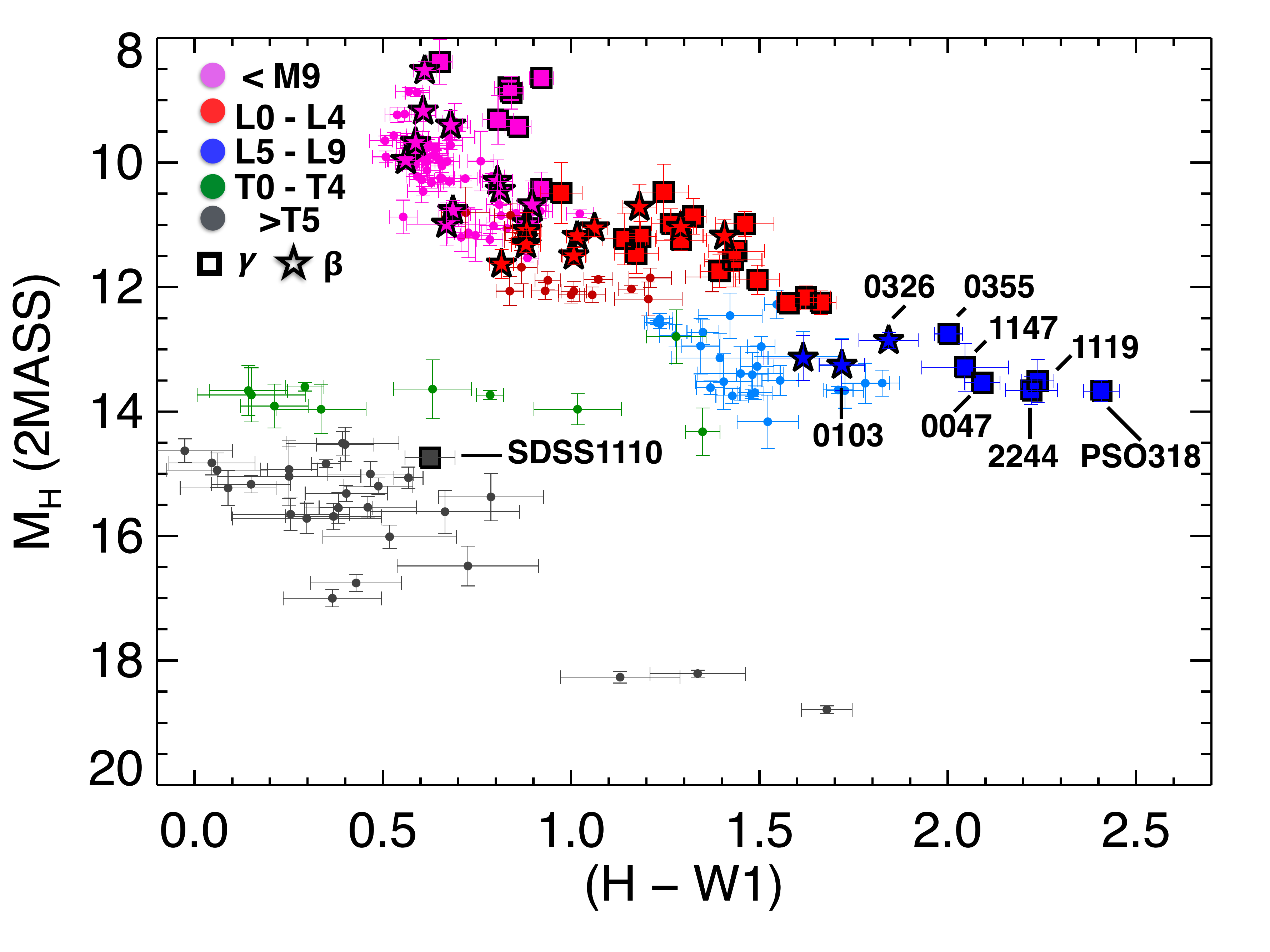}
{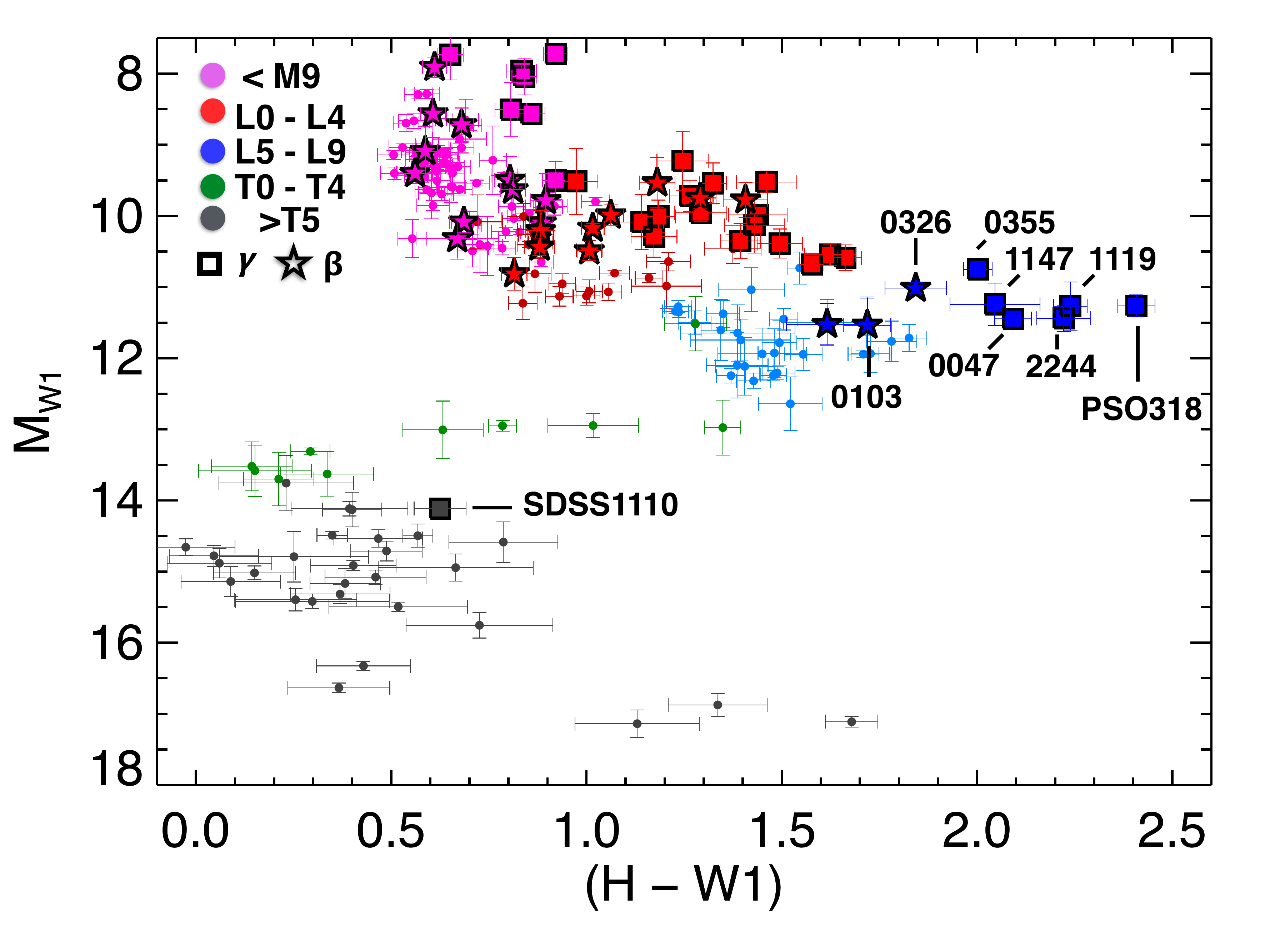}
\end{center}
\caption{The (H-W1) versus $M_{H}$ (left) and $M_{W1}$ (right) color magnitude diagram. Symbols are as described in Figure~\ref{fig:JvJmH}.  
\label{fig:HvHmW1}} 
\end{figure*}

\begin{figure*}[!ht]
\begin{center}
\epsscale{1.0}
\plottwo{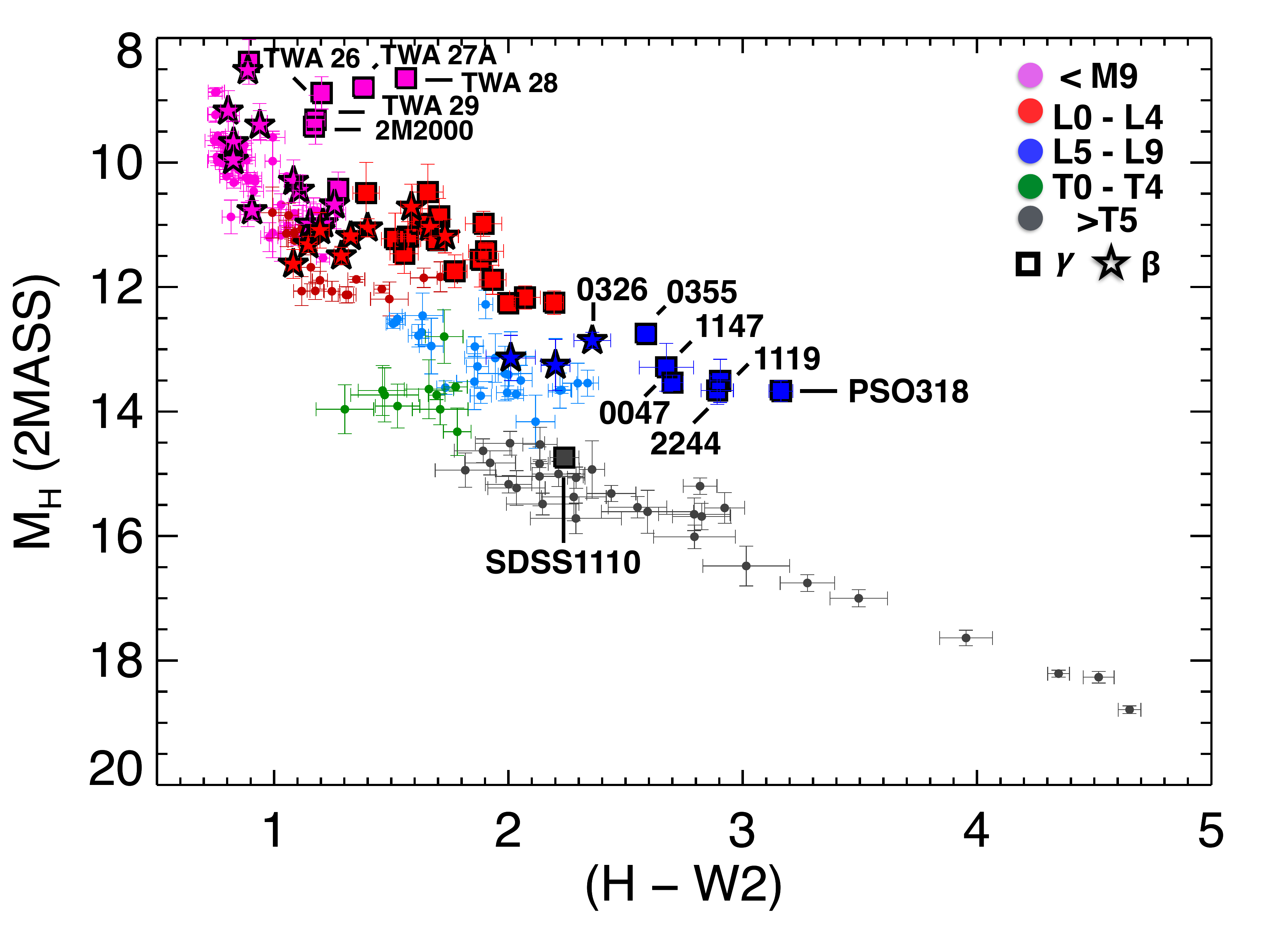}
{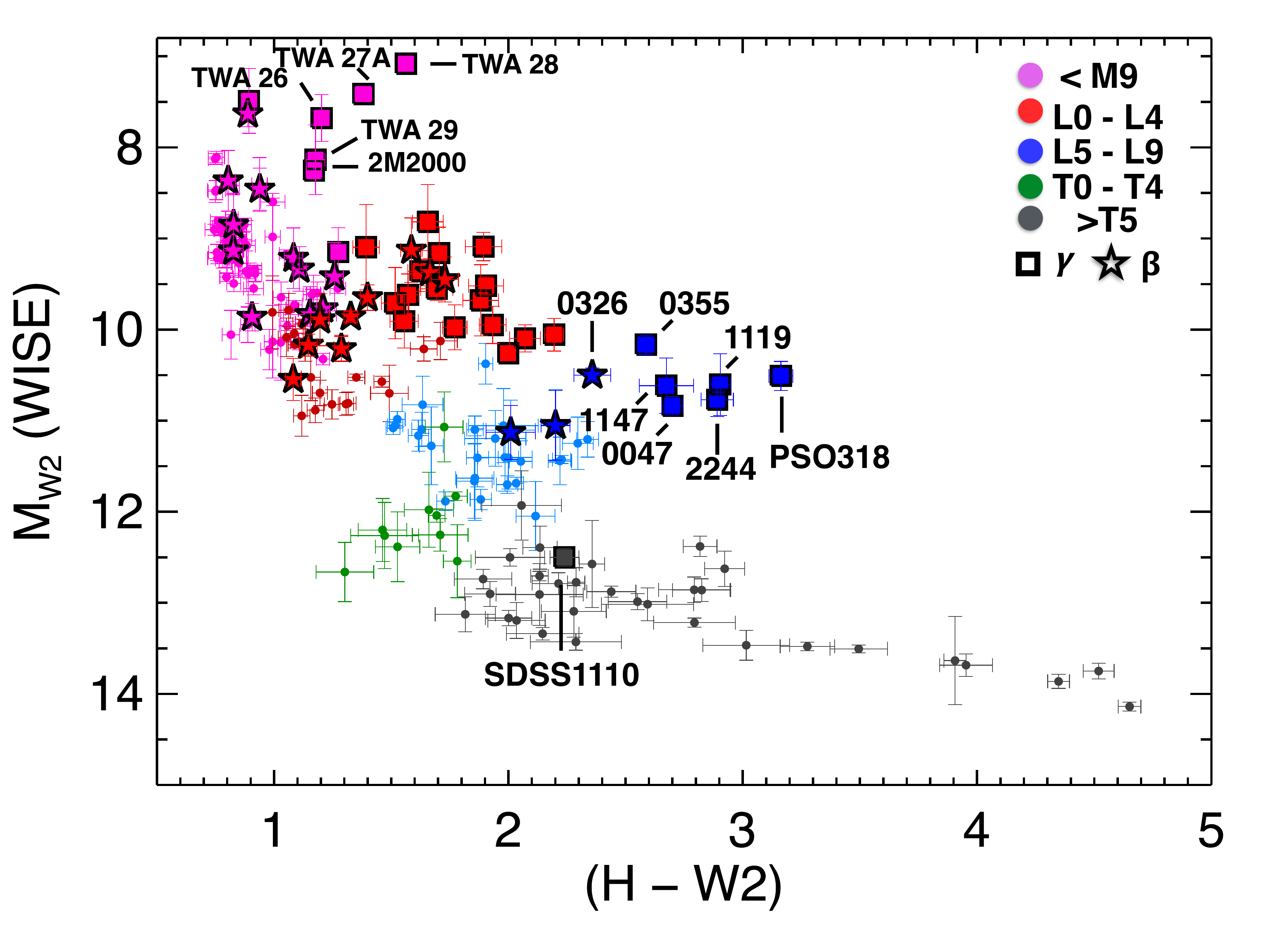}
\end{center}
\caption{The ($H$-$W2$) versus $M_{H}$ (left) and $M_{W2}$ (right) color magnitude diagram. Symbols are as described in Figure~\ref{fig:JvJmH}. 
\label{fig:HvHmW2}} 
\end{figure*}

\begin{figure*}[!ht]
\begin{center}
\epsscale{1.0}
\plottwo{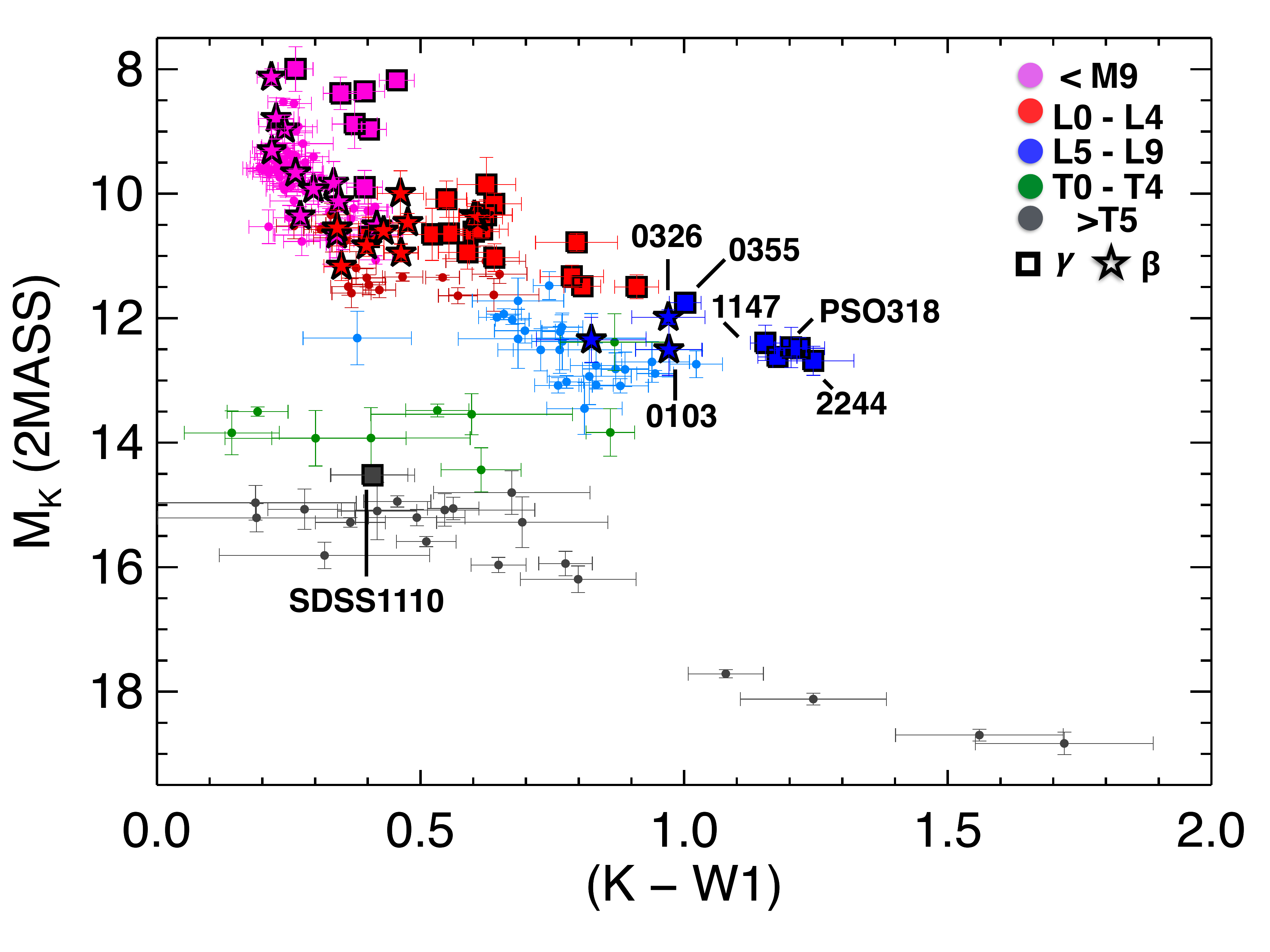}
{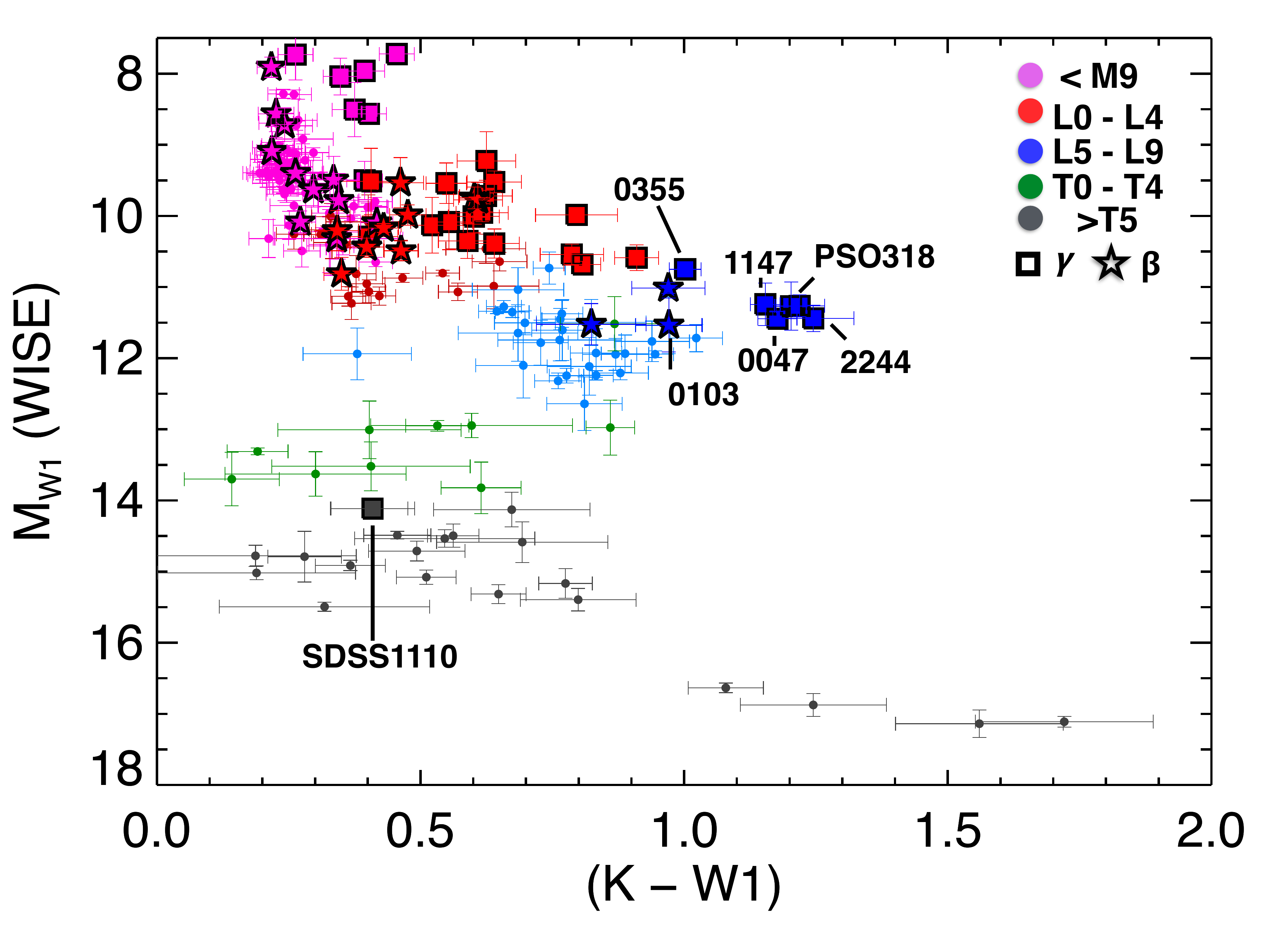}
\end{center}
\caption{The ($K$-$W1$) versus $M_{K}$ (left) and $M_{W1}$ (right) color magnitude diagram. Symbols are as described in Figure~\ref{fig:JvJmH}.  
\label{fig:KvKmW1}} 
\end{figure*}

\begin{figure*}[!ht]
\begin{center}
\epsscale{1.0}
\plottwo{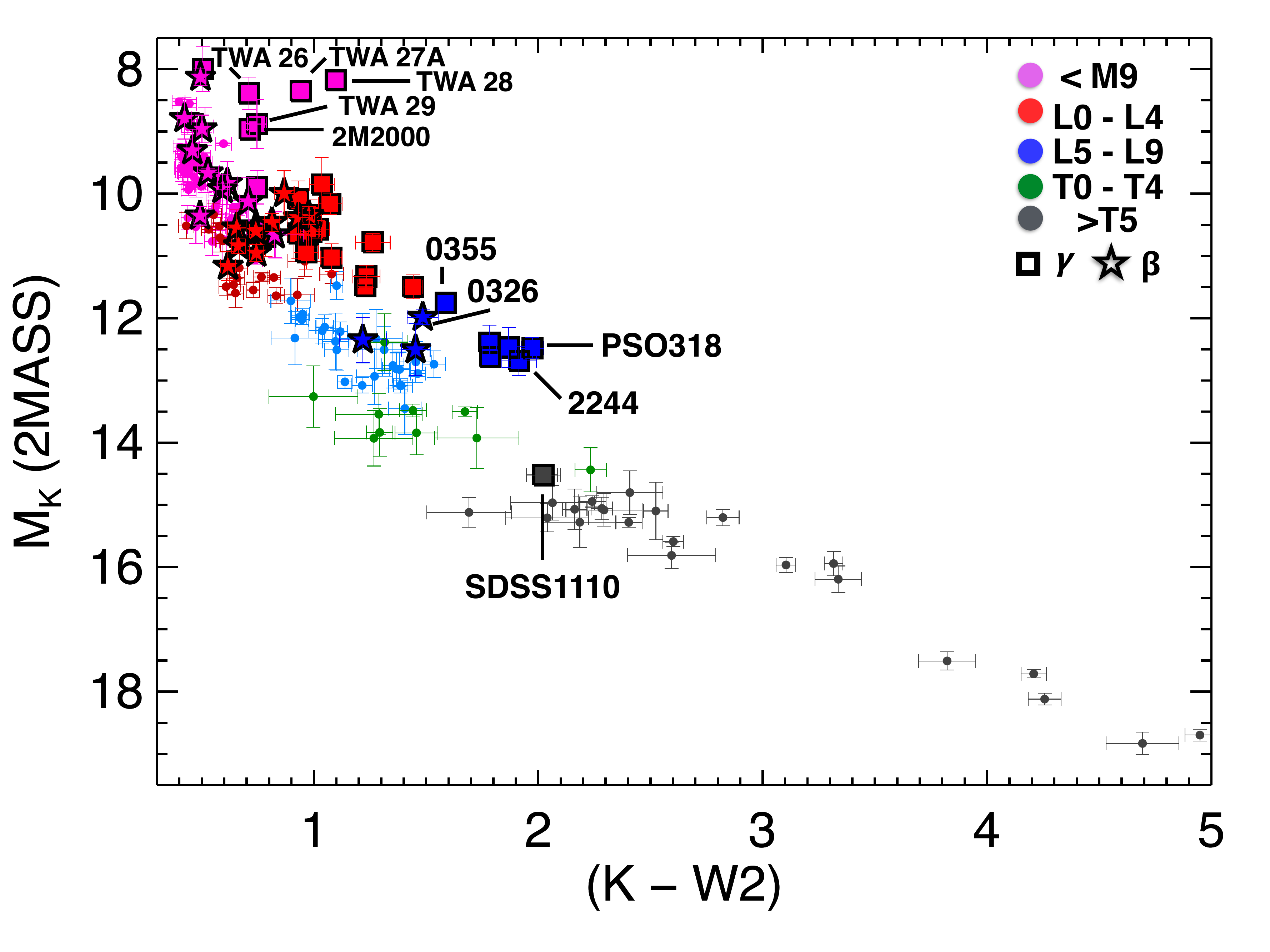}
{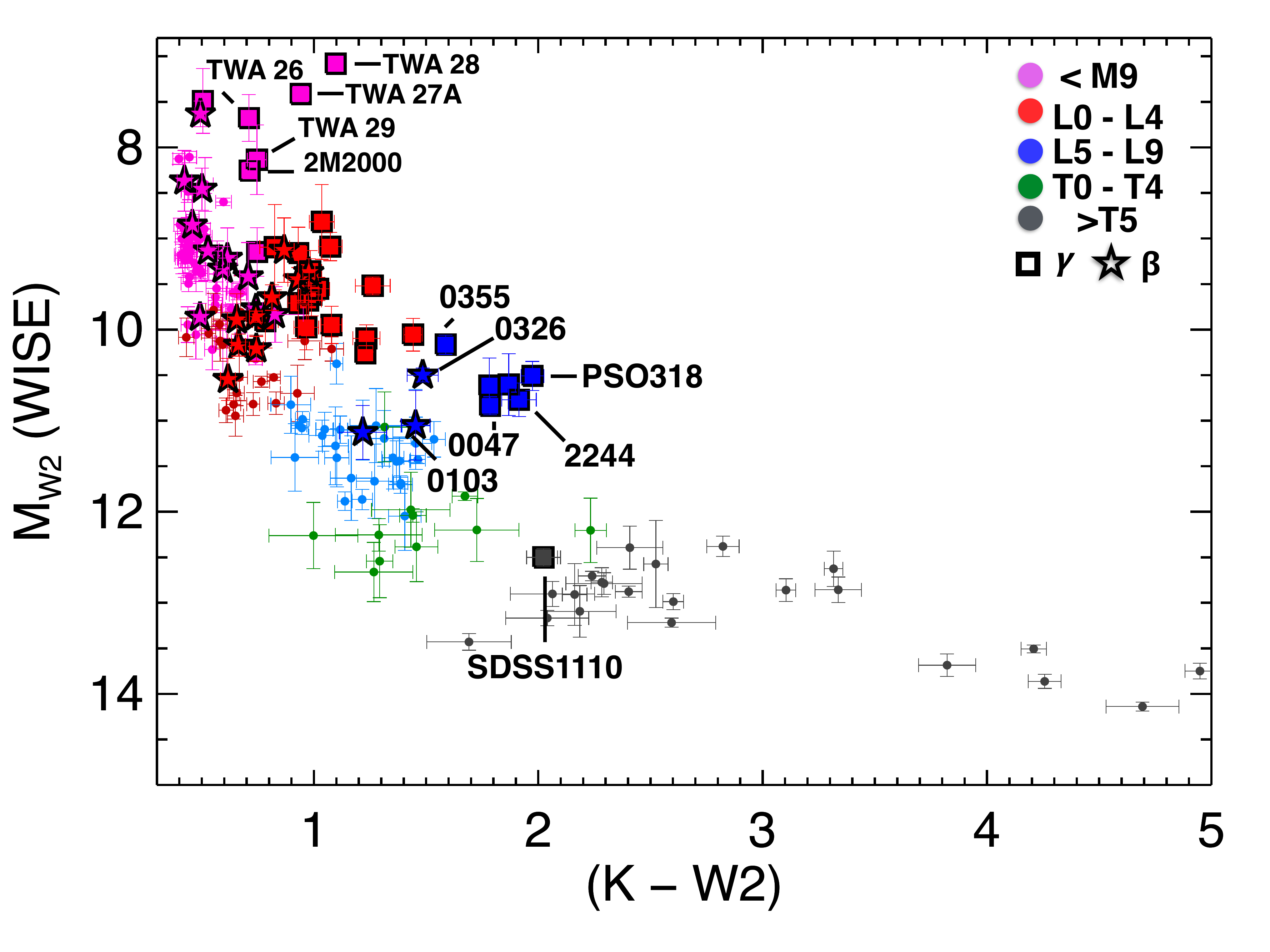}
\end{center}
\caption{The ($K$-$W2$) versus $M_{K}$ (left) and $M_{W2}$ (right) color magnitude diagram. Symbols are as described in Figure~\ref{fig:JvJmH}. 
\label{fig:KvKmW2}} 
\end{figure*}

\begin{figure*}[!ht]
\begin{center}
\epsscale{1.0}
\plottwo{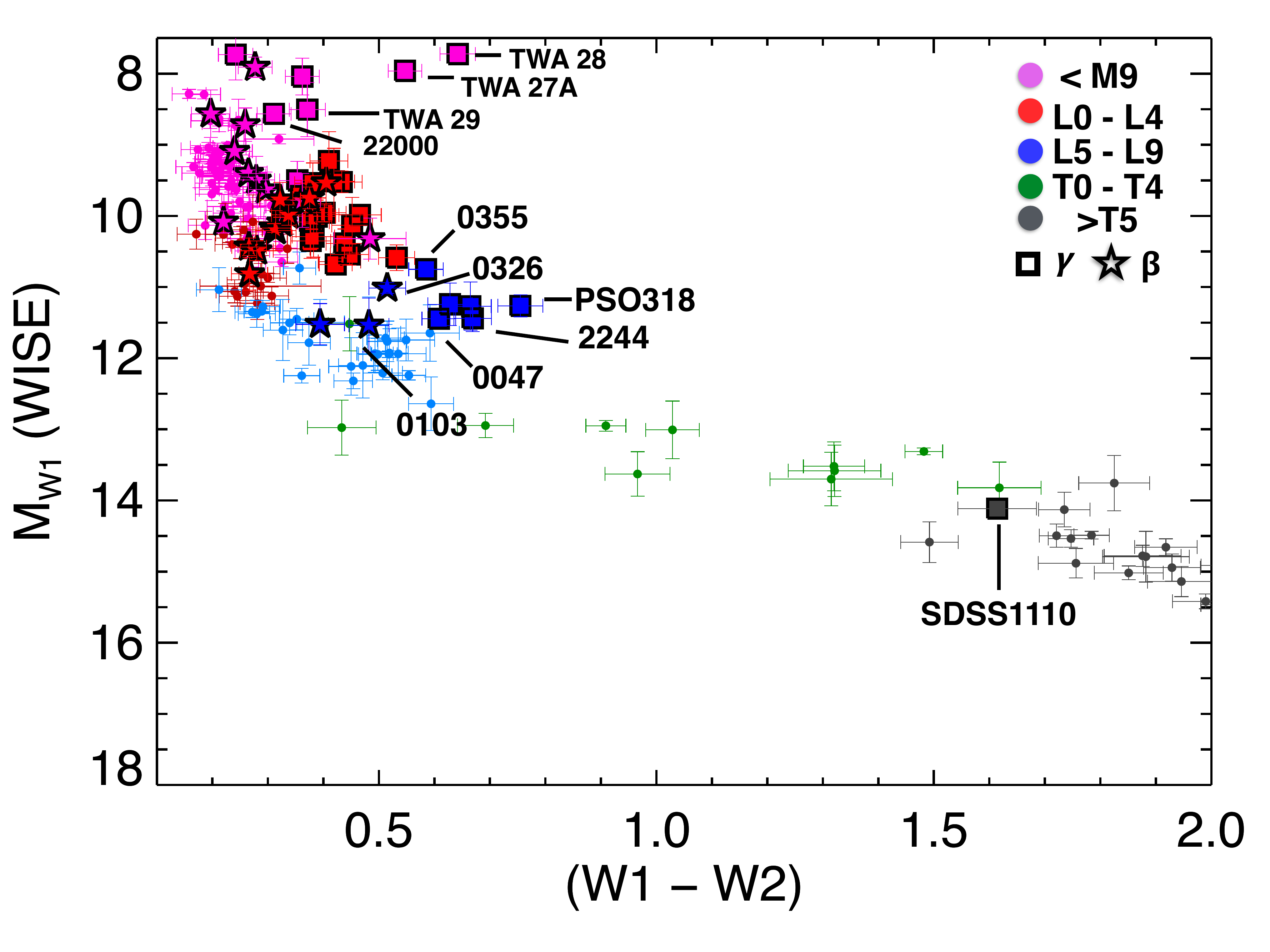}
{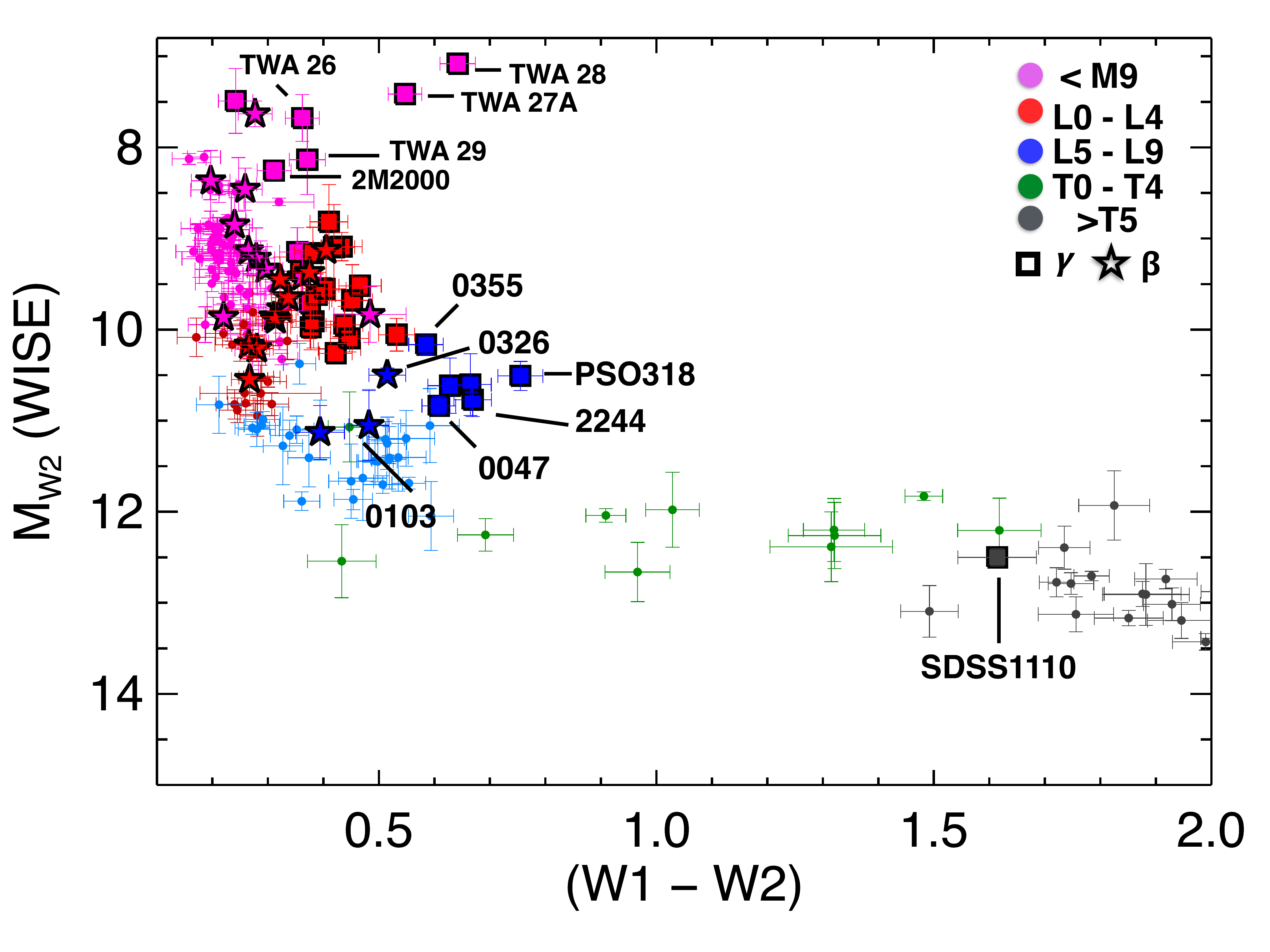}
\end{center}
\caption{The ($W1$-$W2$) versus $M_{W1}$ (left) and $M_{W2}$ (right) color magnitude diagram. Symbols are as described in Figure~\ref{fig:JvJmH}.  
\label{fig:W1vW1mW2}} 
\end{figure*}

Color magnitude diagrams have been discussed at length in the literature as a diagnostic of temperature, gravity, metallicity, and atmosphere properties of the brown dwarf population (e.g. \citealt{Liu13}, \citealt{Faherty12, Faherty13}, \citealt{Filippazzo15}, G15, \citealt{Dupuy12}, \citealt{Patten06}).    Figures~\ref{fig:JvJmH} - \ref{fig:W1vW1mW2} show the full suite of infrared color magnitude diagrams using JHK (2MASS) and W1W2 (WISE) photometry for the field parallax sample omitting binaries, subdwarfs, spectrally peculiar sources and those with absolute magnitude uncertainties $>$ 0.5 in any band.  On each plot we color code objects by spectral ranges of $<$ M9, L0-L4, L5-L9, T0-T4, and $>$T5 and we highlight the low surface gravity objects by their gravity classification.  The latest type sources in our sample are labeled on each plot.  On select plots, we have also labeled the M dwarf members of $\beta$ Pictoris and TW Hydrae.    

For completeness of the discussion, we have included the one confirmed isolated T dwarf member of a  moving group in the analysis (SDSS1110, T5.5, \citealt{Gagne15a}) as well as the L7 wide companion VHS 1256 B (\citealt{Gauza15}).  While the latest-type L dwarfs push the elbow of the L/T transition to an extreme red/faint color/magnitude, SDSS 1110 falls near spectral equivalents on all color magnitude diagrams.   As discussed in \citealt{Gagne15} and \citealt{Filippazzo15}, this source appears $\sim$150 K cooler than equivalents but does not exhibit the extreme color-magnitude properties as seen with the L dwarfs.  

Looking at a given absolute magnitude across all plots and comparing field objects to low surface gravity objects, we find that the latter can be more than a 1 magnitude redder. The most extreme behavior can be seen on Figures~\ref{fig:JvJmW2A} and ~\ref{fig:JvJmW2B} which exploits the largest wavelength difference in color ($J-W2$) and as discussed in section~\ref{sec:photometry} is where the low-gravity objects are the most extreme photometric outliers. 

As was discussed in section~\ref{sec:absmags}, there is a distinct difference in the diversity of absolute magnitudes between young M dwarfs and L dwarfs.  Using the color coding as a visual queue in Figures \ref{fig:JvJmH} -~\ref{fig:W1vW1mW2}, the M dwarfs fall redder than the field sequence but they also scatter to brighter magnitudes.  For the W2 color difference plots (e.g. $M_{JHK}$ vs ($J$-$W2$), ($H$-$W2$), or ($K$-$W2$)), we label the position of the M dwarfs in TW Hydrae and $\beta$ Pictoris as they are strikingly red and bright at these wavelengths and well separated from the field population. 

Comparatively, the L dwarfs flip in their behavior and can be seen as redder but fainter than field sources.  Focusing on Figures~\ref{fig:JvJmH} -~\ref{fig:JvJmK}, and~\ref{fig:HvHmK} which use $JHK$ photometry only, the latest type sources (e.g. PSO318, 0047+6803, 2244+2043, 1119-1137, 1147-2040) are not only redder but they also drive the elbow of the L/T transition $\sim$ 1 magnitude fainter than the field (notably in $M_{J}$). Moving to longer wavelengths, this behavior reverses.  By Figures that evaluate colors against $M_{W1}$ ($\sim$ 3.4$\mu$ m) or $M_{W2}$ (e.g. Figure~\ref{fig:JvJmW1} or~\ref{fig:HvHmW2}) the latest type sources are consistent with or slightly brighter than the elbow.  Hence as discussed in section~\ref{sec:absmags} and \citet{Filippazzo15}, the flux redistribution of young L dwarfs seems to hinge very close to the $K$ band.  Indeed the color magnitude diagram that appears to smoothly and monotonically transition objects from late- M to T dwarfs is the $M_{K}$ vs ($K$-$W2$) plot in Figure~\ref{fig:KvKmW2}.  There is a 10 magnitude difference between the warmest to the coolest objects and the magnitude seems to monotonically decrease with reddening color showing only a hint of an L/T transition elbow at $M_{K}$=13/$M_{W2}$=12.  On this plot, the low-surface gravity sequence lies $\sim$ 1 magnitude brighter than the field with the exception of a small fraction of the sample appearing normal (including the AB Doradus T dwarf SDSS1110).

The L/T transition induces a turning elbow on most color magnitude plots. This feature is brought on when the clouds dissipate as one moves from warmer L dwarfs to cooler T dwarfs and CH$_{4}$ begins to dominate as an opacity source all conspiring to drive the source colors blueward.  The demonstration that these young sources are redder and fainter than the field sequence in the near-infrared indicates that the clouds must persist through lower temperatures  (fainter) and represent an extreme version of the conditions present for field age equivalents (redder).   The brightening of sources at W1 and W2 at extreme red colors likely holds clues to the composition and structure of the clouds as it is a reflection of the flux redistribution to longer wavelengths as discussed in section~\ref{sec:absmags} above.  

For colors and magnitudes evaluated across JH or K and W1 or W2 (e.g. Figure~\ref{fig:HvHmW1} which shows $M_{H}$ vs H-W1), the latest type objects pull the low-gravity sequence redder than the field while maintaining a small spread in absolute magnitude from  0103+1935 through PSO318.  Interestingly this  indicates similarities in these sources not readily apparent in current spectral data. 

Lastly, on Figures~\ref{fig:JvJmH} - \ref{fig:W1vW1mW2} we have given $\gamma$ and $\beta$ classified sources different symbols to investigate whether trends between the two could be identified.  Throughout, there is a hint that the sequence of $\gamma$ classified objects is redder than that of the $\beta$ sources.  This is most prominent on Figures ~\ref{fig:JvJmW1} -  ~\ref{fig:JvJmW2A} where only four $\beta$ L dwarfs rival $\gamma$ sources in color and/or magnitude.  On other figures such as ~\ref{fig:JvJmH}, there appears to be more mixing between the two gravity classifications.  As has been stated throughout this work, spectral type and the corresponding gravity classification are difficult to evaluate and can differ between optical and infrared spectra or from low resolution to medium resolution data.  The data as viewed in this work, seems to indicate that the $\gamma$ classified sources are distinct on color magnitude diagrams from the $\beta$ classified sources with some mixing likely due to a non-uniform spectral typing methodology. 

\subsection{The Bolometric Luminosities and Effective Temperatures of Young Objects\label{sec:bols} }
One expects that younger brown dwarfs should have inflated radii compared to equivalent temperature sources given that they are still contracting.  Consequently, one would expect that $\gamma$ and $\beta$ objects would be overluminous when compared to their field age equivalents.  A rough estimate using the \citet{Burrows01} evolutionary models indicates that 10 Myr (50 Myr) objects with masses ranging from 10 to 75 M$_{Jup}$ have radii that are 25\% to 75\% (13\% to 50\%) larger than 1 -3 Gyr dwarfs with equivalent temperatures. Since $L_{bol}$ scales as $R^{2}$, one might expect that this age difference translates into younger objects being 1.5x - 3.0x (1.3x - 2.3x) overluminous compared to the field.    

Initially, studies categorized low surface gravity brown dwarfs as ``underluminous" compared to field sources based on examining near-infrared absolute magnitude trends alone (e.g. \citealt{Faherty12}).  However as discussed in section~\ref{sec:absmags}, flux shifts to longer wavelengths beyond the $H$ and $K$ bands at lower gravities, so that some absolute magnitude bands might be fainter but $L_{bol}$'s are not underluminous compared to the field (\citealt{Faherty13}, \citealt{Filippazzo15}).   In fact, \citet{Filippazzo15} carefully evaluated $L_{bol}$ values for all brown dwarfs with parallaxes (or kinematic distances in the case of high likelihood moving group members) and presented up to date relations between observables and calculated $L_{bol}$s.  In that work, $\beta$ and $\gamma$ objects were found to split along the M/L transition whereby M dwarfs were overluminous and L dwarfs were within to slightly below the sequence when compared to field objects.  

To investigate $L_{bol}$ trends among the age calibrated sample, we first calculate values for sources reported here-in using the technique described in \citet{Filippazzo15}.  In that work, the authors integrate under the combined optical and near-infrared photometry and spectra as well as the WISE photometry and mid infrared spectra where available.  As described in previous sections, we use parallaxes where available but supplement with kinematic distances when a source was regarded as a high likelihood member to a group (see values in parentheses in Tables~\ref{tab:Members} and ~\ref{tab:kinematics}).   All $L_{bol}$, T$_{eff}$, and Mass values are listed in Table ~\ref{tab:LbolTeffMass} as well as Table~\ref{tab:Members} for members only.

\begin{figure*}[t!]
\center
\includegraphics[angle=0,width=1.0\textwidth]{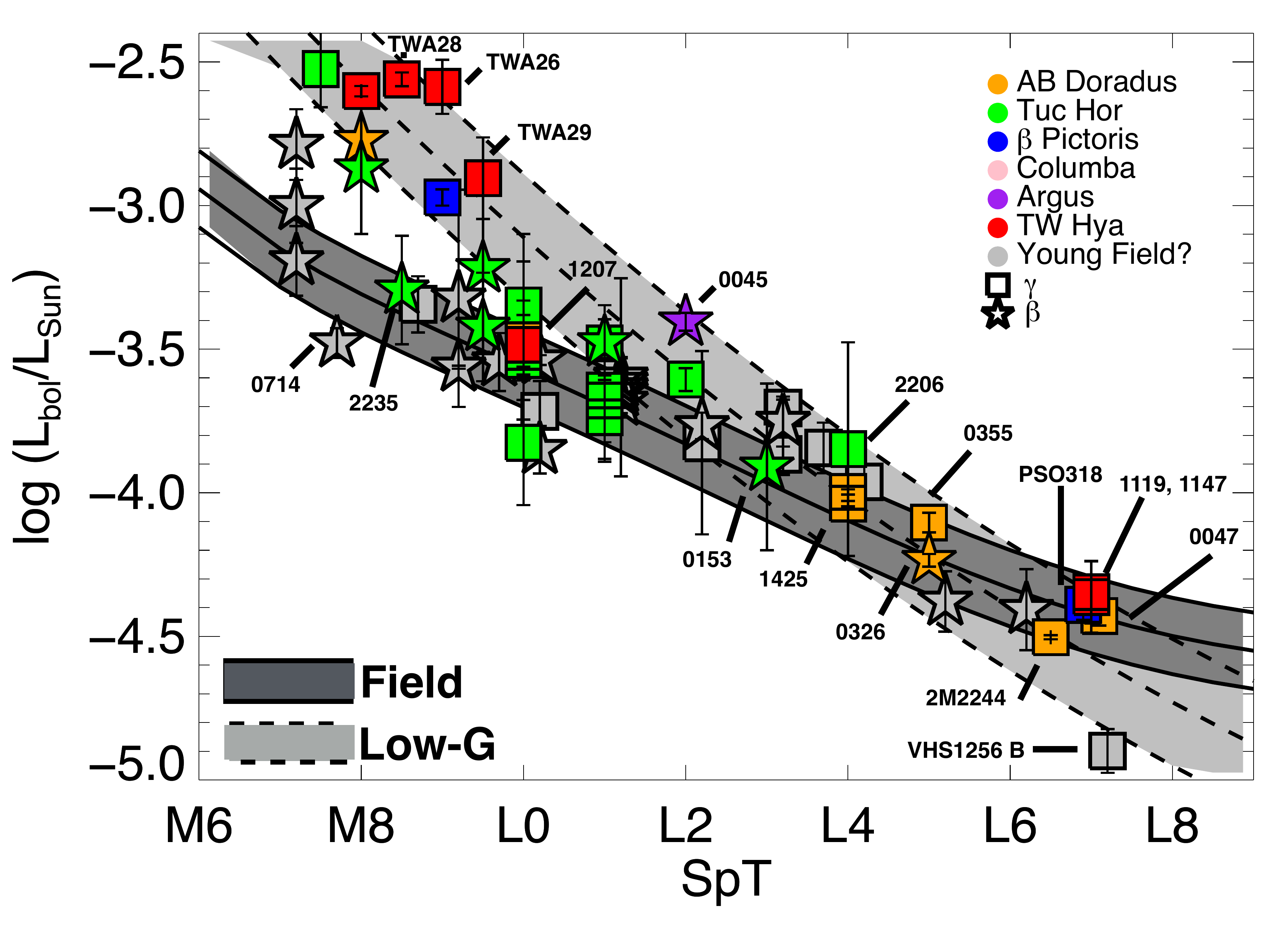}
\caption{The spectral type versus bolometric luminosity plot.  The field polynomial and residuals from \citet{Filippazzo15} is represented by the grey area.  Over-plotted are objects in this work with measured parallaxes or estimated kinematic distances from high confidence group membership.  $L_{bol}$ values were calculated as described in \citet{Filippazzo15}.  Symbols distinguish very low ($\gamma$) from intermediate ($\beta$) gravity sources.  Objects are color coded by group membership. }
\label{fig:Lbol}
\end{figure*}

Figure~\ref{fig:Lbol} shows L$_{bol}$ as a function of spectral type for all objects compared to the field polynomial from \citet{Filippazzo15}.  Also overplotted is the polynomial fit for objects in groups (labeled as GRP in Table~\ref{tab:polynomials}). We highlight bona fide and high-likelihood brown dwarf moving group members as well as the unassociated $\gamma$ and $\beta$ objects with differing symbols and colors. 

Focusing on the age calibrated sample, we find that late-type M dwarfs assigned to groups are overluminous compared to the field.  The L dwarfs assigned to a group are mixed, with the majority falling within the field polynomial relations and a small number falling slightly above.  Comparing group-to-group differences, the 5-15 Myr TW Hydrae M8 and M9 objects are $\sim$5x more luminous than the field while the 30-50 Myr Tucana Horologium source is $\sim$2x more luminous.  Interestingly, the TW Hydrae, $\beta$ Pictoris, and AB Doradus L7's have near equal $L_{bol}$ values, all near the mean value for their subtype.  
 
Comparing the $\gamma$ and $\beta$ gravity sources without membership to the field and high-likelihood group members, we find that -- with the exception of the M7.5 0335+2342 which is highly suspect to be a $\beta$ Pictoris member (see note in Table ~\ref{tab:Members}) -- all non-members fall within the polynomial relations for the field.  This trend indicates that the late- M non-member $\gamma$ and $\beta$ sources may not be young (see discussion in subsection~\ref{sec:nonmembers} and section~\ref{sec:unmatched}).  Indeed, 1022+0200 is a late-M non-member with full kinematics.  When we compare the UV velocity of this source to that of the \citet{Eggen89} criterion for young stars, we find it falls outside of it hinting that it may be drawn from the older disk population.   It remains unclear how to interpret the non-member L dwarfs as this trend is in line with group assigned equivalents.  

\begin{figure*}[t!]
\center
\includegraphics[angle=0,width=1.0\textwidth]{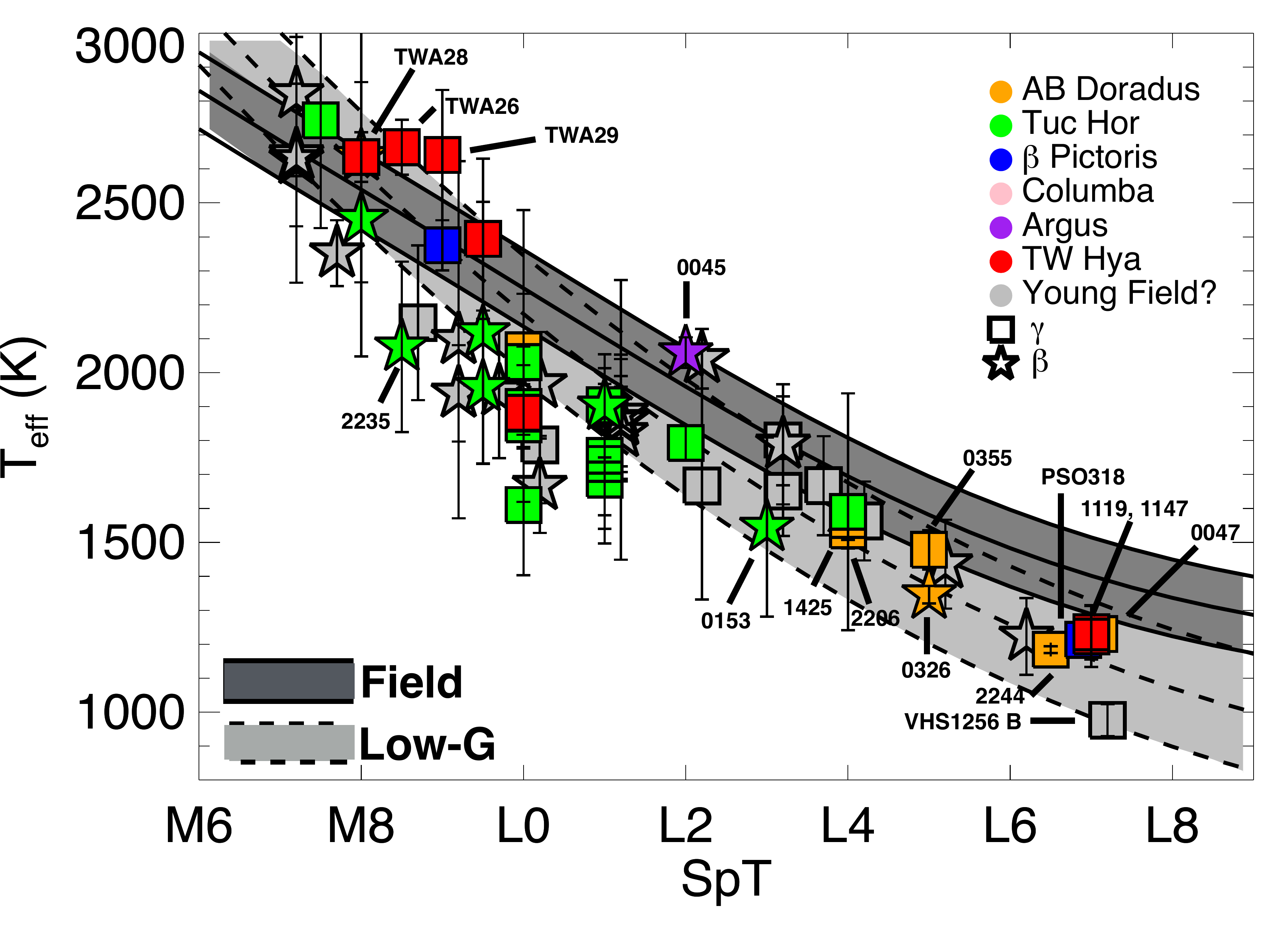}
\caption{The spectral type versus $T_{eff}$ plot.  Symbols are as described in Figure~\ref{fig:Lbol}.}
\label{fig:Teff}
\end{figure*}

The result of $L_{bol}$'s for $\gamma$ and $\beta$ gravity L dwarfs looking like field sources or in some cases underluminous, implies that they are cooler than their field spectral equivalents. In other words, low-gravity or atmospheric conditions potentially induced at a younger age, mimic spectral features of a warmer object.  As young sources are typed on a scheme anchored by field objects, they are incorrectly grouped with warmer sources.  They certainly stand out in photometric, spectroscopic, and band by band absolute magnitude comparisons with field sources (e.g. all of section~\ref{sec:diversity}).  However the low gravity sequence does not logically or easily follow off of the field sequence.  Figure~\ref{fig:Teff} shows the T$_{eff}$ values for $\gamma$ and $\beta$ sources calculated using the method described in \citet{Filippazzo15} along with the polynomial and residuals for field objects (from \citealt{Filippazzo15}) and group members (from Table~\ref{tab:polynomials}).  As noted in \citet{Filippazzo15} -- with the exception of a few -- while M dwarfs are similar if not warmer for their given spectral subtype, the L dwarfs are up to 100-300K cooler.  Examining the 5-130 Myr age calibrated objects within this sample, we can not isolate a trend of younger objects being increasingly cooler than older equivalent sources.  For example, PSO318 and 0047+6803 are equivalent temperatures even though there is a $\sim$100 Myr difference in age.

\subsection{Unmatched objects with Signatures of Youth\label{sec:unmatched}}
Among the 152 objects in this sample with reported spectral signatures of youth in either the optical or the near-infrared, we confidently find that 39 ($\sim$25\%) are high-likelihood or bona fide members of nearby moving groups.  There are 92 ($\sim$61\%) dubbed ambiguous either because their kinematics overlap with more than one group (including the old field) and they need better astrometric measurements to differentiate, or there is not strong enough evidence with the current astrometry to definitively call it high-likelihood or bona fide.  There are 21 objects ($\sim$ 14\%) for which we have enough information to declare them as non-members to any known group assessed in this work.  Among the non-members,  there are 4 optically classified (3 infrared classified) M dwarfs  and 15 optically classified (17 infrared classified) L dwarfs  with 12 optically (10 infrared) classified $\gamma$ objects  and 5 optically (8 infrared) classified $\beta$ objects.  Several of these objects have extreme infrared colors (see Figures~\ref{fig:JmH} -~\ref{fig:W1mW2}).  For instance, 1615+4953 is classified in the optical as an L4$\gamma$ and rivals the most exciting late-type objects in its deviant infrared colors.  However, current kinematics do not show a high probability of membership in any group despite its having a proper motion and radial velocity.    Similarly, 0435-1441 is strikingly red in $JHK$, shows both optical and infrared signatures of youth, and its spectrum needs to be de-reddened (E(B-V)=1.8).  While it is in the direction of the nearby star-forming region MBM 20, the distance noted for that cluster (112 - 161 pc) would drive unrealistic absolute magnitudes and $L_{bol}$ values for this source.

The current census of young, but, unassociated late- M and L dwarfs is similar to what has emerged in studies of early M dwarfs with multiple signatures of youth (e.g. X-ray, UV and IR-excess, Lithium).  A significant portion of objects in the studies of \citet{Rodriguez13} and \citet{Shkolnik11,Shkolnik12} are strong candidates for being 5 -130 Myr objects via their spectral and photometric properties but their kinematics are inconclusive and their age cannot be determined by a group assignment.  Likely, there are a number of groups waiting to be uncovered that may account for this overabundance of young, low mass objects.  Alternatively, after 5 - 130 Myr sources have been dynamically moved from their origins such that tracing back their history to any collection of objects is beyond our capability.   

\section{Comparisons with Directly Imaged Exoplanets}
Several of the objects discussed herein are in the same moving groups as directly imaged exoplanets or planetary mass companions.  For example, there are two bona fide or high-likelihood brown dwarfs in the $\beta$ Pictoris moving group, home to the 11-13 Jupiter mass planet $\beta$ Pictoris b and the newly discovered 2 - 3 Jupiter mass planet 51 Eri b (\citealt{Bonnefoy13, Bonnefoy14}, \citealt{Males14}, \citealt{Macintosh15}), 20 in the Tucana Horologium  association which houses the 10 - 14 M$_{Jup}$ planetary mass companion AB Pictoris b (\citealt{Chauvin05}, \citealt{Bonnefoy10}) as well as the 12 - 15 M$_{Jup}$ planetary mass companion 2M0219 b (\citealt{Artigau15}), and seven in the TW Hydrae Association which is the home of the 3 - 7 M$_{Jup}$ planetary mass companion 2M1207 b (\citealt{Chauvin04}, \citealt{Patience12}).   As such, the brown dwarfs discussed herein should be considered siblings of the directly imaged planets as one can assume that they are co-eval, and share formation conditions and dynamical histories.  The mode of formation for the brown dwarfs versus the directly imaged exoplanets remains a question but will likely drive distinct differences in the observables of each population.  In this section, we look to place the young brown dwarf sample in context with related exoplanet members. 

\subsection{Similarities of Brown Dwarfs and Imaged Exoplanets on Color Magnitude Diagrams}
Young isolated brown dwarfs are far easier to accumulate data on than directly imaged exoplanet equivalents because they do not have a bright star to block when observing.  The collection of currently known giant exoplanets, generally have only infrared ($J,H,K,L',M'$) photometric measurements.  Near-infrared spectroscopy is possible for some although this requires considerable telescope time and advanced instrumentation (e.g. \citealt{Oppenheimer13}, \citealt{Macintosh15}, \citealt{Hinkley15}, \citealt{Bonnefoy13}, \citealt{Patience12}).

\begin{figure*}[!ht]
\begin{center}
\epsscale{1.0}
\plotone{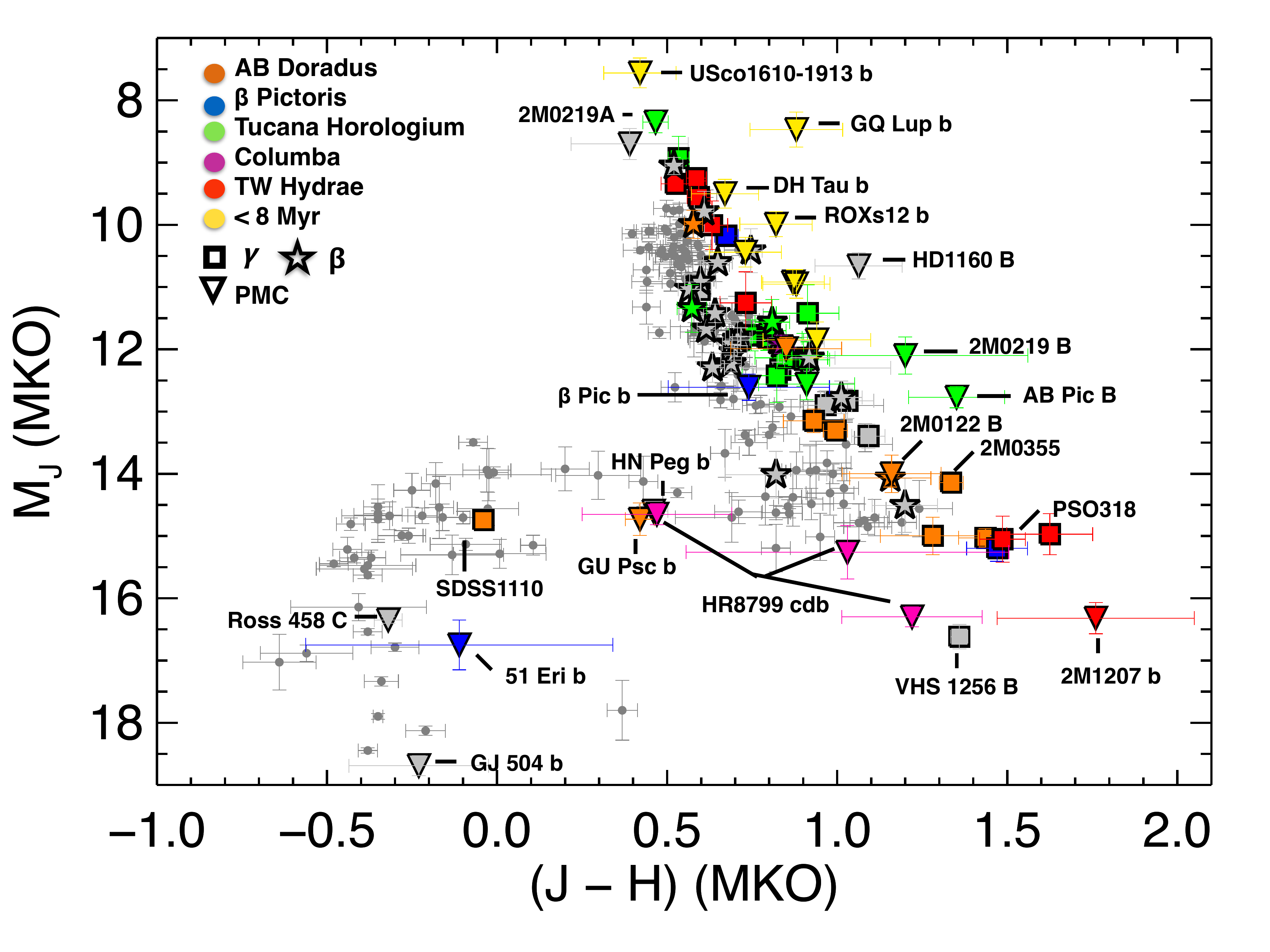}
\end{center}
\caption{The ($J$-$H$) versus $M_{J}$ color magnitude diagram for brown dwarfs and directly imaged planetary mass companions.  All photometry is on the MKO system.  For the brown dwarfs lacking MKO L' photometry, WISE W1 mags were converted using a polynomial listed in Table~\ref{tab:polynomials}. Objects have been color coded by nearby moving group membership and those of interest discussed in detail within the text have been labeled.  
\label{fig:JvJmHwPLANETSA}} 
\end{figure*}

\begin{figure*}[!ht]
\begin{center}
\epsscale{1.0}
\plotone{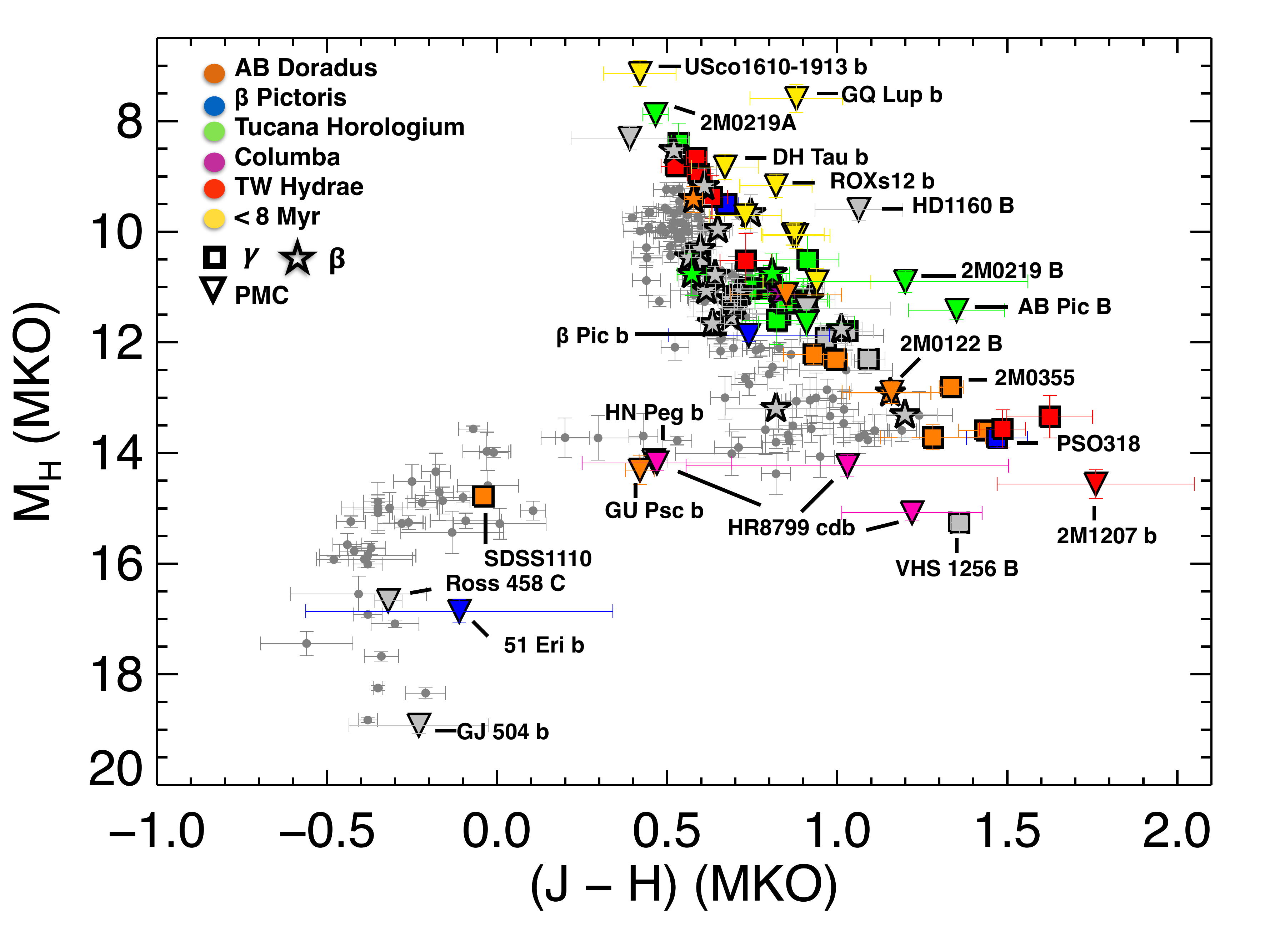}
\end{center}
\caption{The ($J$-$H$) versus $M_{H}$ (right) color magnitude diagram for brown dwarfs and directly imaged planetary mass companions. Symbols are as described in Figure~\ref{fig:JvJmHwPLANETSA}.  
\label{fig:JvJmHwPLANETSB}} 
\end{figure*}

\begin{figure*}[!ht]
\begin{center}
\epsscale{1.0}
\plottwo{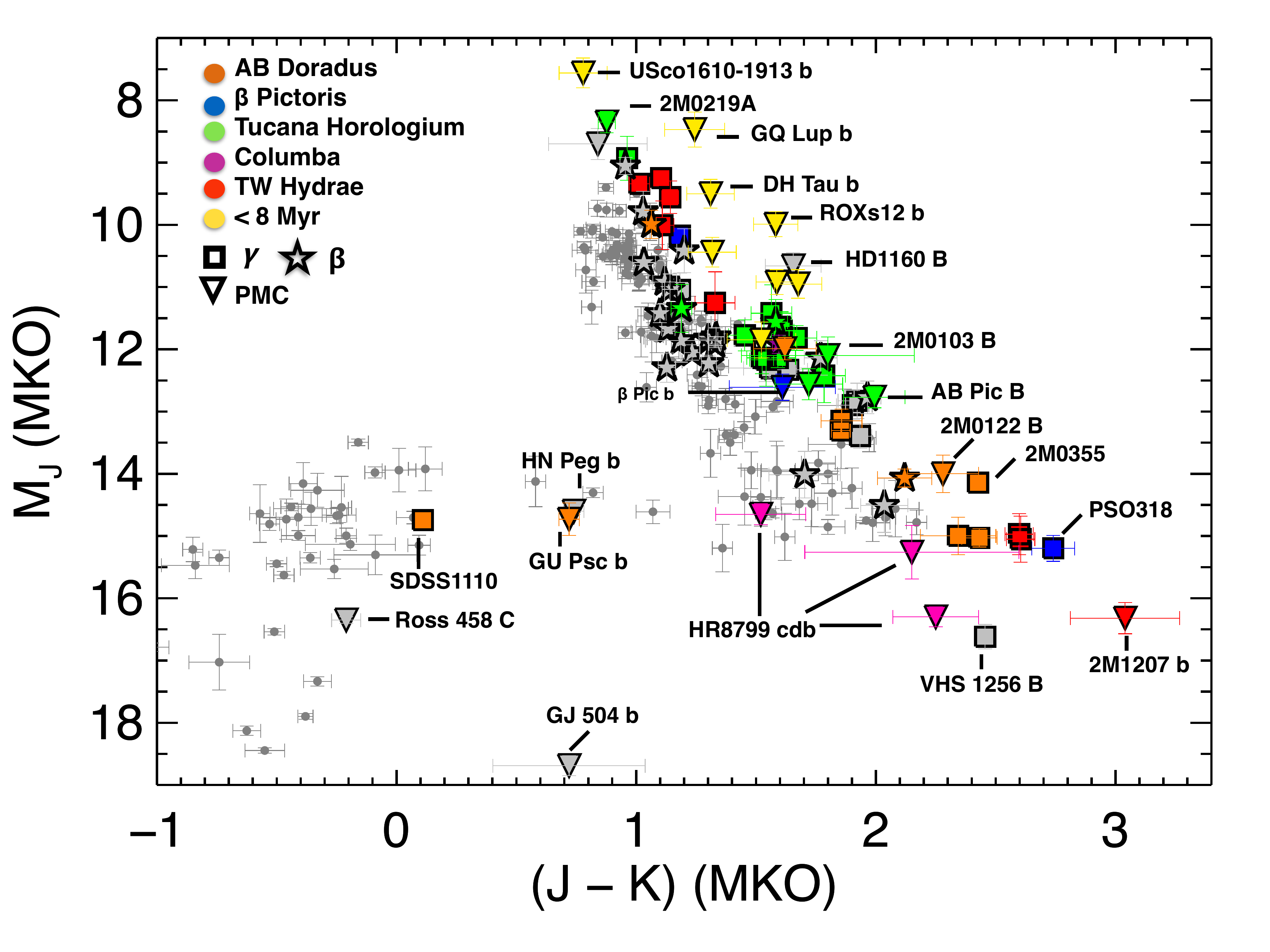}
{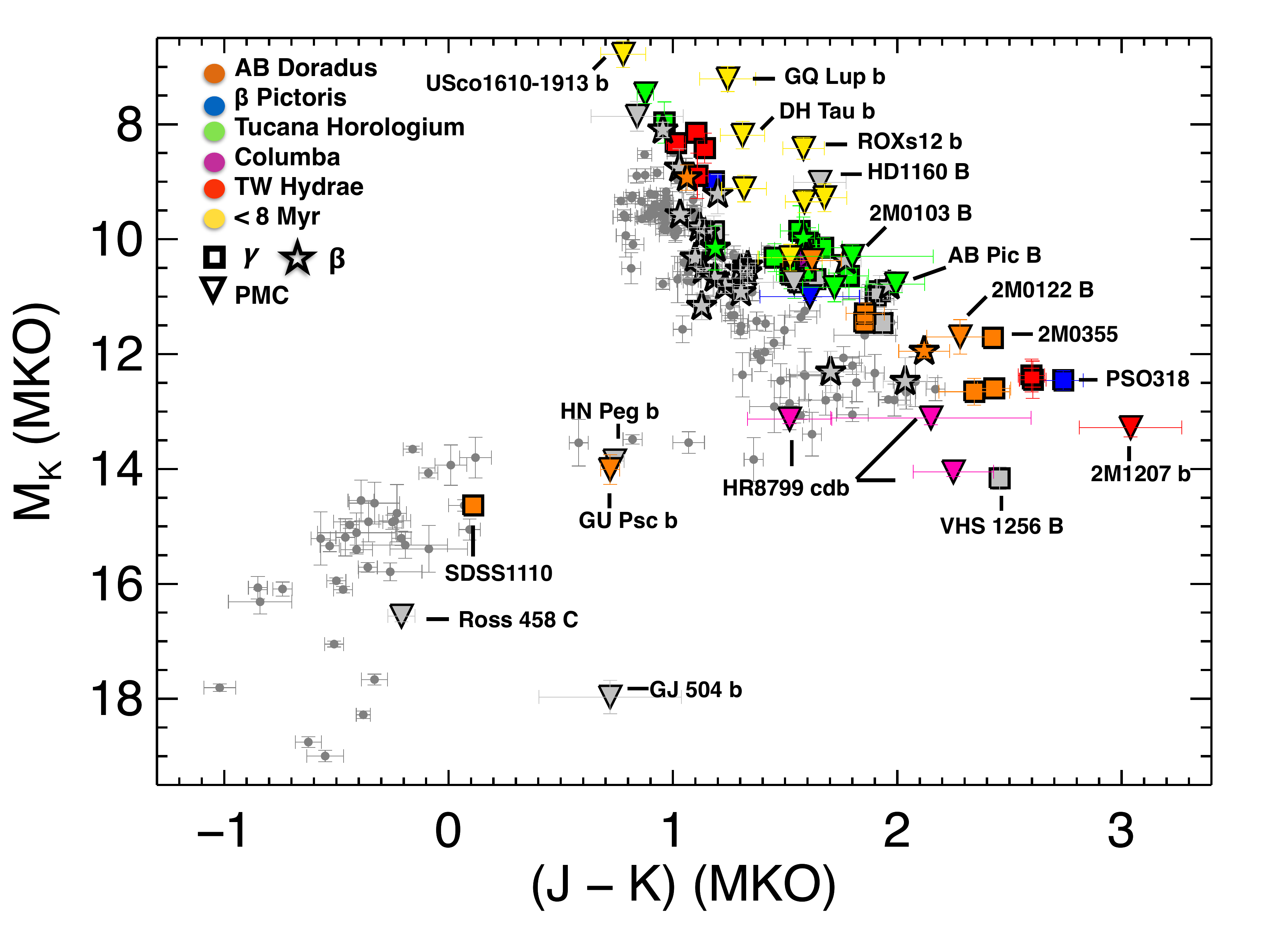}
\end{center}
\caption{The ($J$-$K$) versus $M_{J}$ (left) and $M_{K}$ (right) color magnitude diagram for brown dwarfs and directly imaged planetary mass companions. Symbols are as described in Figure~\ref{fig:JvJmHwPLANETSA}.    
\label{fig:JvJmKwPLANETS}} 
\end{figure*}

\begin{figure*}[!ht]
\begin{center}
\epsscale{1.0}
\plottwo{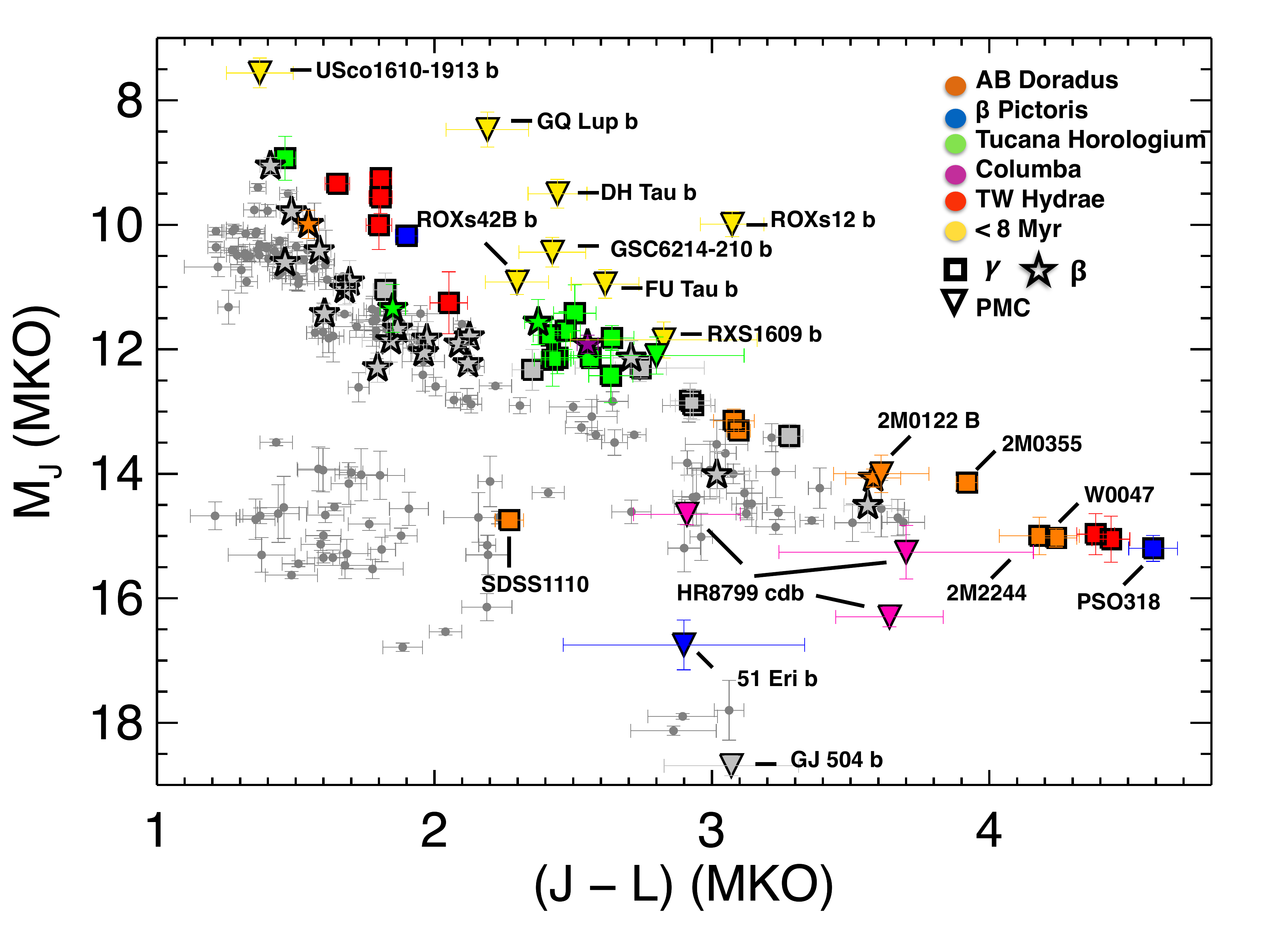}
{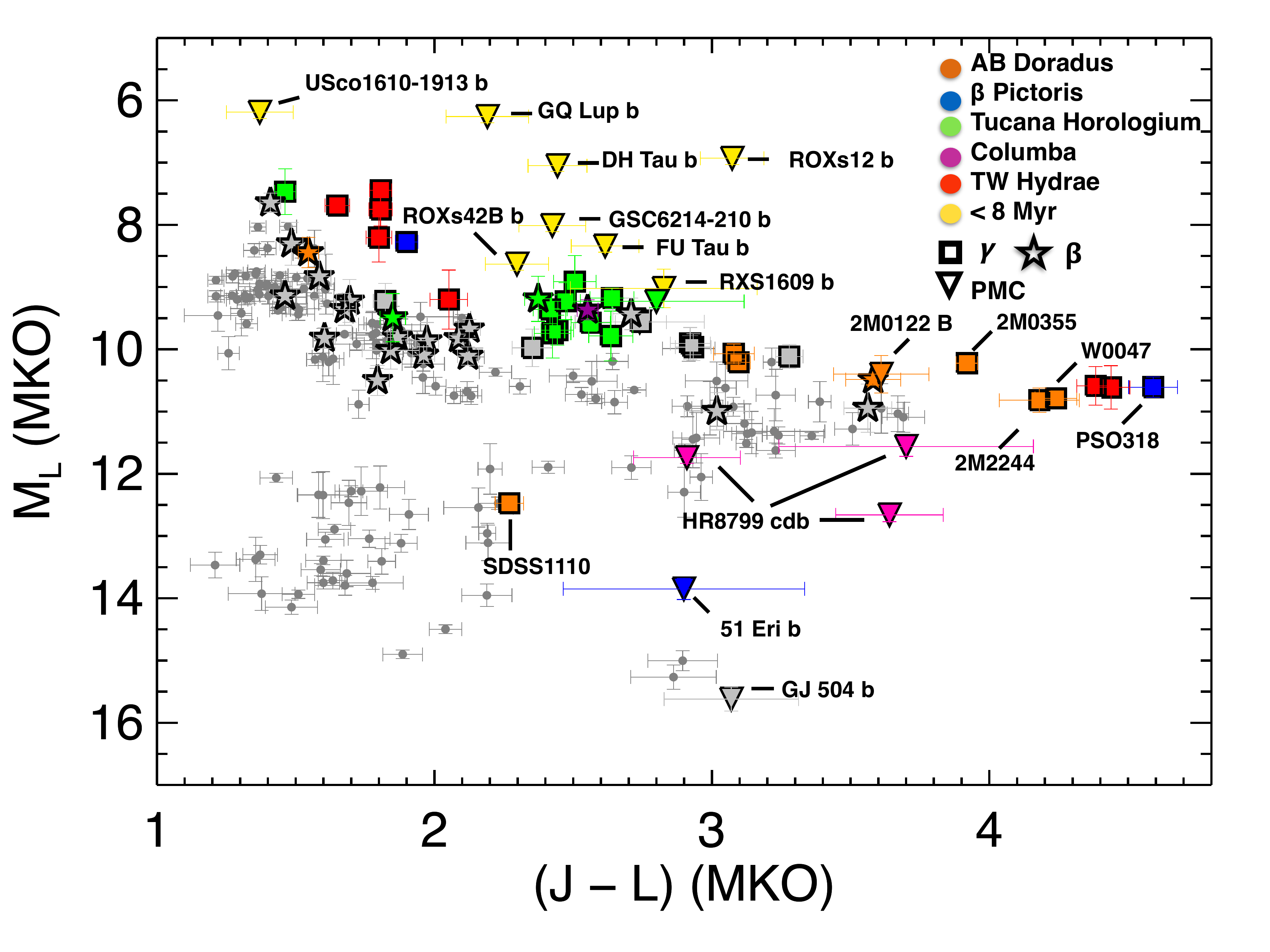}
\end{center}
\caption{The ($J$-$L$) versus $M_{J}$ (left) and $M_{L}$ (right) color magnitude diagram for brown dwarfs and directly imaged planetary mass companions. Symbols are as described in Figure~\ref{fig:JvJmHwPLANETSA}.   
\label{fig:JvJmLwPLANETS}} 
\end{figure*}

\begin{figure*}[!ht]
\begin{center}
\epsscale{1.0}
\plottwo{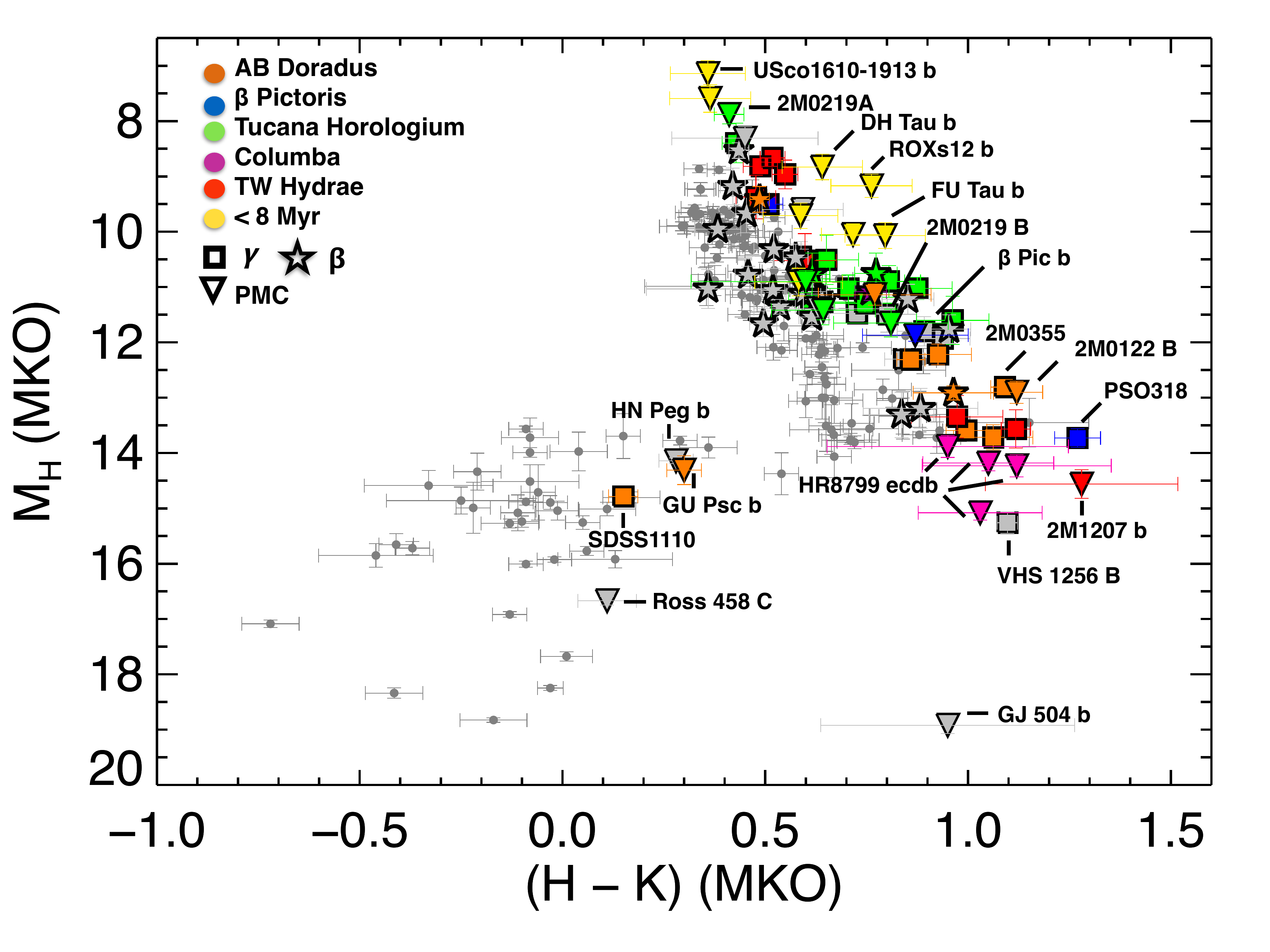}
{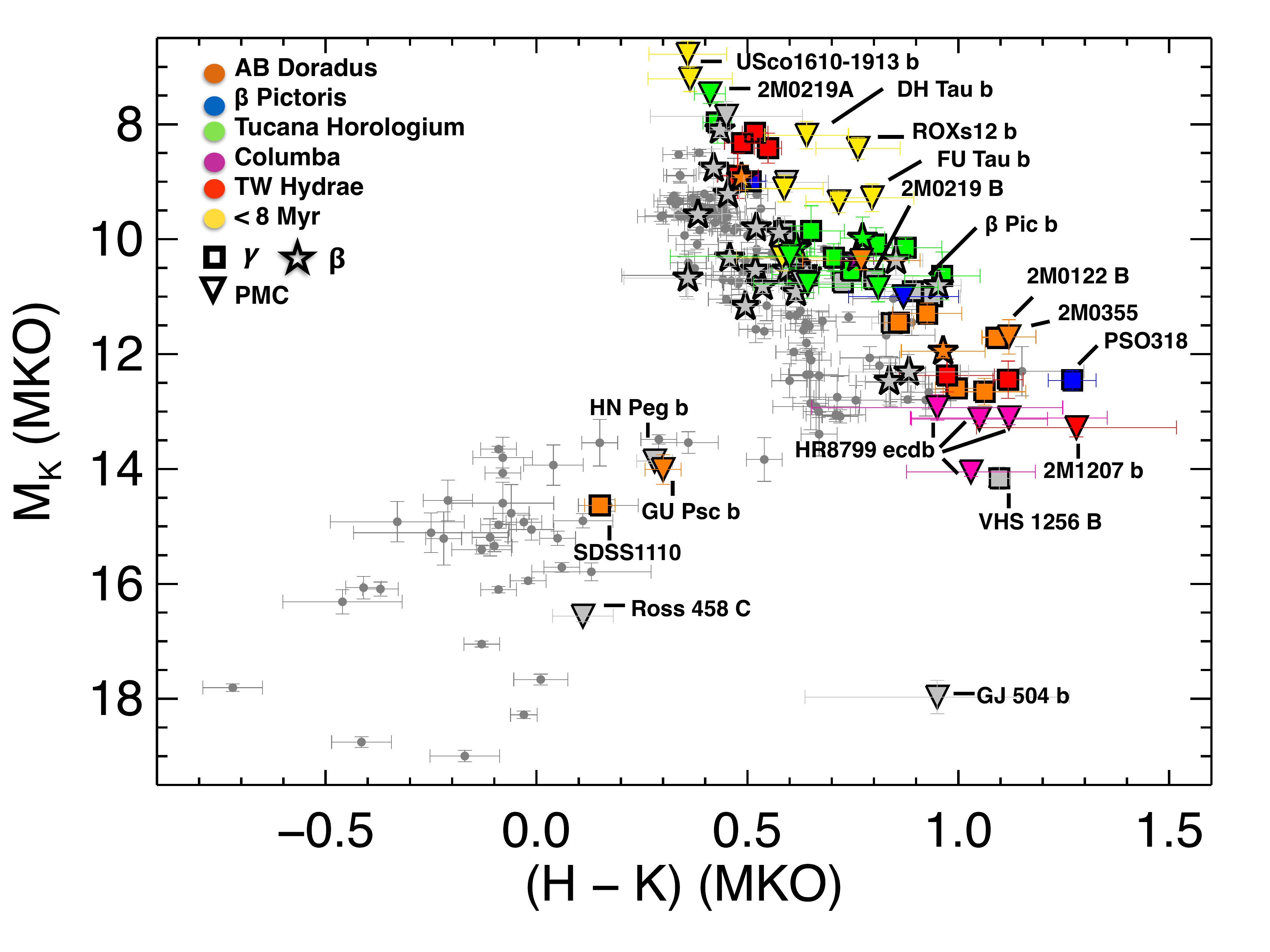}
\end{center}
\caption{The ($H$-$K$) versus $M_{H}$ (left) and $M_{K}$ (right) color magnitude diagram for brown dwarfs and directly imaged planetary mass companions. Symbols are as described in Figure~\ref{fig:JvJmHwPLANETSA}.   
\label{fig:HvHmKwPLANETS}} 
\end{figure*}

\begin{figure*}[!ht]
\begin{center}
\epsscale{1.0}
\plottwo{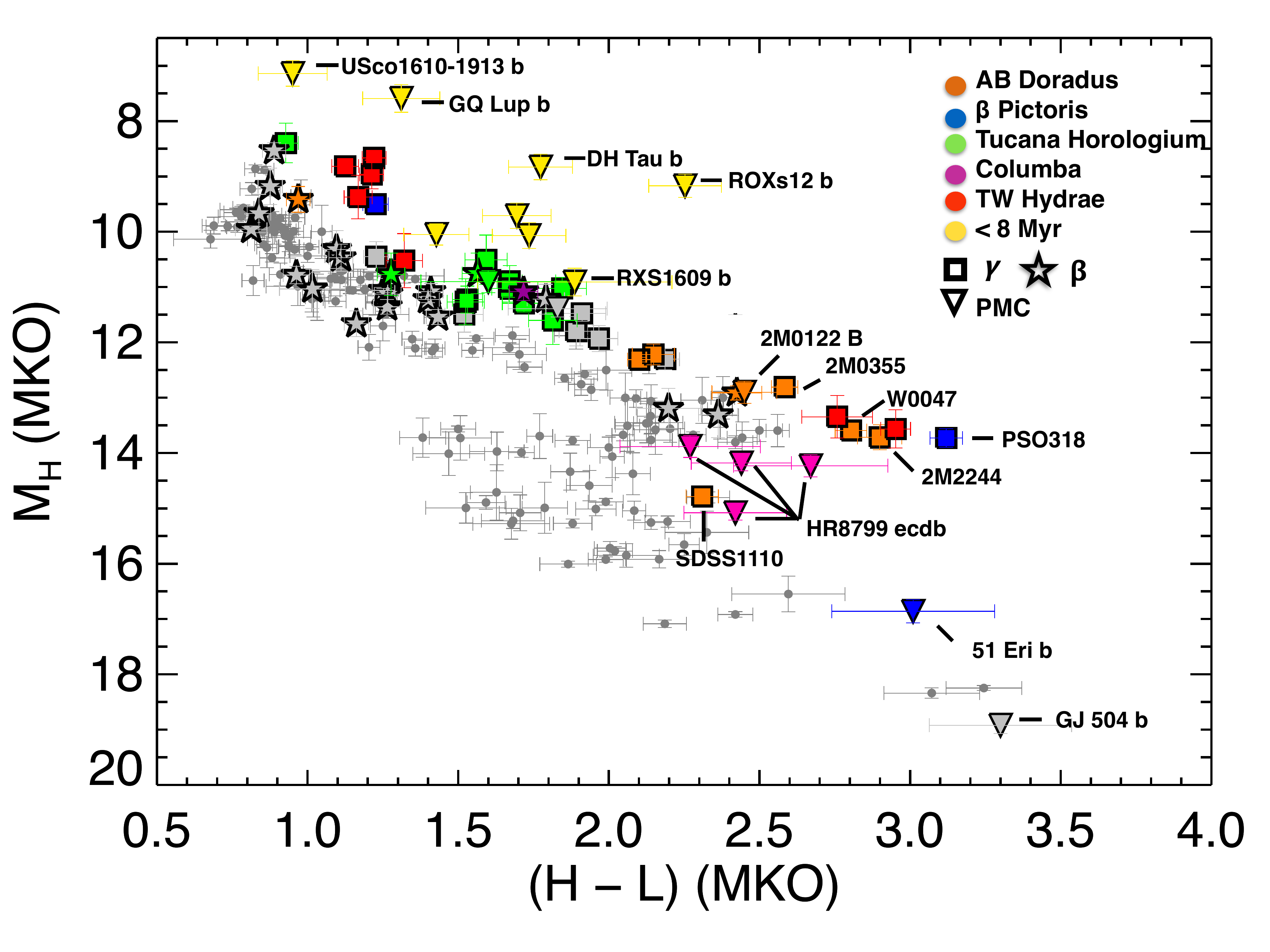}
{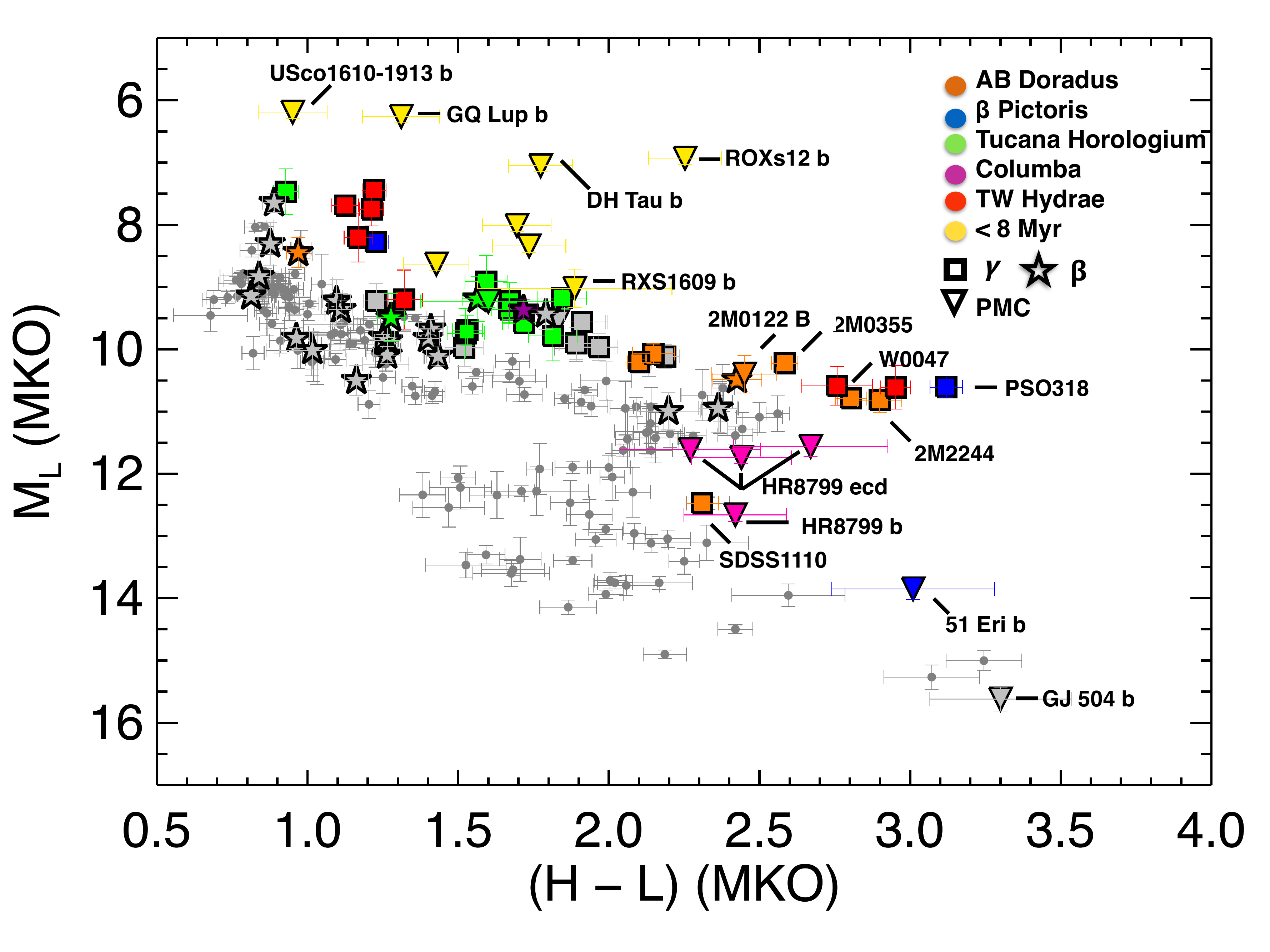}
\end{center}
\caption{The ($H$-$L$) versus $M_{H}$ (left) and $M_{L}$ (right) color magnitude diagram for brown dwarfs and directly imaged planetary mass companions. Symbols are as described in Figure~\ref{fig:JvJmHwPLANETSA}.   
\label{fig:HvHmLwPLANETS}} 
\end{figure*}

\begin{figure*}[!ht]
\begin{center}
\epsscale{1.0}
\plottwo{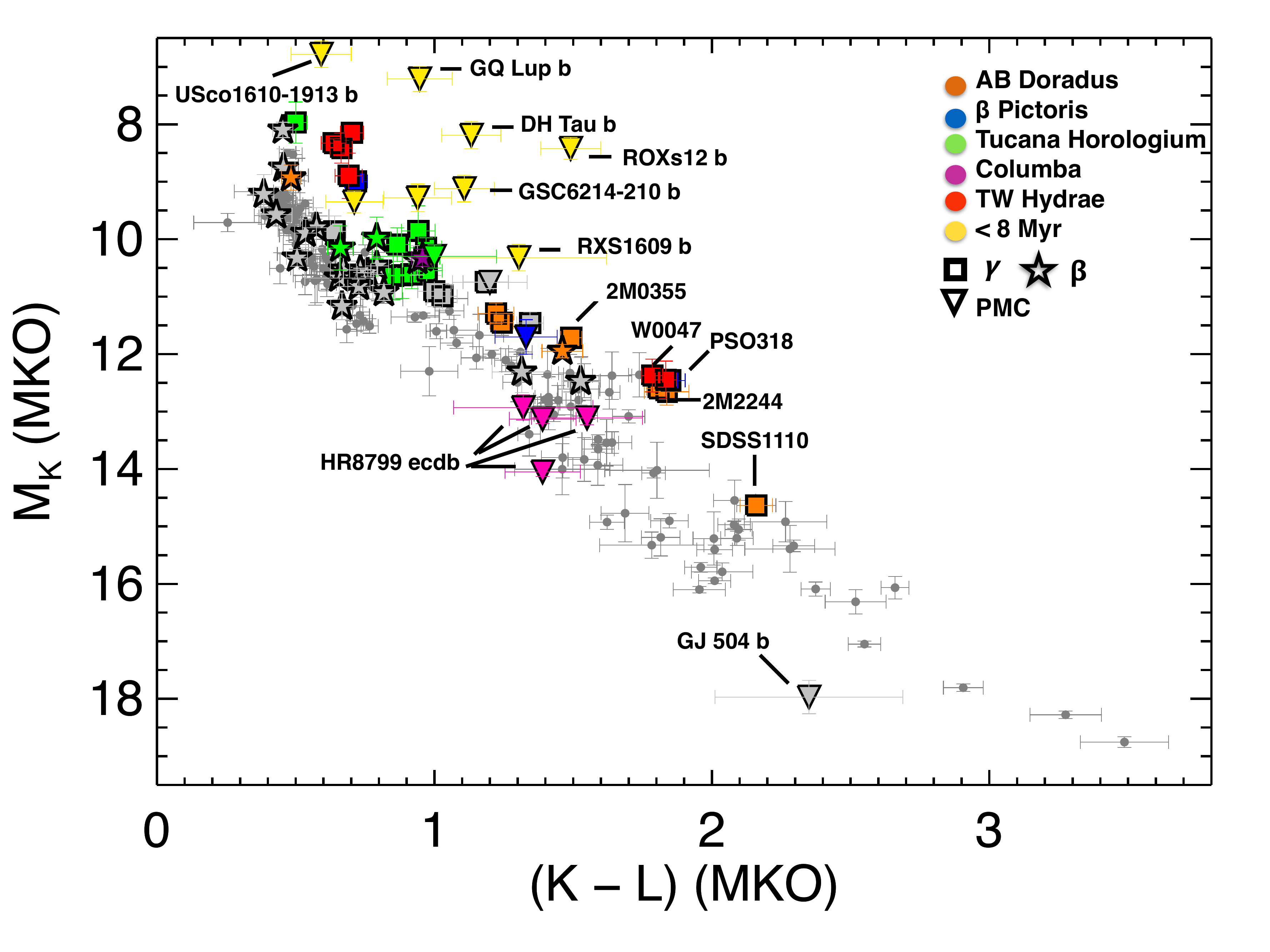}
{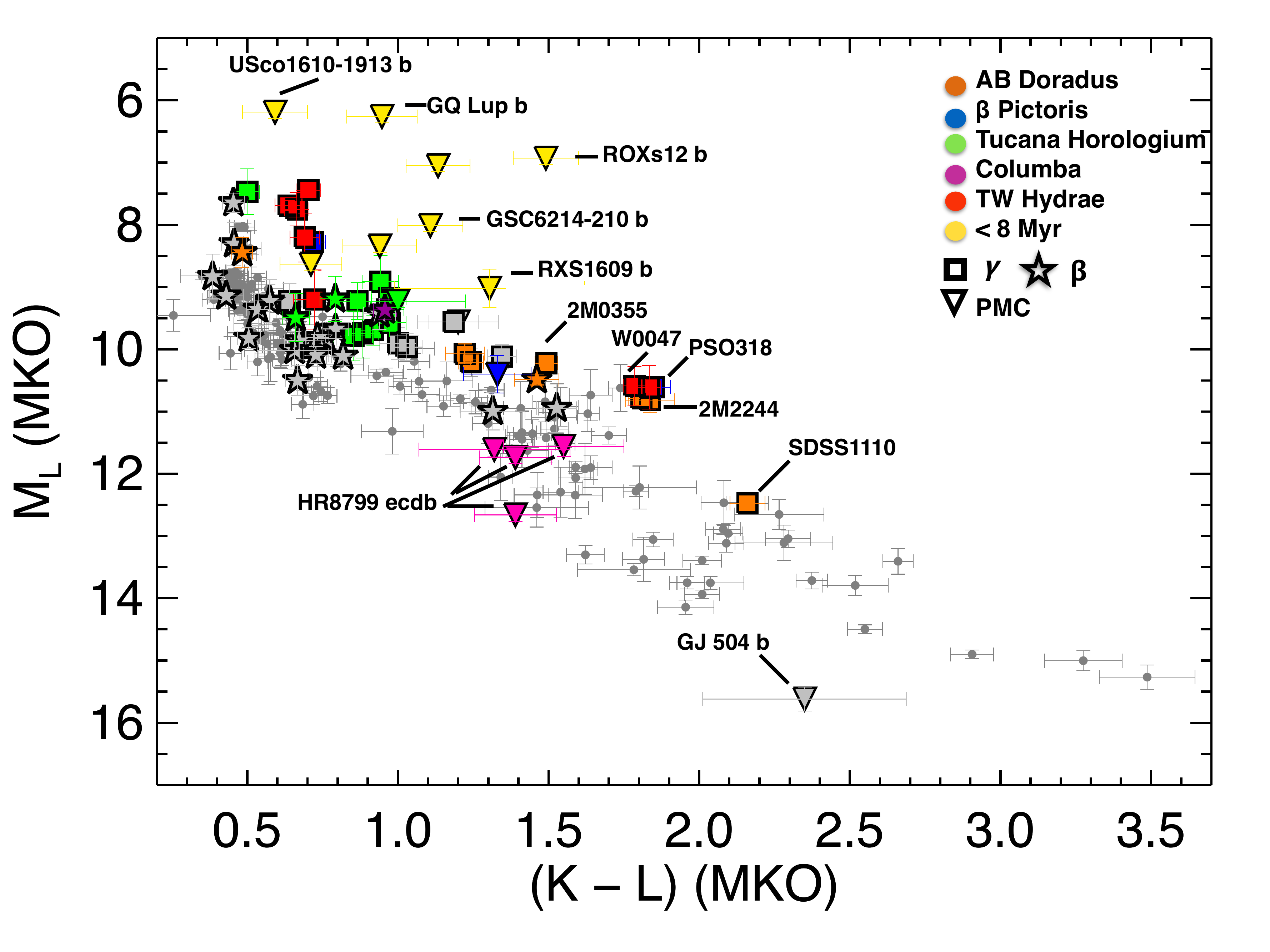}
\end{center}
\caption{The ($K$-$L$) versus $M_{K}$ (left) and $M_{L}$ (right) color magnitude diagram for brown dwarfs and directly imaged planetary mass companions. Symbols are as described in Figure~\ref{fig:JvJmHwPLANETSA}.   
\label{fig:KvKmLwPLANETS}} 
\end{figure*}

Figures~\ref{fig:JvJmHwPLANETSA} -~\ref{fig:KvKmLwPLANETS} show a full suite of near-infrared color magnitude diagrams with the same brown dwarf sample as in Figures~\ref{fig:JvJmH} -~\ref{fig:W1vW1mW2} however we now compliment each with directly imaged exoplanets and color code sources by their respective moving groups. For L' band photometry of the brown dwarfs, we have used the small sample of MLT sources with measured MKO L' -- primarily from \citet{Golimowski04} -- to convert the WISE W1 band photometry which has comparable wavelength coverage.  The polynomial relation used for converting between bands is listed in Table~\ref{tab:polynomials}.

As with the young brown dwarfs discussed herein, an observable feature of note for the exoplanets is that they are redder and fainter than the field brown dwarf population in near-infrared color magnitude diagrams.  This is exemplified by the positions of HR8799 b and 2M1207 b  both of which sit $\sim$ 1 mag below the L dwarf sequence in Figures ~\ref{fig:JvJmHwPLANETSA} ~\ref{fig:JvJmHwPLANETSB}, and ~\ref{fig:JvJmKwPLANETS}.  To explain their position on near-infrared color magnitude diagrams, several authors have proposed thick or high-lying  photospheric clouds in their atmospheres (\citealt{Bowler10}, \citealt{Madhusudhan11}, \citealt{Hinz10}, \citealt{Marley12}, \citealt{Skemer12}, \citealt{Currie11}). An alternative theory proposed by \citet{Tremblin16} has recently emerged that proposes cloudless atmospheres with thermo-chemical instabilities may invoke some of the features seen here-in. Further investigation of those models is required before we can appropriately comment on how well they may reproduce the large sample presented in this work.  

The young brown dwarf sequence in many ways mirrors that of the directly imaged exoplanets but for warmer (or older) objects.  From the warmest (at M7) to the coolest (at L7) the low-gravity brown dwarfs are redder and fainter than their field counterparts.  As can be seen on each Figure, the low-gravity sequence appears to logically extend through many of the directly imaged exoplanets.  Interestingly, the youngest exoplanets (those in Taurus, Upper Scorpius, and $\rho$ Ophiuchus; \citealt{Kraus14a}, \citealt{Currie14}) are redder than either the field or the low-gravity sequences indicating that -- assuming formation differences do not drastically alter the available compositions -- the atmosphere or gravity effects are most pronounced close to formation.  

By color combinations using L' band, the youngest planetary mass objects such as ROXs12 b, DH Tau b, and GQ Lup b are nearly 2 magnitudes redder and $>$ 2 magnitudes brighter than the field sequence.  Conversely, the lowest temperature planets around HR8799 fall within or close to the T dwarf sequence.  The exception is HR8799 b which has a singificantly redder ($J$-$L$) color than the field sequence as well as the young brown dwarf sequence.   It is also $\sim$ 2 magnitudes fainter in $M_{L}$ than the low-gravity sequence which extends redward in Figure~\ref{fig:JvJmLwPLANETS} with minimal scatter in $M_{L}$. \citet{Skemer12} have noted that the 3.3$\mu$m photometry (not plotted) for the HR8799 planets is brighter than predicted by evolutionary or atmosphere models.  As in \citet{Barman11}, \citet{Skemer12} use thick cloudy non-equilibrium chemistry models and remove CH$_{4}$ to fit the data.  Similarly, none of the isolated late-type L dwarfs labeled on Figures~\ref{fig:JvJmHwPLANETSA} -~\ref{fig:KvKmLwPLANETS} have CH$_{4}$ in their near infrared spectra even though 1119-1137, 1147-2040, PSO318 and 0047+6803 have calculated $T_{eff}$s which should allow for detectable CH$_{4}$.  Interestingly only the planetary mass companion 2M1207 b rivals the far reaching red sequence of the young brown dwarfs, yet that source is lacking an L' or equivalent band detection.  Hence at this point the late-type young brown dwarf sequence prevails over the exoplanet sequence in their extreme L' colors even as the earliest type youngest planetary mass sources prevail at slightly bluer magnitudes.     

The latest-type low-gravity brown dwarfs, and the known directly imaged exoplanets push the L/T transition to cooler temperatures and redder colors.  Interestingly we have two sets of objects in the same group that span either side of the famed L/T transition.  0047+6803 ($T_{eff}$=1227$\pm$ 30 K, Mass= 9 - 15 $M_{Jup}$) and SDSS1110  ($T_{eff}$= 926 $\pm$ 18, Mass= 7 - 11 $M_{Jup}$) are both in the $\sim$110 - 130 Myr AB Doradus  moving group.  PSO318 ($T_{eff}$= 1210 $\pm$ 41, Mass= 5 -  9 $M_{Jup}$) and 51 Eri b ($T_{eff}$= 675 $\pm$ 75, Mass= 2 - 12$M_{Jup}$) are both in the $\sim$ 25 Myr $\beta$ Pictoris moving group.  The two sets differ by $\sim$100 Myr in age.  Comparing 0047+6803 to PSO318 we find they have similar spectral types, have T$_{eff}$ within 1$\sigma$ but may differ in mass by up to $\sim$10 $M_{Jup}$ .  Both push the L/T transition redder on multiple color magnitude diagrams in Figures~\ref{fig:JvJmH} -~\ref{fig:W1vW1mW2} and Figures~\ref{fig:JvJmHwPLANETSA} -~\ref{fig:KvKmLwPLANETS}.  Overall PSO318 and 0047+6803 have similar absolute magnitudes (see Figures~\ref{fig:MJ} -~\ref{fig:MW3}) although PSO318 can be significantly redder in specific colors.   51 Eri b and SDSS1110 are thought to have similar spectral types but differ in mass and $T_{eff}$ by as much as 344 K and 5 $M_{Jup}$ respectively.  Interestingly, 51 Eri b is $\sim$ 2 mag fainter in $M_{JHL}$ than SDSS 1110. Comparing its position on Figures~\ref{fig:JvJmHwPLANETSA},~\ref{fig:JvJmLwPLANETS} and~\ref{fig:HvHmLwPLANETS}, 51 Eri b appears much more like the T8 Ross 458 C thought to be 150 - 800 Myr (\citealt{Burgasser10}).  Regardless for both groups we see that by the time we have reached the mid to late-T dwarf phase, sources are back on or very close to, the field sequence.  As cloud clearing is thought to happen at the L/T transition, we surmise this is further evidence that much of the diversity seen among the young, warm exoplanet and brown dwarf population is atmosphere related.  

\begin{figure*}[t!]
\center
\includegraphics[angle=0,width=1.0\textwidth]{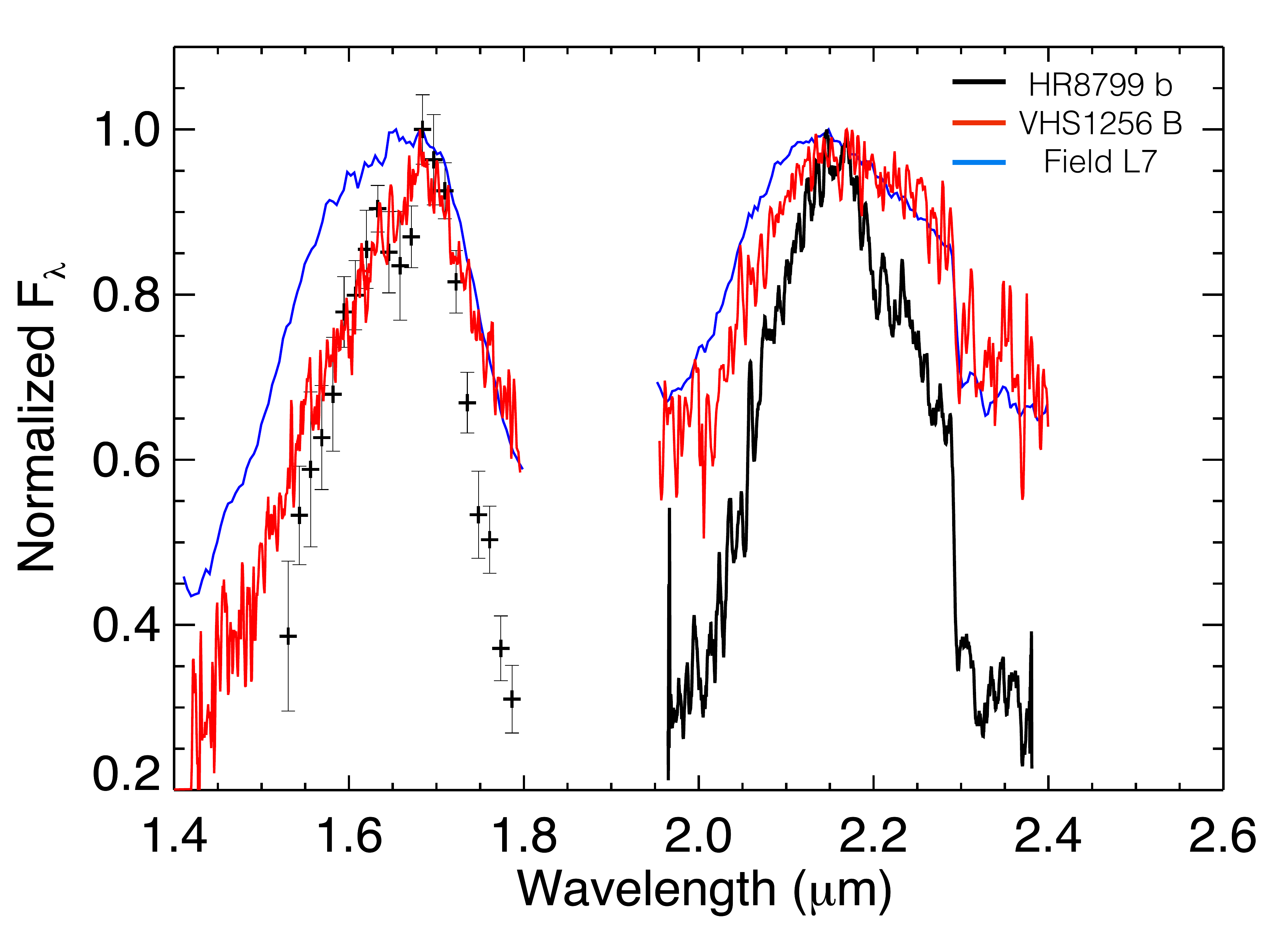}
\caption{The near-infrared spectrum comparison of HR8799 b (black, from \citealt{Oppenheimer13}, \citealt{Barman11}), VHS 1256 B (red, from \citealt{Gauza15}), and a field L7 (\citet{Gagne15b}).  }
\label{fig:planetcomparison}
\end{figure*}

A note of caution when looking for similarities on color magnitude diagrams between brown dwarfs and giant exoplanets comes in the way of comparing the newly discovered L7 companion VHS1256 B to HR8799 b.  In \citet{Gauza15}, the authors noted a similarity of this companion and the giant exoplanet. On Figure~\ref{fig:JvJmKwPLANETS}, the two have enticingly similar near-infrared values.  However, Figure~\ref{fig:planetcomparison} shows the near-infrared spectrum of each object as well as a field L7 to demonstrate strong differences in H-band.  Clearly there is some commonality between the two sources, but color magnitude combinations alone do not give a full enough picture to draw conclusions.   

The directly imaged companion GJ 504 b is another example of how we require multiple color magnitude diagram plots to begin exploring the potential characteristics of a single object (\citealt{Kuzuhara13}).  GJ 504 b is nearly 1 magnitude redder than late-type T dwarfs in $M_{JHK}$ vs ($J$-$K$)  or ($H$-$K$) diagrams (Figures~\ref{fig:JvJmKwPLANETS} and~\ref{fig:HvHmKwPLANETS}), but it appears normal in  $M_{JH}$ vs ($J$-$H$) plots (Figure~\ref{fig:JvJmHwPLANETSA}).  Recent work by \citet{Fuhrmann15} suggest the primary may not be young therefore this source may not be planetary mass.  Regardless it is a low mass T dwarf orbiting at $<$ 50 AU, potentially formed in a disk around a nearby star hence it may be characteristically different than equivalent temperature T dwarfs (see e.g. \citealt{Skemer15}).

\subsubsection{Bolometric Luminosities}
Comparing the bolometric luminosities (L$_{bol}$) across the sample of field objects, new bona fide or high-likelihood moving group members, and directly imaged exoplanets allows us to investigate how the flux varies across the sample.   For the directly imaged exoplanets, we use the L$_{bol}$ values reported in the literature (see \citealt{Males14}, \citealt{Bonnefoy14}, \citealt{Currie14}). 
 
\begin{figure*}[t!]
\center
\includegraphics[angle=0,width=1.0\textwidth]{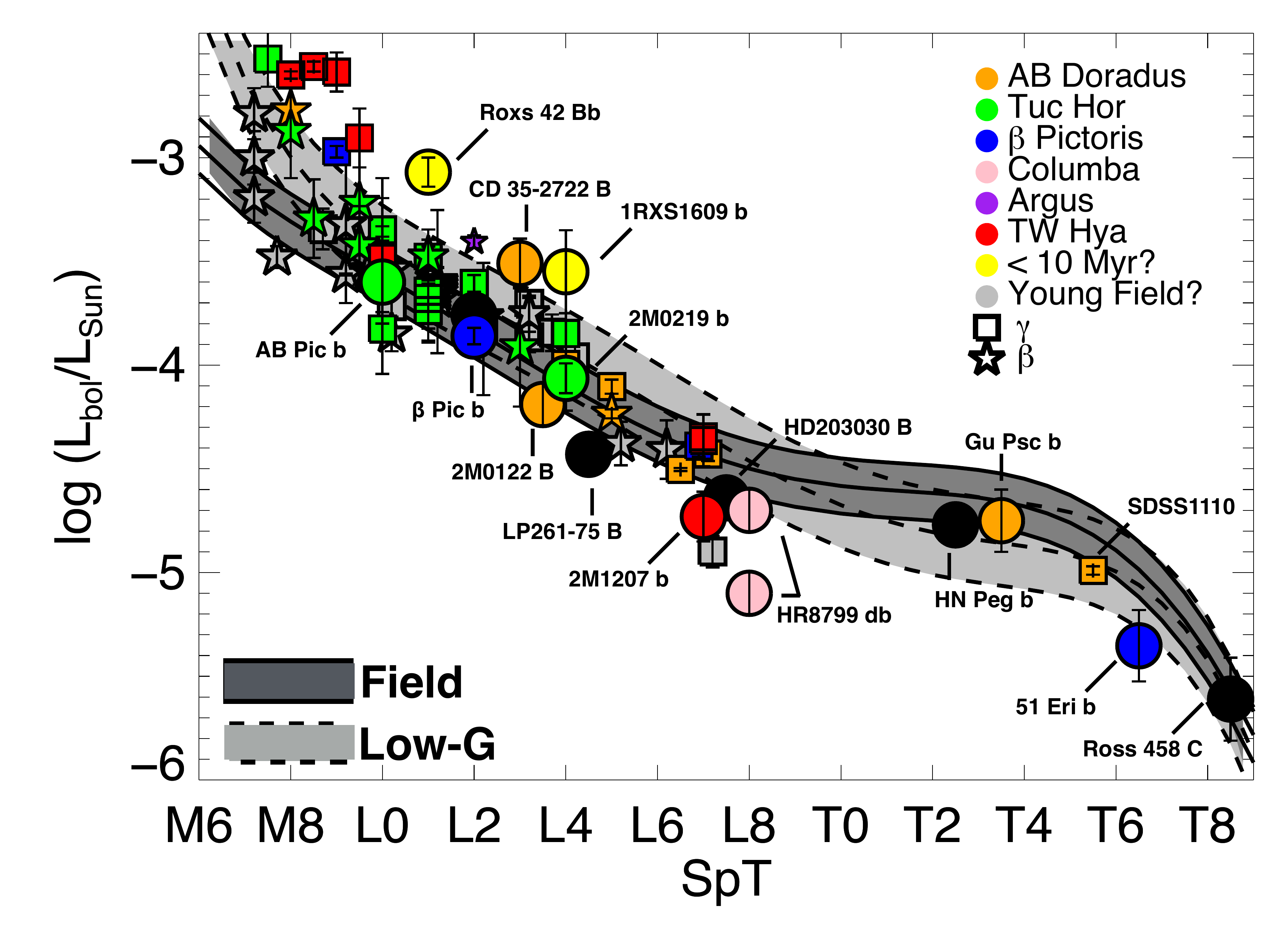}
\caption{The spectral type versus bolometric luminosity plot for brown dwarfs and directly imaged planets.  $L_{bol}$ values and spectral types for planets have been taken from the literature with the exception of HR8799bd and 2M1207b which we delegate as L8 objects to represent their nature as late-type objects.  Symbols are as in Figure~\ref{fig:Lbol}. }
\label{fig:Lbol2}
\end{figure*}

As discussed in both section~\ref{sec:bols} and \citet{Filippazzo15}, the young M dwarfs are overluminous for their spectral type while the young L's are normal to slightly underluminous. Several authors have noted this peculiarity among the directly imaged exoplanets as well (e.g. \citealt{Bowler13}, \citealt{Males14}).  Examining the two populations together allows us to investigate whether there is an obvious age-associated correlation within the scatter.  The youngest objects in Figure~\ref{fig:Lbol2} are the directly imaged exoplanets such as ROXs 42B b, and 1RXS1609 b which belong to star forming regions at just a few Myr of age.  These sources appear overluminous in comparison to equivalent sources regardless of how late their spectral type (e.g. 1RXS1609b which is thought to resemble an L4).  The latest type planets that have direct, comparable data -- 2M1207 b,  HR8799 b, 51 Eri b -- are all $\sim$ 1 $\sigma$ or more below the field sequence indicating that they are either far cooler than the latest L dwarfs or there is unaccounted flux in the bolometric luminosity calculations.  Interestingly, this appears to be where the young brown dwarf and the directly imaged exoplanet comparisons diverge.  The latest type exoplanets in equivalent age groups to the isolated brown dwarfs, are either far cooler than any currently discovered young brown dwarf equivalent, or the physical conditions diverge (e.g. atmosphere conditions, chemistry, other).

 \subsubsection{Masses from combining Evolutionary Models with L$_{Bol}$, and Age}
\begin{figure*}[t!]
\center
\includegraphics[angle=0,width=1.0\textwidth]{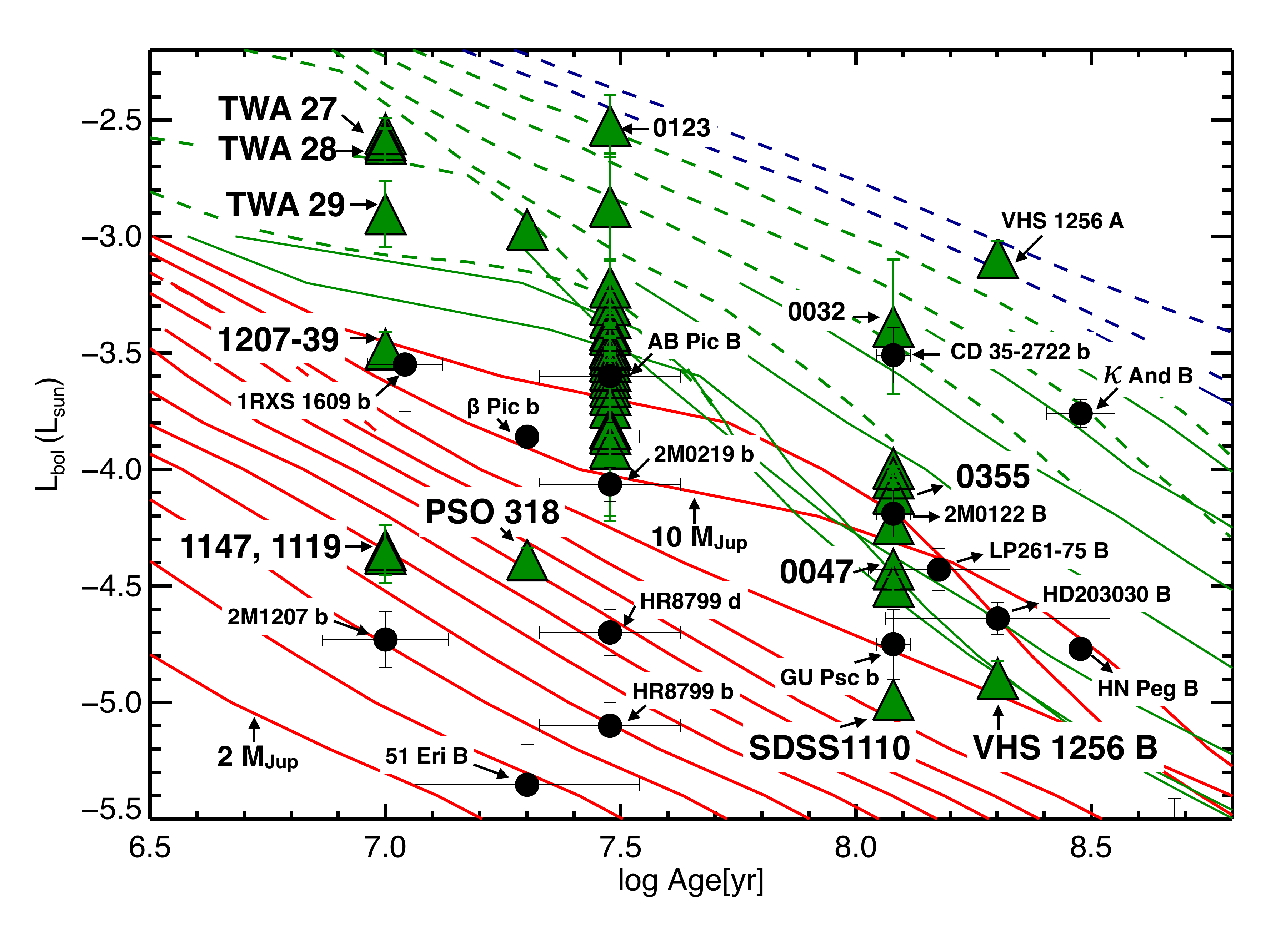}
\caption{The age versus bolometric luminosity plot with model isochrone tracks at constant mass from \citet{Saumon08} (solid lines) and \citet{Baraffe15} (dashed lines).  We have color coded $<$13 $M_{Jup}$ tracks in red, 13 $M_{Jup}$ $<$ M $<$ 75 $M_{Jup}$ tracks in green and $>$ 75 $M_{Jup}$ blue.  Over-plotted are both the young brown dwarfs discussed in this work and directly imaged exoplanets with measured quantities.}
\label{fig:Lbolage}
\end{figure*}

In Figure~\ref{fig:Lbolage}, we combine the L$_{bol}$ values with the ages of the moving group members to estimate masses. In Table~\ref{tab:LbolTeffMass} we also list masses calculated from the spectral energy distribution analysis as described in \citet{Filippazzo15}.  Over-plotted on Figure~\ref{fig:Lbolage} with the young brown dwarfs are directly imaged planetary and brown dwarf mass companions with $L_{bol}$ values collected from the literature.  The models from \citet{Saumon08}(solid) and \citet{Baraffe15} (dashed) are shown with lines of equal mass color coded as stars ($>$ 75 M$_{Jup}$, blue) brown dwarfs ($>$ 13 M$_{Jup}$, green) and planets ($<$ 13 M$_{Jup}$, red).  

It is unclear how each one of the objects in this sample formed however, using the 13  M$_{Jup}$ boundary as a mass distinguisher between brown-dwarf and planet-type objects, we find that there are close to 9 solitary objects with masses $<$ 13 M$_{Jup}$.  Several of those sources lie in an ambiguous area where low mass brown dwarfs and planetary mass objects cross (30  - 130 Myr between masses of 10 - 20 $M_{Jup}$).

With the exception of Y type objects whose age is still undetermined (e.g. W0855, \citealt{Luhman14}), 1119-1137, 1147-2040, and PSO318 are the lowest mass isolated sources categorized.  PSO 318 has a lower  luminosity than $\beta$ Pictoris b, the 11 - 13 M$_{Jup}$ planet in the same association while 1119-1137 and 1147-2040 are significantly more luminous than their sibling exoplanet 2M1207b.  Interestingly, the AB Doradus equivalently typed L7 member, 0047+6803, is higher mass than all of these sources even though its bolometric luminosity is comparable.  Similarly, 0355+1133 which is in the same group as 0047+6803, shares photometric anomalies with PSO318 (near-infrared and mid infrared color) and spectral anomalies with 2M1207b yet it is much higher mass.  

 The overlap in masses of the directly imaged exoplanets and isolated brown dwarfs invites questions of formation given co-evolving groups.  Whether the latter was formed via star formation processes or ejected after planetary formation processes is yet to be seen and requires further investigation. 

\section{Conclusions}
In this work we investigate the kinematics and fundamental properties of a sample of 152 suspected young brown dwarfs.  We present near-infrared spectra and confirm low-surface gravity features for 43 of the objects designating them either intermediate ($\beta$) or very low ($\gamma$) gravity sources.  We report 18 new parallaxes (10 field objects for comparison and 8 low surface gravity), 19 new proper motions, and 38 new radial velocities and investigate the likelihood of membership in a nearby moving group.  We use four kinematic membership codes (1) BANYAN I, (2) BANYAN II, (3) LACEwING, and (4) Convergent method, as well as a visual check of the available space motion for each target against known members of well known nearby kinematic groups to determine the likelihood of co-membership for our sources.  We categorize objects as (1) bona fide --BM, (2) high-likelihood --HLM, (2) ambiguous --AM, or (3) non-member --NM  of nearby moving groups. We find 39 sources are bona fide or high-likelihood members of known associations (8 in AB Doradus, 1 in Argus, 2 in $\beta$ Pictoris, 1 in Columba, 7 in TW Hydrae, and 20 in Tucana Horologium). A further 92 objects have an ambiguous status and 21 objects are not members of any known group evaluated in this work. 

Examining the distribution of gravity classifications between different groups we find that the youngest association (TW Hydrae) has only very low-gravity  ($\gamma$) sources associated with it but slightly older groups such as Tucana Horologium (9 optically classified, 10 infrared classified $\gamma$ objects and 3 optically classified, 5 infrared classified $\beta$ objects) and AB Doradus (4 optically classified, 6 infrared classified $\gamma$ objects and 1 optically classified, 2 infrared classified $\beta$ objects) show a mix of both intermediate ($\beta$) and very low-gravity sources.  This diversity is evidence that classically delegated gravity features in the spectra of brown dwarfs are influenced by other parameters such as metallicity or (more likely) atmospheric conditions.  

We investigate colors for the full sample across the suite of MKO, 2MASS and WISE photometry  ($J,H,K_{s},W1,W2,W3$).  In color versus spectral type diagrams, we find that the $\gamma$ and $\beta$ classified objects are distinct from the field ($>$ 3 Gyr sources).  They are most deviant from the field sequence in the ($J$-$W2$) color where they are an average of 2$\sigma$ redder than the subtype mean.  They are least deviant in the ($J$-$H$) color where they are an average of 0.7$\sigma$ redder than the subtype mean.  Based on the 5 -130 Myr age calibrated sample, we conclude that the extent of deviation in infrared color is not indicative of the age of the source (meaning redder does not translate to younger).  In any given color a $\gamma$ or a $\beta$ object -- whether confirmed in a group or not  -- may mark the extreme outlier for a given subtype.  We find that the L0 dwarfs in Tucana Horologium (expected to have the same, T$_{eff}$, age and metallicity) deviate from the field sequence of infrared colors by between 1 and 4 $\sigma$ (depending on the particular color examined).  Assuming clouds are the source of the diversity, we conclude that there is a variation in cloud properties between otherwise similar objects.

Examination of the absolute magnitudes for the parallax sample indicates a clear flux redistribution for low- and intermediate-gravity brown dwarfs (compared to field brown dwarfs) from near-infrared to wavelengths at (and longer than) the WISE W3 band.  There is also a clear correlation of this trend with spectral subtype. The M dwarfs are 1-2$\sigma$ brighter than field equivalents at $J$ band but  4-5$\sigma$ brighter at $W3$.  The L dwarfs are 1-2$\sigma$ fainter at $J$ band but  1-2$\sigma$ brighter at $W3$.  Clouds, which are a far more dominant opacity source for L dwarfs, are likely the cause.  

Sources that are not confirmed in groups do not all trace the same behavior in absolute magnitude or color indicating that some sources may not be young.  Variations in atmospheric conditions or metallicity likely drive the diversity.   

On color-magnitude diagrams, the low-surface gravity brown dwarfs  pull the elbow of the field L/T transition significantly redder and fainter with the most extreme case being the $M_{J}$ vs ($J$-$W2$)  plot where young objects are up to 1 mag redder than the field sequence.  Conversely, the $M_{K}$ versus ($K$-$W2$) plot shows a 10 magnitude difference between the warmest and coolest brown dwarfs yet seems to monotonically decrease in magnitude with reddening color.  On this figure there is little evidence for an L/T transition elbow and the young objects form a secondary sequence that is $\sim$1 mag redder than the field sequence.  Interestingly as we move to longer wavelengths the effect reverses and the latest type objects pull the elbow of the L/T transition back up as they are equivalent or slightly brighter than field equivalents at $M_{W1,W2}$.  Comparing the sequence of $\gamma$ and $\beta$ classified sources on CMD's compared to the field, we find a hint that the two are distinct with the former redder than the latter.  This trend is clearest on the $M_{J}$ versus ($J$-$W2$) figure although there is still a small mix of $\gamma$ and $\beta$ sources at extreme colors for a given absolute magnitude.  

Comparing the low-gravity sample with directly imaged exoplanets on color magnitude diagrams we find that the former sequence logically extends through the latest type planets on multiple color magnitude diagrams.  The small collection of hot, planetary mass objects in star forming regions such as $\rho$ Ophiuchus and Taurus are strikingly red, bright, and luminous compared to either the field sequence or the low-gravity objects indicating that the atmosphere and/or gravity effects that drive the population diversity may be pronounced close to formation.   Comparing $\beta$ Pictoris members 51 Eri b (mid T dwarf) and PSO318 (late L dwarf) with AB Doradus members SDSS1110 (mid T dwarf) and 0047+6803 (late L dwarf) we find that even though the members differ by $\sim$ 100 Myr, the late-type L's similarly push the elbow of the L/T transition redder and fainter whereas the T dwarfs appear on or very close to the field sequence.  We surmise that this behavior, seen in two sets of objects at different ages across the L/T transition where cloud clearing is thought to be significant, is evidence that much of the diversity seen among young warm exoplanet and brown dwarfs is atmosphere related.

\clearpage

\clearpage
\afterpage{\clearpage}

\clearpage
\onecolumngrid
\LongTables
\newpage
\begin{landscape}
\tabletypesize{\scriptsize}
\begin{deluxetable}{lllllllllll}
\tablecaption{Photometric Properties of Low Surface Gravity Dwarfs}
\tablehead{
\colhead{2MASS Designation}    &
\colhead{SpT}&
\colhead{SpT}&
\colhead{J}       &
\colhead{H}       &
\colhead{K}       &
\colhead{W1}      &
\colhead{W2}      &
\colhead{W3}      &
\colhead{W4}      \\
 &
 \colhead{ OpT}&
 \colhead{IR}   &
      \colhead{2MASS}&
   \colhead{2MASS}&
      \colhead{2MASS}&
         \colhead{WISE}&
            \colhead{WISE}&
               \colhead{WISE}&
                              \colhead{WISE}&
                              \colhead{Ref\tablenotemark{e}}\\
\colhead{(1)}     &
\colhead{(2)}     &
\colhead{(3)}     &
\colhead{(4)}     &
\colhead{(5)}     &
\colhead{(6)}     &
\colhead{(7)}     &
\colhead{(8)}     &
\colhead{(9)}     &
\colhead{(10)}     &
\colhead{(11)}    }
\startdata

00011217+1535355	&	--	--	&	L4	$\beta$	&	15.522	$\pm$	0.061	&	14.505	$\pm$	0.052	&	13.71	$\pm$	0.043	&	12.938	$\pm$	0.023	&	12.517	$\pm$	0.026	&	11.498	$\pm$	0.177	&	8.953	$\pm$	--	                                      &                            5 	\\
00040288-6410358	&	L1	$\gamma$	&	L1	$\gamma$\tablenotemark{a}	&	15.786	$\pm$	0.071	&	14.831	$\pm$	0.073	&	14.010	$\pm$	0.045	&	13.370	$\pm$	0.025	&	12.937	$\pm$	0.027	&	12.178	$\pm$	0.244	&	9.161	$\pm$	--	                              &                            15 , 27		\\
00182834-6703130	&	--	--	&	L0	$\gamma$	&	15.457	$\pm$	0.057	&	14.48	$\pm$	0.061	&	13.711	$\pm$	0.039	&	13.171	$\pm$	0.025	&	12.768	$\pm$	0.026	&	12.763	$\pm$	0.445	&	9.046	$\pm$	--	                              &                            5 	\\
00191296-6226005	&	--	--	&	L1	$\gamma$	&	15.64	$\pm$	0.06	&	14.618	$\pm$	0.054	&	13.957	$\pm$	0.051	&	13.351	$\pm$	0.025	&	12.883	$\pm$	0.026	&	12.378	$\pm$	0.262	&	9.114	$\pm$	--	                              &                            5 	\\
00192626+4614078	&	M8	--	&	M8	$\beta$	&	12.603	$\pm$	0.017	&	11.940	$\pm$	0.021	&	11.502	$\pm$	0.011	&	11.260	$\pm$	0.023	&	11.001	$\pm$	0.020	&	10.857	$\pm$	0.070	&	9.401	$\pm$	--	                                      &                            7, 5 	\\
00274197+0503417	&	M9.5	$\beta$	&	L0	$\beta$\tablenotemark{a}	&	16.189	$\pm$	0.092	&	15.288	$\pm$	0.099	&	14.960	$\pm$	0.115	&	14.619	$\pm$	0.036	&	14.135	$\pm$	0.054	&	12.239	$\pm$	--	&	8.890	$\pm$	--	                                      &                            6, 27, 36 	\\
00303013-1450333	&	L7	--	&	L4-L6	$\beta$	&	16.278	$\pm$	0.111	&	15.273	$\pm$	0.1	&	14.481	$\pm$	0.1	&	13.657	$\pm$	0.028	&	13.263	$\pm$	0.034	&	12.275	$\pm$	--	&	9.149	$\pm$	--	                                      &                            18, 5 	\\
\enddata
\tablecomments{References: 1=\citet{Reid08}, 2=\citet{Cruz09}, 3=\citet{Cruz07}, 4=\citet{Schmidt07}, 5=\citet{Gagne15b}, 6=\citet{Allers13}, 7=\citet{Cruz03}, 8=\citet{Faherty12}, 9=\citet{Liu13}, 10=\citet{Gizis02}, 11=\citet{Looper07a}
12=\citet{Reid02}, 13=\citet{Faherty09}, 14=\citet{Kirkpatrick08}, 15=\citet{Kirkpatrick10}, 16=\citet{Burgasser10a}, 17=\citet{Reid07}, 18=\citet{Kirkpatrick00}, 19=\citet{Gagne14a}, 20=\citet{Teegarden03}, 21=\citet{Faherty13}, 22=\citet{Gagne14b}, 23=\citet{Scholz05}
24=\citet{Salim03}, 25=\citet{Burgasser08}, 26=\citet{Crifo05}, 27=This paper, 28=\citet{Kirkpatrick06}, 29=\citet{Looper08}, 30=\citet{Sahlmann16}, 31=\citet{Schneider16}, 32=\citet{Kellogg16}, 33=\citet{Kellogg15}, 34=\citet{Gizis12}, 35=\citet{Gizis15}, 36=Cruz et al. in prep}
\tablenotetext{a}{These sources have new infrared spectra presented in this paper.  In the majority of cases we use the infrared spectral type and gravity classification diagnosed in this work.  If an object had SpeX data we default to the resultant type and classification with that data.}
\tablenotetext{b}{This source is referred to as PSO318 for the remainder of the text}
\tablenotetext{c}{The sources PSO318, 2M1119, and 2M1147 have photometry reported on the MKO system that we have converted to 2MASS using the relations in \citet{Stephens09}.  In the case of  PSO 318, JHK was converted to MKO whereas for 2M119, J and H were converted and for 2M1147 only J was converted. }
\tablenotetext{d}{Teegardens Star. }
\tablenotetext{e}{References are for the spectral data and gravity analysis (if different then the original spectral data reference).}
\tablenotetext{f}{Referred to as TVLM831-154910 in \citet{Tinney95}.}
\label{tab:photometry}
\end{deluxetable}
\clearpage
\end{landscape}

\tabletypesize{\scriptsize}
\begin{deluxetable}{lccccccccc}
\tablewidth{0pt}
\tablecolumns{10}
\tablecaption{Details on Parallax Targets and Observations}
\tablehead{
\colhead{2MASS Designation} &
\colhead{SpT}&
\colhead{SpT}&
\colhead{Nights} &
\colhead{Frames\tablenotemark{a}}  &
\colhead{Ref Stars\tablenotemark{a}} &
\colhead{$\Delta$t} &
\colhead{Telescope}&
\colhead{Note\tablenotemark{b}} &
\colhead{Ref}\\
&
\colhead{OpT}&
\colhead{IR}&
&
&
&
\colhead{(yr)}\\
  \colhead{(1)}     &
  \colhead{(2)}     &
  \colhead{(3)}     &
  \colhead{(4)}     &
  \colhead{(5)}     &
  \colhead{(6)}     &
  \colhead{(7)}     &
  \colhead{(8)}    &
\colhead{(9)}  & 
\colhead{(10)}}
\startdata
00452143+1634446	&	L2$\beta$	&	L2$\gamma$	&	4	&	26	&	24	&	3.0	&	MDM	& LG&1,2\\
01410321+1804502	&	L1	&	L4.5	&	4	&	23	&	26	&	3.3	&					MDM	&N&3,4,\\
02132880+4444453	&	L1.5	&	---	&	5	&	34	&	81	&	3.0	&				MDM	&N&5\\
02235464-5815067	&	L0$\gamma$	&	---	&	5	&	48	&	26	&	4.9	&			DuPont	&LG&1\\
02411151-0326587	&	L0$\gamma$	&	L1$\gamma$	&	6	&	55	&	20	&	4.9	&	DuPont	&LG&2,13\\
02535980+3206373	&	M7$\beta$	&	M6	&	5	&	28	&	53	&	3.9	&		MDM	&LG&14\\
03140344+1603056	&	L0	&	---	&	5	&	31	&	37	&	3.0	&					MDM	&N&6\\
05002100+0330501	&	L4	&	---	&	5	&	30	&	61	&	2.8	&					MDM	&N&6\\
\enddata 
\tablenotetext{a} {The number of reference stars and individual image frames used in the parallax solution}
\tablenotetext{b} {LG is a low surface gravity object and N is a field object}
\tablecomments{References: 1=\citet{Cruz09}, 2=\citet{Allers13}, 3=\citet{Wilson03}, 4=\citet{Cruz07}, 5=\citet{Cruz03}, 6=\citet{Reid08}, 7=\citet{Phan-Bao08}, 8=\citet{Hawley02}, 9=\citet{Kendall04}, 10=\citet{Gizis00}, 11=\citet{Burgasser08}, 12=\citet{Bowler10a}, 13=\citet{Kirkpatrick08}, 14=This paper, 15=\citet{Salim03} }
\label{tab:parallaxes}
\end{deluxetable}

\tabletypesize{\scriptsize}
\begin{deluxetable}{lllllllllll}
\tablewidth{0pt}
\tablecolumns{7}
\tablecaption{Results of Parallax Program}

\tablehead{
\colhead{Name} &
\colhead{SpT}&
\colhead{SpT}&
\colhead{$\mu_{\alpha}$\tablenotemark{a}} &
\colhead{$\mu_{\delta}$\tablenotemark{a}}  &
\colhead{$\pi_{abs}$} 
\\
&
\colhead{OpT}&
\colhead{IR}&
\colhead{$\arcsec yr^{-1}$}&
\colhead{$\arcsec yr^{-1}$}&
\colhead{mas}
\\
  \colhead{(1)}     &
  \colhead{(2)}     &
  \colhead{(3)}     &
  \colhead{(4)}     &
  \colhead{(5)}     &
   \colhead{(6)}}
\startdata
00452143+1634446	&	L2$\beta$	&	L2$\gamma$	&	0.355	$\pm$	0.01	&	-0.04	$\pm$	0.01	&	62.5	$\pm$	3.7	\\
01410321+1804502	&	L1	&	L4.5	&	0.41	$\pm$	0.01	&	-0.047	$\pm$	0.01		&41.0	$\pm$	2.8	\\
02132880+4444453	&	L1.5	&	---	&	-0.054	$\pm$	0.01	&	-0.147	$\pm$	0.01		&50.0	$\pm$	2.1	\\
02235464-5815067	&	L0$\gamma$	&	---	&	0.0986	$\pm$	0.0008	&	-0.0182	$\pm$	0.0009	&27.4	$\pm$	2.6	\\
02411151-0326587	&	L0$\gamma$	&	L1$\gamma$	&	0.0737	$\pm$	0.001	&	-0.0242	$\pm$	0.0019	&	26.7	$\pm$	3.3	\\
02535980+3206373	&	M7$\beta$	&	M6	&	0.087	$\pm$	0.01	&	-0.096	$\pm$	0.01		&17.7	$\pm$	2.5	\\
03140344+1603056	&	L0	&	---	&	-0.247	$\pm$	0.01	&	-0.05	$\pm$	0.01	&	69.0	$\pm$	2.4	\\
05002100+0330501	&	L4	&	---	&	0.008	$\pm$	0.01	&	-0.353	$\pm$	0.01	&	75.2	$\pm$	3.7	\\
\enddata 
\tablenotetext{a}{The proper motions from MDM are not corrected to absolute and have zero-point uncertainties of $\sim$ 10 mas yr$^{-1}$; see \citet{Weinberger13} for a discussion
of the du Pont proper motions.}
\label{tab:distances}
\end{deluxetable} 

\clearpage
\tabletypesize{\scriptsize}
\begin{deluxetable}{lrrrrrr}
\tablewidth{0pt}
\tablecolumns{7}
\tablecaption{Literature Parallax Comparisons }
\tablehead{
\colhead{2MASS Designation} &
\colhead{SpT}&
\colhead{SpT}&
\colhead{$\mu_{\alpha}$} &
\colhead{$\mu_{\delta}$}  &
\colhead{$\pi$}& 
\colhead{Ref} \\
&
&
&
\colhead{$\arcsec yr^{-1}$}&
\colhead{$\arcsec yr^{-1}$}&
\colhead{mas}&
\colhead{ }\\
  \colhead{(1)}     &
  \colhead{(2)}     &
  \colhead{(3)}     &
  \colhead{(4)}     &
  \colhead{(5)}     &
  \colhead{(6)}    &
  \colhead{(7)} }
\startdata
00452143+1634446	&	L2$\beta$	&	L2$\gamma$	&	0.355	$\pm$	0.01	&	-0.04	$\pm$	0.01	&	62.5	$\pm$	3.7&	1	\\
				&		&&						0.3562	$\pm$	0.00137	&	-0.035	$\pm$	0.0109	&	57.3	$\pm$	2.0&	2	\\
\hline
\\
02411151-0326587	&	L0$\gamma$	&	L1$\gamma$	&	0.0737	$\pm$	0.001	&	-0.0242	$\pm$	0.0019	&	26.7	$\pm$	3.3&	1	\\
				&		&&						0.084	$\pm$	0.0117	&	-0.0224	$\pm$	0.0086	&	21.4	$\pm$	2.6&	2	\\
\hline
\\
10224821+5825453	&	L1$\beta$	&	L1$\beta$	&	-0.807	$\pm$	0.01	&	-0.73	$\pm$	0.01	&	54.3	$\pm$	2.5&	1	\\
				&		&&						-0.799	$\pm$	0.0064	&	-0.7438	$\pm$	0.0132	&	46.3	$\pm$	1.3&	2	\\
\hline
\\
14162408+1348263	&	L6	&	L6	&	0.088	$\pm$	0.01	&	0.136	$\pm$	0.01	&	107.5	$\pm$	3.5&	1	\\
				&		&&						0.0951	$\pm$	0.003	&	0.1303	$\pm$	0.003	&	109.9	$\pm$	1.8&	2	\\
\hline
\\
15525906+2948485	&	L0$\beta$	&	L0$\beta$	&	-0.162	$\pm$	0.01	&	-0.06	$\pm$	0.01	&	48.8	$\pm$	2.7&	1	\\
				&		&&						-0.1541	$\pm$	0.0053	&	-0.0622	$\pm$	0.0106	&	47.7	$\pm$	0.9&	3	\\
\enddata 
\tablecomments{References: (1) This work (2) Osorio et al. 2014 (3) Dupuy \& Liu 2012} 
\label{tab:parallaxes-duplicates}
\end{deluxetable}

\clearpage
\tabletypesize{\scriptsize}
\begin{deluxetable}{lllllllllllll}
\tablecaption{New Proper Motion Measurements\label{tab:PMs}}
\tablehead{
  \colhead{2MASS}        &
  \colhead{SpT}        &
  \colhead{SpT}        &
  \colhead{$\mu_{RA}$}        &
  \colhead{$\mu_{DEC}$}        &
  \colhead{Ref} &
  \colhead{$\Delta$t}  \\
   \colhead{ }        &
  \colhead{OpT}        &
  \colhead{NIR}        &
  \colhead{$\arcsec$ yr$^{-1}$}        &
  \colhead{$\arcsec$ yr$^{-1}$}        &
  \colhead{ }   &
  \colhead{yr }   \\
  \colhead{(1)}     &
  \colhead{(2)}     &
  \colhead{(3)}     &
  \colhead{(4)}     &
  \colhead{(5)}     & 
  \colhead{(6)}     &
  \colhead{(7)}  \\  }
\startdata

00040288-6410358	&	L1$\gamma$	&	L1$\gamma$	&	0.064	$\pm$	0.012		&	-0.047	$\pm$	0.012	&		WISE-2MASS 	& 	10.02	\\
00374306-5846229	&	L0$\gamma$	&	---		&	0.057	$\pm$	0.01			&	0.017	$\pm$	0.005	&	FourStar	&	14.15	\\
01174748-3403258	&	L1$\beta$	&	L1$\beta$	&	0.084	$\pm$	0.015		&	-0.045	$\pm$	0.008	&		FourStar	&	15.08	\\
01262109+1428057	&	L4$\gamma$	&	L2$\gamma$	&	0.07	$\pm$	0.012		&	-0.008	$\pm$	0.012	&		WISE-2MASS	&	9.68	\\
01415823-4633574	&	L0$\gamma$	&	L0$\gamma$	&	0.105	$\pm$	0.01		&	-0.049	$\pm$	0.01		&	FourStar	&	14.29	\\
02103857-3015313	&	L0$\gamma$	&	L0$\gamma$	&	0.145	$\pm$	0.036	&	-0.04		$\pm$	0.007	&		FourStar	&	14.39	\\
02340093-6442068	&	L0$\gamma$	&	L0$\beta$$\gamma$	&	0.088	$\pm$	0.012	&	-0.015	$\pm$	0.012	&		WISE-2MASS	&	10.53	\\
03032042-7312300	&	L2$\gamma$	&	---		&	0.043	$\pm$	0.012		&	0.003	$\pm$	0.012	&		WISE-2MASS	&	10.48	\\
\enddata
\end{deluxetable}
\clearpage

\clearpage

\tabletypesize{\scriptsize}
\begin{deluxetable}{lllllllllllll}
\tablecaption{near-infrared Spectral Data\label{tab:spectra}}
\tablehead{
  \colhead{2MASS}        &
  \colhead{Date-Observed}        &
  \colhead{Instrument}        &
  \colhead{Mode}        &
  \colhead{Slit}        &
  \colhead{Int time} &
  \colhead{Images}  \\
  \colhead{(1)}     &
  \colhead{(2)}     &
  \colhead{(3)}     &
  \colhead{(4)}     &
  \colhead{(5)}     & 
  \colhead{(6)}     &
    \colhead{(7)}  \\  }
\startdata

00040288-6410358	&	11-Sept-2014	&	FIRE	&	Echelle	&	0.6	&	900	&	2	&\\
00274197+0503417	&	08-Aug-2014	&	FIRE	&	Echelle	&	0.6	&	900	&	2	&\\
00452143+1634446	&	08-Aug-2014	&	FIRE	&	Echelle	&	0.6	&	900	&	2	&\\
00550564+0134365&	04-Sept-2003	&	SpeX	&	Prism	&	0.8	&	180	&	2	&\\
01174748-3403258	&	12-Sept-2014	&	FIRE	&	Echelle	&	0.6	&	900	&	2	&\\
01174748-3403258(2)	&	11-Sept-2014	&	FIRE	&	Echelle	&	0.6	&	600	&	2	&\\
01244599-5745379	&	11-Sept-2014	&	FIRE	&	Echelle	&	0.6	&	900	&	2	&\\
02103857-3015313	&	11-Sept-2014	&	FIRE	&	Echelle	&	0.6	&	750	&	2	&\\
\enddata
\tablecomments{}
\end{deluxetable}
\clearpage

\begin{landscape}
\tabletypesize{\scriptsize}
\begin{deluxetable}{lrclccllllll}
\tablecaption{Radial Velocities \label{tab:spectroscopic}}
\tablehead{
\colhead{Name}              &
\colhead{Telescope}         &
\colhead{Resolving Power}   &
\colhead{Date-obs}          &
\colhead{Standards Used}    &
\colhead{R.V.}              &
\colhead{R.V. final}        \\
\colhead{(1)}     &
\colhead{(2)}     &
\colhead{(3)}     &
\colhead{(4)}     &
\colhead{(5)}     &
\colhead{(6)}     &
\colhead{(7)}     }
\startdata
\hline
00325584-4405058 & Magellan Clay MIKE     	& 25000 & 01-Nov-2006 &  4,5  &   12.95$\pm$ 1.92 & 12.95$\pm$ 1.92                   \\  
00332386-1521309 & Gemini South Phoenix   	& 50000 & 28-Oct-2009 &  6,7  &   -6.37$\pm$ 0.40 & -6.37$\pm$ 0.40                  \\
00374306-5846229 & Gemini South Phoenix  	& 50000 & 20-Nov-2007 &  6,7  &    6.63$\pm$ 0.08 &  6.62$\pm$ 0.07                   \\
00374306-5846229 & Gemini South Phoenix  	& 50000 & 23-Dec-2007 &  6,7  &    6.01$\pm$ 0.74 &                                   \\
00452143+1634446 & Keck II NIRSPEC      	& 20000 & 15-Sep-2008 &  1,3  &    3.16$\pm$ 0.83 &  3.16$\pm$ 0.83                   \\
00464841+0715177 & Keck II NIRSPEC      	& 20000 & 14-Sep-2008 & 1,2,3 &   -2.75$\pm$ 0.27 & -2.75$\pm$ 0.27                   \\
00550564+0134365 & Keck II NIRSPEC    		& 20000 & 14-Sep-2008 & 1,2,3 &   -1.21$\pm$ 0.38 & -1.21$\pm$ 0.38                   \\
01415823-4633574 & Magellan Clay MIKE   	& 25000 & 02-Nov-2006 &  4,5  &    6.41$\pm$ 1.56 &  6.41$\pm$ 1.56                   \\
\enddata
\tablecomments{Data from 2MASS \citet{Cutri03} and ALLWISE \citet{Cutri13}. Standard stars are as follows: 1.) 2MASS J18212815+1414010 (+9.78 \citealt{Blake10}), 2.) 2MASS J00452143+1634446 (+3.29 \citealt{Blake10}), 3.) 2MASS J22244381-0158521 (-37.55 \citealt{Blake10}), 4.) LHS~2924 (-37.4 \citealt{Mohanty03}), 5.) BRI~1222-1221 (-4.8 \citealt{Mohanty03}), 6.) 2MASS J11553952-3727350 (+45.0 \citealt{Seifahrt10}), 7.) 2MASS J05233822-1403022 (+11.82 \citealt{Blake07})}
\tablenotetext{a}{Low Quality Spectrum}
\tablenotetext{b}{No Useable Spectrum}
\tablenotetext{c}{No Spectroscopic Signs of Youth}
\tablenotetext{d}{This object is listed in \citet{Filippazzo15} with a distance and reference to Faherty+ in prep.  The value in that paper was spectrophotometric and should not be regarded as a parallax.} 
\end{deluxetable}
\clearpage
\end{landscape}

\tabletypesize{\scriptsize}
\begin{deluxetable}{rrl|ll|ll|ll|ll}
\tablecaption{Comparison of new and previously published RVs. \label{tab:spectroscopic2}}
\startdata
\setlength{\tabcolsep}{0.02in}
\tablehead{
 \colhead{Name}      &
 \colhead{Telescope} &
 \colhead{RV$_{ours}$} &
 \colhead{RV$_{others}$} &
 \colhead{RV}        &
 \colhead{RV$_{others}$} &
 \colhead{RV}        &
 \colhead{RV$_{others}$} &
 \colhead{RV}        &
 \colhead{RV$_{others}$} &
 \colhead{RV} \\
 \colhead{}          &
 \colhead{}          &
 \colhead{$km s^{-1}$} &
 \colhead{$km s^{-1}$} &
 \colhead{ref.}      &
 \colhead{$km s^{-1}$} &
 \colhead{ref.}      &
 \colhead{$km s^{-1}$} &
 \colhead{ref.}      &
 \colhead{$km s^{-1}$} &
 \colhead{ref.}      \\
 \colhead{(1)}       &
 \colhead{(2)}       &
 \colhead{(3)}       &
 \colhead{(4)}       &
 \colhead{(5)}       &
 \colhead{(6)}       &
 \colhead{(7)}       &
 \colhead{(8)}       &
 \colhead{(9)}       &
 \colhead{(10)}      &
 \colhead{(11)}      }
\startdata
00242463-0158201 & Gemini South Phoenix & $+$11.65$\pm$1.60 & $+$10.4$\pm$3    & 3 & $+$16$\pm$2     & 7 &  &  &  & \\
00452143+1634446 & Keck II NIRSPEC    & $+$3.16$\pm$0.83    & $+$3.29$\pm$0.43  & 1 &  &  &  &  &  &  \\
01415823-4633574& Magellan Clay MIKE & $+$6.41$\pm$1.56     & $+$12$\pm$15      & 2 &  &  &  &  &  &  \\
03393521-3525440 & Magellan Clay MIKE    & $+$7.43$\pm$0.72    & $+$7.6$\pm$3      & 3 & $+$6$\pm$2 & 8 & $+$5.8$\pm$2.4 & 5 & $+$10$\pm$2 & 9 \\
04433761+0002051 & Magellan Clay MIKE & $+$16.97$\pm$0.76   & $+$17.1$\pm$3     & 3 &  &  &  &  &  &  \\
05233822-1403022 & Gemini South Phoenix & $+$12.48$\pm$0.41 & $+$12.21$\pm$0.21 & 1 &  &  &  &  &  &  \\
06085283-2753583 & Magellan Clay MIKE & $+$26.35$\pm$0.07   & $+$24$\pm$1       & 4 &  &  &  &  &  &  \\
1224522-123835 & Magellan Clay MIKE & $-$2.87$\pm$0.59    & $-$5.8$\pm$3      & 3 & $-$4.8$\pm$2    & 7 &  &  &  & \\
\enddata
\tablecomments{References: 1. \citet{Blake10}, 2. \citet{Kirkpatrick06}, 3. \citet{Reiners09}, 4. \citet{Rice10}, 5. \citet{Burgasser15}, 6. \citet{Galvez-Ortiz10}, 7. \citet{Mohanty03}, 8. \citet{Basri00}, 9. \citet{Reid02}}

\end{deluxetable}

\tabletypesize{\scriptsize}
\begin{deluxetable}{ccccccccc}
\tablecaption{Adopted near-infrared Spectral Types\label{tab:spectra2}}
\tablehead{
  \colhead{Name}        &
  \colhead{SpT}        &
  \colhead{SpT}        &
    \colhead{Gravity Score}&
    \colhead{Gravity Type}&
    \colhead{Gravity Score}&
    \colhead{Gravity Type}\\
    \colhead{} &
    \colhead{Adopted} &
    \colhead{Allers13} &
    \colhead{MedRes} &
    \colhead{MedRes} &
    \colhead{LowRes} &
    \colhead{LowRes}\\
 \colhead{(1)}     &
  \colhead{(2)}     &
  \colhead{(3)}     &
  \colhead{(4)}     &
  \colhead{(5)}     &
    \colhead{(6)}    &
     \colhead{(7)}\\ }
\startdata
00040288-6410358	&	L1 $\gamma$	&	L2.3	$\pm$	0.3	&	2221	&	VL-G	&	2211	&	VL-G	\\
00274197+0503417	&	M8 $\beta$	&	M9.0	$\pm$	0.8	&	2n11	&	INT-G	&	2n01	&	INT-G	\\
00452143+1634446	&	L0 $\beta$	&	L1.5	$\pm$	0.4	&	1211	&	INT-G	&	1211	&	INT-G	\\
00550564+0134365&	L2$\gamma$    &	L1.0 $\pm$ 0.9& --- & --- & 2122& VL-G\\
01174748-3403258	&	L1 $\gamma$	&	L2.2	$\pm$	0.5	&	1211	&	INT-G	&	1211	&	INT-G	\\
01174748-3403258(2) &	L1 $\gamma$	&	L1.9	$\pm$	0.5	&	2212	&	VL-G	&	2222	&	VL-G	\\
01244599-5745379	&	L0 $\gamma$	&	L0.7	$\pm$	0.5	&	2222	&	VL-G	&	2222	&	VL-G	\\
02103857-3015313	&	L0 $\gamma$	&	L1.0	$\pm$	0.3	&	2211	&	VL-G	&	2211	&	VL-G	\\
\enddata
\tablenotetext{a}{Reddening E(B-V)=1.8 and SPT = M7 γ are a good solution, but spt/reddening is almost degenerate.}
\end{deluxetable}
\clearpage

\clearpage
\LongTables
\clearpage
\tabletypesize{\scriptsize}
\begin{landscape}
\begin{deluxetable}{lllllllllllllllll}
\tablecaption{Kinematic data on Low Surface Gravity Dwarfs}
\tablehead{
  \colhead{2MASS Designation}      &
  \colhead{SpT}&
  \colhead{SpT}&
    \colhead{$\mu_{RA\cos{DEC}}$} &
  \colhead{$\mu_{DEC}$}       &
  \colhead{$\pi$}   &
    \colhead{RV}   &
    \colhead{(Ref)}\\
&
  \colhead{(OpT)}     &
  \colhead{(IR)}     &
    \colhead{($\arcsec$ yr$^{-1}$)}   &
    \colhead{($\arcsec$ yr$^{-1}$)}   &
  \colhead{(mas)}   &
\colhead{(km s$^{-1}$)}&
\\
  \colhead{(1)}     &
  \colhead{(2)}     &
  \colhead{(3)}     &
  \colhead{(4)}     &
  \colhead{(5)}     &
  \colhead{(6)}     &
  \colhead{(7)}     &
  \colhead{(8)}     }
  
\startdata
00011217+1535355&----&L4$\beta$&0.1352$\pm$0.0107&-0.1696$\pm$0.0137&--&--&10\\
00040288-6410358&L1$\gamma$&L1$\gamma$\tablenotemark{a}&0.064$\pm$0.012&-0.047$\pm$0.012&(17$\pm$1)&(6.07$\pm$2.89)&1\\
00182834-6703130&----&L0$\gamma$&0.066$\pm$0.004&-0.0523$\pm$0.0126&--&--&1,10\\
00191296-6226005&----&L1$\gamma$&0.0541$\pm$0.0047&-0.0345$\pm$0.0121&(21$\pm$6)&(6.7$\pm$2.5)&1,10\\
00192626+4614078&      M8--&M8$\beta$&0.14000$\pm$0.06&-0.1$\pm$0.05&(31$\pm$3)&-19.5$\pm$2&    2,4                                                   \\
00274197+0503417&M9.5$\beta$&L0$\beta$&0.0105$\pm$0.0004&-0.0008$\pm$0.0003&13.8$\pm$1.6&--&3\\
00303013-1450333&L7--&L4-L6$\beta$&0.2497$\pm$0.0141&-0.0053$\pm$0.015&37.42$\pm$4.50&--&10,11\\
00325584-4405058&L0$\gamma$&L0$\beta$&0.12830$\pm$0.0034&-0.0934$\pm$0.003&21.6$\pm$7.2&12.95$\pm$1.92&1,5\\
\enddata
\tablecomments{References: 1=This Paper,2=\citet{Faherty09},3=\citet{Dahn02},4=\citet{Reiners09a},5=\citet{Marocco13},6=\citet{Zapatero-Osorio14},7=\citet{Gizis15},8=\citet{Faherty12},9=\citet{Riedel15},10=\citet{Gagne15b},11=\citet{Vrba04},12=\citet{Shkolnik12},13=\citet{Dieterich14},14=\citet{Blake10},15=\cite{Weinberger13},16=\cite{Mohanty03},17=\citet{Gagne14a},18=\citet{Teixeira08},19=\citet{Ducourant14},20=\citet{Schneider14},21=\citet{Liu13}, 22=\citet{Sahlmann16}, 23=\citet{Allers16}, 24=\citet{Henry06}, 25=\citet{Tinney95}, 26=\citet{Dittmann14}, 27=\citet{Kellogg16}, 28=\citet{Schneider16}}
\tablenotetext{a}{These sources have new infrared spectra presented in this paper.  In the majority of cases we use the infrared spectral type and gravity classification diagnosed in this work.  If an object had SpeX data we default to the resultant type and classification with that data.}
\tablenotetext{b}{These sources were listed in \citet{Filippazzo15} as members of associations but as has been noted in Table ~\ref{tab:memberships}, we have downgraded them to ambiguos young objects}
\label{tab:kinematics}
\end{deluxetable}
\clearpage
\end{landscape}
\tabletypesize{\scriptsize}
\begin{deluxetable}{lrrrrrr}
\tablecaption{Kinematics for target dwarfs\label{tab:uvw}}
\tablehead{
\colhead{Name}    &
\colhead{U}       &
\colhead{V}       &
\colhead{W}       &
\colhead{X}       &
\colhead{Y}       &
\colhead{Z}       \\
\colhead{(1)}     &
\colhead{(2)}     &
\colhead{(3)}     &
\colhead{(4)}     &
\colhead{(5)}     &
\colhead{(6)}     &
\colhead{(7)}     }
00325584-4405058	&	-10.77	$\pm$	3.85	&	-32.08	$\pm$	8.55	&	-8.34	$\pm$	2.26	&	8.96	$\pm$	2.59	&	-9.24	$\pm$	2.67	&	-41.15	$\pm$	11.89	\\
003323.86-1521309	&	-52.85	$\pm$	5.68	&	-26.91	$\pm$	3.93	&	3.23	$\pm$	0.86	&	-1.86	$\pm$	0.18	&	8.46	$\pm$	0.82	&	-39.09	$\pm$	3.8	\\
00452143+1634446	&	-22.62	$\pm$	1.47	&	-14.31	$\pm$	1.21	&	-5.02	$\pm$	0.84	&	-5.66	$\pm$	0.33	&	9.48	$\pm$	0.55	&	-11.54	$\pm$	0.67	\\
00470038+6803543	&	-8.58	$\pm$	1.06	&	-27.83	$\pm$	1.22	&	-13.53	$\pm$	0.49	&	-6.52	$\pm$	0.24	&	10.23	$\pm$	0.37	&	1.1	$\pm$	0.04	\\
01231125-6921379	&	-7.92	$\pm$	1.96	&	-20.39	$\pm$	2.71	&	-1.82	$\pm$	2.42	&	14.9	$\pm$	2.11	&	-27.1	$\pm$	3.84	&	-33.76	$\pm$	4.78	\\
02235464-5815067	&	-7.94	$\pm$	0.86	&	-18.46	$\pm$	1.19	&	-1.11	$\pm$	0.7	&	4.29	$\pm$	0.4	&	-20.41	$\pm$	1.88	&	-29.71	$\pm$	2.74	\\
02411151-0326587	&	-9.68	$\pm$	4.71	&	-11.49	$\pm$	1.49	&	-1.39	$\pm$	6.56	&	-21.39	$\pm$	2.55	&	1.63	$\pm$	0.19	&	-30.27	$\pm$	3.6	\\
03350208+2342356	&	-16.91	$\pm$	1.77	&	-11.46	$\pm$	2.11	&	-8.08	$\pm$	1.95	&	-36.8	$\pm$	2.02	&	10.13	$\pm$	0.55	&	-18.31	$\pm$	1	\\
\enddata
\tablecomments{Data is only presented for targets with measured parallax, or measured parallax and radial velocity.  While the actual uncertainties are best described by a radially-oriented ellipsoid, they are given here for comparison to other values.}
\end{deluxetable}
\clearpage

\begin{landscape}
\tabletypesize{\scriptsize}
\begin{deluxetable}{lll|lll|ll|ll|ll|l}
\tablecaption{Moving Group membership probabilities\label{tab:memberships}}
\tablehead{
  \colhead{}             &
  \colhead{}        &
  \colhead{}             &
  \colhead{BANYAN II}    &
\colhead{}  &
 \colhead{} &
    \colhead{LACEwING}     &
  \colhead{}             &
  \colhead{Convergence}  &
  \colhead{}             &
  \colhead{BANYAN I}     &
  \colhead{}             &
\colhead{}   \\
  \colhead{Name}    &
  \colhead{SpT}&
  \colhead{SpT}&
  \colhead{Prob.}&
  \colhead{Contam.}&
  \colhead{Group}&
  \colhead{Prob.}&
    \colhead{Group}&
\colhead{Prob.}&
    \colhead{Group}&
\colhead{Prob.}&
    \colhead{Group} &
    \colhead{Decision}\\
  \colhead{(1)}     &
  \colhead{(2)}     &
  \colhead{(3)}     &
  \colhead{(4)}     &
  \colhead{(5)}     &
  \colhead{(6)}     &
  \colhead{(7)}     &
  \colhead{(8)}     &
  \colhead{(9)}     &
  \colhead{(10)}    &
  \colhead{(11)}    &
  \colhead{(12)}   &
  \colhead{(13)} }
\startdata
\hline
00011217+1535355		&		--	--		&		L4	$\beta$		&		97.4		&		1.1		&		AB DOR		&		35.12		&		AB DOR		&		43.1		&		AB DOR		&		99.04		&		AB DOR		&		AM	\\
00040288-6410358		&		L1	$\gamma$		&		L1	$\gamma$\tablenotemark{a}		&		99.6		&		0		&		TUC-HOR		&		38.76		&		TUC-HOR		&		100		&		TUC-HOR		&		100		&		TUC-HOR		&		HLM	\\
00182834-6703130		&		--	--		&		L0	$\gamma$		&		99.9		&		$<$ 0.1		&		TUC-HOR		&		48.71		&		TUC-HOR		&		76.7		&		TUC-HOR		&		99.83		&		TUC-HOR		&		AM	\\
00191296-6226005		&		--	--		&		L1	$\gamma$		&		99.7		&		$<$ 0.1		&		TUC-HOR		&		0		&		NONE		&		99.7		&		TUC-HOR		&		89.44		&		TUC-HOR		&		HLM	\\
00192626+4614078		&		M8	--		&		M8	$\beta$		&		72		&		13.3		&		AB DOR		&		66.42		&		AB DOR		&		99		&		AB DOR		&		99		&		AB DOR		&		HLM	\\
00274197+0503417		&		M9.5	$\beta$		&		L0	$\beta$\tablenotemark{a}		&		0		&		100		&		ARGUS		&		0		&		NONE		&		100		&		CHA-NEAR		&		100		&		OLD		&		AM	\\
00303013-1450333		&		L7	--		&		L4-L6	$\beta$		&		26.5		&		2.6		&		ARGUS		&		29.65		&		AB DOR		&		98.1		&		CHA-NEAR		&		97.87		&		ARGUS		&		AM	\\
00325584-4405058	&		L0	$\gamma$		&		L0	$\beta$		&		67.3		&		1		&		TUC-HOR		&		91.3		&		AB DOR		&		96		&		AB DOR		&		100		&		AB DOR		&		HLM	\\
\enddata
\tablenotetext{a}{These sources have new infrared spectra presented in this paper.  In the majority of cases we use the infrared spectral type and gravity classification diagnosed in this work.  If an object had SpeX data we default to the resultant type and classification with that data.}
\tablenotetext{b}{\citet{Deacon16} use the parallax reported in this work to show that this source is co-moving with the M dwarf TYC 9486-927-1.  In that work they report a likelihood of membership in the $\beta$ pictoris moving group, however we find this unlikely with the given kinematics.  The membership of this interesting wide system remains unknown.}
\tablenotetext{c}{These sources were listed in \citet{Filippazzo15} as members of associations but as has been noted in Table ~\ref{tab:memberships}, we have downgraded them to ambiguos young objects}
\tablenotetext{d}{The source 0335+2342 is listed as a bonafide member of $\beta$ Pictoris in \citet{Gagne14} using a 2MASS to WISE proper motion.  Aside from the \citet{Shkolnik12} proper motion used in this work,  this source also has a PPMXL proper motion.  Depending on which value is used in the kinematic analysis, the probability of membership varies.  Based on its position on the various photometric and absolute magnitude diagrams in this work, we suspect this source is a $\beta$ Pictoris member.  However refined kinematics will confirm this.  Furthermore, \citet{Allers13} list this object as an M7 VLGsource  however we have chosen to list it as an M7.5 $\beta$ based on the analysis in \citet{Gagne14}.}
\tablecomments{Computations of BANYAN I and II were done with the web calculator, without photometric data. BANYAN II and LACEwING were computed assuming the objects are young. The predicted distance and RV (as predicted by LACEwING for the final group) should be compared to the values in Table \ref{tab:kinematics}.}
\end{deluxetable}
\clearpage
\end{landscape}

\begin{landscape}
\tabletypesize{\scriptsize}
\begin{deluxetable}{llllllllllllllll}
\tablecaption{High Confidence Moving Group Members\label{tab:Members}}
\tablehead{
  \colhead{2MASS}        &
  \colhead{SpT}        &
  \colhead{SpT}        &
  \colhead{$\mu_{RA}$}        &
  \colhead{$\mu_{DEC}$}        &
    \colhead{$\pi$}        &
  \colhead{RV}    &
    \colhead{$L_{Bol}$}&
  \colhead{$T_{eff}$}&
  \colhead{Mass\tablenotemark{a}}&
  \colhead{Ref}\\
   \colhead{ }        &
  \colhead{OpT}        &
  \colhead{NIR}        &
  \colhead{$\arcsec$ yr$^{-1}$}        &
  \colhead{$\arcsec$ yr$^{-1}$}        &
  \colhead{mas}        &
  \colhead{km s$^{-1}$}        &
  \colhead{ } &
  \colhead{ }   \\
  \colhead{(1)}     &
  \colhead{(2)}     &
  \colhead{(3)}     &
  \colhead{(4)}     &
  \colhead{(5)}     &
  \colhead{(6)}     &
  \colhead{(7)}     &
  \colhead{(8)}     \\  }
\startdata
{\bf AB DORADUS}\\																				
00192626+4614078	&	      M8	--				&		M8	$\beta$		&		0.14000	$\pm$	0.06		&		-0.1	$\pm$	0.05		&		(31$\pm$3)				&		-19.5	$\pm$	2		&	-2.773$\pm$0.129	&	2637.0$\pm$371.0	&	66.62$\pm$48.99	&	    2,4                                                   \\
00325584-4405058	&	L0	$\gamma$			&			L0	$\beta$		&	0.12830$\pm$0.0034	&	-0.0934$\pm$0.003	&	21.6$\pm$7.2	&	12.95$\pm$1.92	&	-3.388$\pm$0.289	&	2066.0$\pm$413.0	&	41.64$\pm$29.47	&	  1,5                                                   \\
00470038+6803543	&		L7	($\gamma$?)	&	L6-L8	$\gamma$			&	0.38700$\pm$0.004	&	-0.197$\pm$0.004	&	82$\pm$3	&	-20$\pm$1.4	&	-4.429$\pm$0.033	&	1230.0$\pm$27.0	&	11.84$\pm$2.63	&	7\\
03264225-2102057	&		L5	$\beta$	&		L5$\beta$$\gamma$\tablenotemark{a}		&		0.108	$\pm$	0.014		&		-0.146	$\pm$	0.015		&		(41	$\pm$	1)		&		(22.91	$\pm$	2.07)		&	-4.235$\pm$0.022	&	1346.0$\pm$26.0	&	13.77$\pm$2.6	&	1,2\\
03552337+1133437	&		L5	$\gamma$	&	L3-L6	$\gamma$			&	0.22500$\pm$0.0132	&	-0.63$\pm$0.015	&	110.8$\pm$4.3	&	11.92$\pm$0.22	&	-4.104$\pm$0.034	&	1478.0$\pm$58.0	&	21.62$\pm$6.14	&	6,14\\
14252798-3650229	&		L3	--	&		L4$\gamma$			&	-0.28489$\pm$0.0014	&	-0.46308$\pm$0.001	&	86.45$\pm$0.83	&	5.37$\pm$0.25	&	-4.038$\pm$0.009	&	1535.0$\pm$53.0	&	22.52$\pm$6.07	&	13,14\\
22064498-4217208	&		L4	$\gamma$	&	L4	$\gamma$\tablenotemark{a}		&		0.128	$\pm$	0.013		&		-0.181	$\pm$	0.008		&		(35	$\pm$	2)		&		(7.6	$\pm$	2.0)		&	-3.997$\pm$0.009	&	1566.0$\pm$59.0	&	23.15$\pm$6.37	&	1\\
\enddata
\tablecomments{\tablecomments{References: See Table ~\ref{tab:kinematics}.}
\tablenotetext{a}{These sources have new infrared spectra presented in this paper.  In the majority of cases we use the infrared spectral type and gravity classification diagnosed in this work.  If an object had SpeX data we default to the resultant type and classification with that data.}
}
\end{deluxetable}
\clearpage
\end{landscape}

\tabletypesize{\scriptsize}
\begin{deluxetable}{llllllllllllllll}
\tablecaption{Fundamental Parameters for Young Sources\label{tab:LbolTeffMass}}
\tablehead{
  \colhead{2MASS}        &
  \colhead{SpT}        &
  \colhead{SpT}        &
    \colhead{$\pi$}        &
    \colhead{$L_{Bol}$\tablenotemark{b}}&
  \colhead{$T_{eff}$\tablenotemark{b}}&
  \colhead{Mass\tablenotemark{b}} \\
   \colhead{ }        &
  \colhead{OpT}        &
  \colhead{NIR}        &
  \colhead{mas}        &
  \colhead{} &
  \colhead{K} &
  \colhead{$M_{Jupiter}$} \\
  \colhead{(1)}     &
  \colhead{(2)}     &
  \colhead{(3)}     &
  \colhead{(4)}     &
  \colhead{(5)}     &
  \colhead{(6)}     &
  \colhead{(7)}     \\  }
\startdata
\hline
00040288-6410358	&	L1 $\gamma$	&	L1 $\gamma$\tablenotemark{a}	&	(17 $\pm$ 1)	&	-3.48$\pm$0.051	&	1904.0$\pm$63.0	&	16.11$\pm$2.9	\\
00191296-6226005	&	-- --	&	L1 $\gamma$	&	(21 $\pm$ 6)	&	-3.644$\pm$0.248	&	1755.0$\pm$258.0	&	14.88$\pm$4.52	\\
00192626+4614078	&	M8 --	&	M8 $\beta$	&	(31$\pm$3)	&	-2.773$\pm$0.129	&	2637.0$\pm$371.0	&	66.62$\pm$48.99	\\
00274197+0503417	&	M9.5$\beta$	&	L0$\beta$	&	13.8$\pm$1.6	&	-3.545$\pm$0.101	&	1945.0$\pm$197.0	&	31.64$\pm$19.25	\\
00303013-1450333	&	L7--	&	L4-L6$\beta$	&	37.42$\pm$4.50	&	-4.378$\pm$0.105	&	1436.0$\pm$131.0	&	50.89$\pm$23.43	\\
00325584-4405058	&	L0$\gamma$	&	L0$\beta$	&	21.6$\pm$7.2	&	-3.388$\pm$0.289	&	2066.0$\pm$413.0	&	41.64$\pm$29.47	\\
003323.86-1521309	&	L4$\beta$	&	L1	&	24.8$\pm$2.5	&	-3.616$\pm$0.088	&	1880.0$\pm$173.0	&	29.68$\pm$17.71	\\
00381489-6403529	&	-- --	&	M9.5 $\beta$	&	(23 $\pm$ 5)	&	-3.423$\pm$0.189	&	1957.0$\pm$226.0	&	22.03$\pm$9.39	\\
\enddata
\tablecomments{\tablecomments{References: See Table ~\ref{tab:kinematics}.}
\tablenotetext{a}{These sources have new infrared spectra presented in this paper.  In the majority of cases we use the infrared spectral type and gravity classification diagnosed in this work.  If an object had SpeX data we default to the resultant type and classification with that data.}
}
\tablenotetext{b}{Lbol, Teff, and Mass are calculated as described in \citet{Filippazzo15}.  Values that are slightly different from that work, have been updated using new data presented in this paper.}
\end{deluxetable}
\clearpage

\begin{landscape}
\tabletypesize{\scriptsize}
\begin{deluxetable}{cccccccccccccccccr}
\tablecaption{Average Infrared Colors of late-type M and L dwarfs\label{tab:meancolors1}}
\tablewidth{0pt}
\tablehead{
\colhead{SpT} &
\colhead{N$_{J-H}$\tablenotemark{a}} &
\colhead{($J$-$H$)$_{avg}$} &
\colhead{$\sigma$(J-H)} &
\colhead{N$_{J-K}$\tablenotemark{a}} &
\colhead{($J$-$K$)$_{avg}$} &
\colhead{$\sigma$(J-K)} &
\colhead{N$_{J-W1}$\tablenotemark{a}} &
\colhead{($J$-$W1$)$_{avg}$} &
\colhead{$\sigma$(J-W1)} &
\colhead{N$_{J-W2}$\tablenotemark{a}} &
\colhead{($J$-$W2$)$_{avg}$} &
\colhead{$\sigma$(J-W2)}&
\colhead{N$_{H-K}$\tablenotemark{a}} &
\colhead{($H$-$K$)$_{avg}$} &
\colhead{$\sigma$(H-K)} \\
\colhead{(1)} &
\colhead{(2)} &
\colhead{(3)} &
\colhead{(4)} &
\colhead{(5)} &
\colhead{(6)} &
\colhead{(7)} &
\colhead{(8)} &
\colhead{(9)} &
\colhead{(10)}&
\colhead{(11)}&
\colhead{(12)}&
\colhead{(13)} &
\colhead{(14)}&
\colhead{(15)}&
\colhead{(16)}&
\\
}
\startdata
M7	&	1998	&	0.61	&	0.09	&	1976	&	0.97	&	0.10	&	1992	&	1.17	&	0.09	&	1990	&	1.37	&	0.10	&	1976	&	0.36	&	0.10\\
M8	&	703	&	0.65	&	0.09	&	698	&	1.06	&	0.11	&	692	&	1.27	&	0.12	&	692	&	1.49	&	0.13	&	698	&	0.41	&	0.09\\
M9	&	430	&	0.69	&	0.10	&	428	&	1.15	&	0.11	&	420	&	1.41	&	0.14	&	420	&	1.65	&	0.17	&	428	&	0.45	&	0.10\\
L0	&	231	&	0.75	&	0.12	&	223	&	1.22	&	0.15	&	228	&	1.55	&	0.17	&	225	&	1.83	&	0.19	&	210	&	0.48	&	0.12\\
L1	&	129	&	0.81	&	0.14	&	125	&	1.35	&	0.19	&	114	&	1.71	&	0.21	&	114	&	1.97	&	0.23	&	127	&	0.54	&	0.13\\
L2	&	65	&	0.91	&	0.14	&	66	&	1.51	&	0.21	&	57	&	1.98	&	0.28	&	57	&	2.27	&	0.32	&	65	&	0.59	&	0.11\\
L3	&	65	&	0.96	&	0.14	&	64	&	1.61	&	0.22	&	59	&	2.11	&	0.29	&	59	&	2.42	&	0.34	&	64	&	0.65	&	0.15\\
L4	&	39	&	1.07	&	0.17	&	38	&	1.74	&	0.25	&	34	&	2.40	&	0.34	&	33	&	2.74	&	0.40	&	38	&	0.67	&	0.14\\
\enddata
\tablenotetext{a}{Only normal (non- low surface gravity, subdwarf, or young) L dwarfs with photometric uncertainty $<$ 0.1 were used in calculating the average.}
\end{deluxetable}
\clearpage
\end{landscape}

\begin{landscape}
\tabletypesize{\scriptsize}
\begin{deluxetable}{cccccccccccccccccr}
\tablecaption{Average Infrared Colors of late-type M and L dwarfs\label{tab:meancolors2}}
\tablewidth{0pt}
\tablehead{
\colhead{SpT} &
\colhead{N$_{H-W1}$\tablenotemark{a}} &
\colhead{($H$-$W1$)$_{avg}$} &
\colhead{$\sigma$(H-W1)} &
\colhead{N$_{H-W2}$\tablenotemark{a}} &
\colhead{($H$-$W2$)$_{avg}$} &
\colhead{$\sigma$(H-W2)} &
\colhead{N$_{K-W1}$\tablenotemark{a}} &
\colhead{($K$-$W1$)$_{avg}$} &
\colhead{$\sigma$(K-W1)} &
\colhead{N$_{K-W2}$\tablenotemark{a}} &
\colhead{($K$-$W2$)$_{avg}$} &
\colhead{$\sigma$(K-W2)}&
\colhead{N$_{W1-W2}$\tablenotemark{a}} &
\colhead{($W1$-$W2$)$_{avg}$} &
\colhead{$\sigma$(W1-W2)} \\
\colhead{(1)} &
\colhead{(2)} &
\colhead{(3)} &
\colhead{(4)} &
\colhead{(5)} &
\colhead{(6)} &
\colhead{(7)} &
\colhead{(8)} &
\colhead{(9)} &
\colhead{(10)}&
\colhead{(11)}&
\colhead{(12)}&
\colhead{(13)} &
\colhead{(14)}&
\colhead{(15)}&
\colhead{(16)}&
\\
}
\startdata
M7	&	1992	&	0.55	&	0.09	&	1988	&	0.76	&	0.10	&	1970	&	0.20	&	0.09	&	1970	&	0.41	&	0.10	&	1991	&	0.21	&	0.06	\\
M8	&	692	&	0.62	&	0.10	&	692	&	0.84	&	0.12	&	687	&	0.22	&	0.09	&	687	&	0.44	&	0.11	&	692	&	0.22	&	0.06	\\
M9	&	420	&	0.72	&	0.11	&	420	&	0.96	&	0.15	&	417	&	0.26	&	0.10	&	417	&	0.51	&	0.14	&	421	&	0.24	&	0.06	\\
L0	&	222	&	0.81	&	0.13	&	219	&	1.08	&	0.17	&	223	&	0.34	&	0.12	&	219	&	0.61	&	0.13	&	229	&	0.27	&	0.08	\\
L1	&	113	&	0.91	&	0.15	&	113	&	1.17	&	0.18	&	112	&	0.37	&	0.11	&	111	&	0.63	&	0.14	&	118	&	0.26	&	0.06	\\
L2	&	56	&	1.05	&	0.17	&	56	&	1.34	&	0.22	&	57	&	0.45	&	0.12	&	57	&	0.74	&	0.16	&	58	&	0.29	&	0.07	\\
L3	&	58	&	1.16	&	0.20	&	58	&	1.47	&	0.24	&	59	&	0.51	&	0.14	&	59	&	0.82	&	0.18	&	63	&	0.31	&	0.06	\\
L4	&	34	&	1.34	&	0.24	&	34	&	1.67	&	0.29	&	34	&	0.68	&	0.14	&	34	&	1.01	&	0.18	&	35	&	0.34	&	0.07	\\
\enddata
\tablenotetext{a}{Only normal (non- low surface gravity, subdwarf, or young) L dwarfs with photometric uncertainty $<$ 0.1 were used in calculating the average.}
\end{deluxetable}
\clearpage
\end{landscape}

\begin{landscape}
\tabletypesize{\scriptsize}
\begin{deluxetable}{cccccccccccccccccr}
\tablecaption{Infrared Colors of low gravity late-type M and L dwarfs\label{tab:meancolors3}}
\tablewidth{0pt}
\tablehead{
\colhead{Name} &
\colhead{SpT} &
\colhead{SpT} &
\colhead{($J$-$H$)} &
\colhead{\# of $\sigma_{(J-H)}$\tablenotemark{a}} &
\colhead{($J$-$K$)} &
\colhead{\# of $\sigma_{(J-K)}$\tablenotemark{a}} &
\colhead{($J$-$W1$)} &
\colhead{\# of $\sigma_{(J-W1)}$\tablenotemark{a}} &
\colhead{($J$-$W2$)} &
\colhead{\# of $\sigma_{(J-W2)}$\tablenotemark{a}} &
\colhead{($H$-$K$)} &
\colhead{\# of $\sigma_{(H-K)}$\tablenotemark{a}} \\
\colhead{(1)} &
\colhead{(2)} &
\colhead{(3)} &
\colhead{(4)} &
\colhead{(5)} &
\colhead{(6)} &
\colhead{(7)} &
\colhead{(8)} &
\colhead{(9)} &
\colhead{(10)}&
\colhead{(11)}&
\colhead{(12)}&
\colhead{(13)}\\
}
\startdata
01231125-6921379	&	M7.5$\gamma$	&	-- --	&	0.61	$\pm$	0.03	&	0.0	&	1.00	$\pm$	0.03	&	0.3	&	1.26	$\pm$	0.03	&	1.0	&	1.50	$\pm$	0.03	&	1.3	&	0.39	$\pm$	0.03	&	0.3	\\
01294256-0823580	&	M5 --	&	M7$\beta$	&	0.57	$\pm$	0.03	&	-0.4	&	0.88	$\pm$	0.03	&	-0.9	&	1.11	$\pm$	0.03	&	-0.7	&	1.33	$\pm$	0.03	&	-0.4	&	0.31	$\pm$	0.03	&	-0.5	\\
02501167-0151295	&	-- --	&	M7$\beta$	&	0.61	$\pm$	0.03	&	0.0	&	0.98	$\pm$	0.03	&	0.1	&	1.20	$\pm$	0.03	&	0.3	&	1.44	$\pm$	0.03	&	0.6	&	0.37	$\pm$	0.03	&	0.1	\\
02530084+1652532	&	-- --	&	M7$\beta$	&	0.51	$\pm$	0.04	&	-1.1	&	0.81	$\pm$	0.05	&	-1.6	&	1.07	$\pm$	0.03	&	-1.1	&	1.34	$\pm$	0.03	&	-0.3	&	0.30	$\pm$	0.06	&	-0.6	\\
02535980+3206373	&	M7$\beta$	&	M6$\beta$	&	0.69	$\pm$	0.03	&	0.8	&	1.07	$\pm$	0.03	&	1.0	&	1.29	$\pm$	0.03	&	1.4	&	1.49	$\pm$	0.03	&	1.2	&	0.38	$\pm$	0.03	&	0.2	\\
03350208+2342356	&	M8.5--	&	M7.5$\beta$	&	0.60	$\pm$	0.03	&	-0.2	&	0.99	$\pm$	0.02	&	0.2	&	1.21	$\pm$	0.03	&	0.4	&	1.48	$\pm$	0.03	&	1.1	&	0.39	$\pm$	0.02	&	0.3	\\
05181131-3101529	&	M6.5--	&	M7$\beta$	&	0.64	$\pm$	0.03	&	0.4	&	0.98	$\pm$	0.03	&	0.1	&	1.24	$\pm$	0.03	&	0.7	&	1.48	$\pm$	0.03	&	1.1	&	0.33	$\pm$	0.03	&	-0.3	\\
05264316-1824315	&	-- --	&	M7$\beta$	&	0.52	$\pm$	0.03	&	-1.0	&	0.91	$\pm$	0.03	&	-0.6	&	1.16	$\pm$	0.03	&	-0.1	&	1.41	$\pm$	0.03	&	0.4	&	0.39	$\pm$	0.03	&	0.3	\\
\enddata
\tablenotetext{a}{Values are the \# of $\sigma$ (as reported in Table  ~\ref{tab:meancolors1})  from the field sequence that each object differs.  A negative (-) number indicates that the color was blueward of the field sequence average whereas a positive (+) number indicates that the color was redward of the field sequence average.}
\end{deluxetable}
\clearpage
\end{landscape}

\begin{landscape}
\tabletypesize{\scriptsize}
\begin{deluxetable}{lllllllllllllll}
\tablecaption{ Infrared Colors of low gravity late-type M and L dwarfs\label{tab:meancolors4}}
\tablewidth{0pt}
\tablehead{
\colhead{Name} &
\colhead{SpT} &
\colhead{SpT} &
\colhead{($H$-$W1$)} &
\colhead{\# of $\sigma_{(H-W1)}$\tablenotemark{a}} &
\colhead{($H$-$W2$)} &
\colhead{\# of $\sigma_{(H-W2)}$\tablenotemark{a}} &
\colhead{($K$-$W1$)} &
\colhead{\# of $\sigma_{(K-W1)}$\tablenotemark{a}} &
\colhead{($K$-$W2$)} &
\colhead{\# of $\sigma_{(K-W2)}$\tablenotemark{a}} &
\colhead{$(W1$-$W2$)} &
\colhead{\# of $\sigma_{(W1-W2)}$\tablenotemark{a}} \\
\colhead{(1)} &
\colhead{(2)} &
\colhead{(3)} &
\colhead{(4)} &
\colhead{(5)} &
\colhead{(6)} &
\colhead{(7)} &
\colhead{(8)} &
\colhead{(9)} &
\colhead{(10)}&
\colhead{(11)}&
\colhead{(12)}&
\colhead{(13)}\\
}
\startdata
01231125-6921379	&	M7.5$\gamma$	&	-- --	&	0.65	$\pm$	0.03	&	1.1	&	0.89	$\pm$	0.03	&	1.3	&	0.26	$\pm$	0.03	&	0.7	&	0.51	$\pm$	0.03	&	1.0	&	0.24	$\pm$	0.03	&	0.5	\\
01294256-0823580	&	M5--	&	M7$\beta$	&	0.54	$\pm$	0.03	&	-0.1	&	0.76	$\pm$	0.03	&	0.0	&	0.23	$\pm$	0.03	&	0.3	&	0.44	$\pm$	0.03	&	0.3	&	0.22	$\pm$	0.03	&	0.1	\\
02501167-0151295	&	-- --	&	M7$\beta$	&	0.59	$\pm$	0.03	&	0.4	&	0.83	$\pm$	0.03	&	0.7	&	0.22	$\pm$	0.03	&	0.2	&	0.46	$\pm$	0.03	&	0.5	&	0.24	$\pm$	0.03	&	0.5	\\
02530084+1652532	&	-- --	&	M7$\beta$	&	0.56	$\pm$	0.05	&	0.1	&	0.83	$\pm$	0.04	&	0.7	&	0.26	$\pm$	0.05	&	0.7	&	0.53	$\pm$	0.05	&	1.2	&	0.27	$\pm$	0.03	&	0.9	\\
02535980+3206373	&	M7$\beta$	&	M6$\beta$	&	0.61	$\pm$	0.03	&	0.6	&	0.80	$\pm$	0.03	&	0.4	&	0.23	$\pm$	0.03	&	0.3	&	0.42	$\pm$	0.03	&	0.1	&	0.20	$\pm$	0.03	&	-0.2	\\
03350208+2342356	&	M8.5--	&	M7.5$\beta$	&	0.61	$\pm$	0.03	&	0.7	&	0.89	$\pm$	0.03	&	1.3	&	0.22	$\pm$	0.03	&	0.2	&	0.49	$\pm$	0.02	&	0.8	&	0.28	$\pm$	0.03	&	1.1	\\
05181131-3101529	&	M6.5--	&	M7$\beta$	&	0.59	$\pm$	0.03	&	0.5	&	0.83	$\pm$	0.03	&	0.7	&	0.26	$\pm$	0.03	&	0.7	&	0.50	$\pm$	0.03	&	0.9	&	0.24	$\pm$	0.03	&	0.5	\\
05264316-1824315	&	-- --	&	M7$\beta$	&	0.64	$\pm$	0.03	&	0.9	&	0.89	$\pm$	0.03	&	1.3	&	0.25	$\pm$	0.03	&	0.5	&	0.50	$\pm$	0.03	&	0.9	&	0.26	$\pm$	0.03	&	0.8	\\
\enddata
\tablenotetext{a}{Values are the \# of $\sigma$ (as reported in Table  ~\ref{tab:meancolors2})  from the field sequence that each object differs.  A negative (-) number indicates that the color was blueward of the field sequence average whereas a positive (+) number indicates that the color was redward of the field sequence average.}
\end{deluxetable}
\clearpage
\end{landscape}

\begin{deluxetable}{cccccccccc}\tablecaption{Coefficients of Polynomial Fits for M6 -T9 Dwarfs \label{tab:polynomials}}[t]
\tablehead{$M_{Filter}$&x&rms&$c_0$&$c_1$&$c_2$&$c_3$&$c_4$&$c_5$&$c_6$}\startdata
$M_{J\text{ FLD}}$&6.0\textless SpT \textless 29.0&0.402&-8.350e+00&7.157e+00&-1.058e+00&7.771e-02&-2.684e-03&3.478e-05&\nodata\\
$M_{J\text{ YNG}}$&7.0\textless SpT \textless 17.0&  0.647 & 4.032e-03  & -1.416e-01  &  2.097e+00        &   8.478e-01&\nodata&\nodata&\nodata\\
$M_{J\text{ GRP}}$&7.0\textless SpT \textless 17.0&  0.660 & -3.825e-03 &  1.370e-01  &  -9.279e-01   &   10.141e+00&\nodata&\nodata&\nodata\\
$M_{H\text{ FLD}}$&6.0\textless SpT \textless 29.0&0.389&-7.496e+00&6.406e+00&-9.174e-01&6.551e-02&-2.217e-03&2.841e-05&\nodata\\
$M_{H\text{ YNG}}$&7.0\textless SpT \textless 17.0&0.634&	  2.642e-03 & -1.049e-01  &   1.753e+00       &   1.207e+00 &\nodata&\nodata&\nodata\\
$M_{H\text{ GRP}}$&7.0\textless SpT \textless 17.0&0.603&	 -3.909e-03 &  1.346e-01  &  -9.347e-01   &   9.728e+00 &\nodata&\nodata&\nodata\\
$M_{Ks\text{ FLD}}$&6.0\textless SpT \textless 29.0&0.537&-6.704e+00&5.970e+00&-8.481e-01&5.978e-02&-1.997e-03&2.540e-05&\nodata\\
$M_{Ks\text{ YNG}}$&7.0\textless SpT \textless 17.0&0.640&	 -1.585e-02 &  7.338e-01  &   4.537e+00    &\nodata&\nodata&\nodata&\nodata\\
\enddata
\tablecomments{Relations use 2MASS or WISE magnitudes.  Polynomial fits to optical M/L dwarfs and NIR T dwarfs (L dwarfs with no optical spectral type have NIR spectral types) excluding subdwarfs, low-gravity dwarfs, and binaries for the field (FLD) relations.  We present polynomials inclusive of (1) all $\gamma$ and $\beta$ sources under the YNG polynomials as well as (2)  only High Likelihood/bonafide moving group members under the GRP polynomials (see Table ~\ref{tab:Members}).  The function is defined as M$_{J,H,Ks, W1, W2, W3, Lbol, Teff}$=$\sum_{i=0}^{n}$ c$_{i}$(SpT)$^{i}$ and is valid for varying spectral types M6-T9 where 6=M6, 10=L0, 20=T0, etc.  An FTEST was used to determine the goodness of fit for each polynomial. In the case of all FLD polynomials, the sample of \citet{Filippazzo15} is used.  In the case of all YNG or GRP polynomials, they are valid from M7 - L7.  We list a second L$_{bol}$ and Teff for the YNG polynomial (captioned YNG2) that includes planetary mass companions (e.g. HN Peg b, Gu Psc b, Ross 458C, etc) from \citet{Filippazzo15} and allows us to extend the polynomial from M7-T8.} 
\tablenotetext{a}{Add W1 photometry to SpT polynomial conversion:  M$_{L\text{converted}}$= W1 + $\sum_{i=0}^{n}$ c$_{i}$(SpT)$^{i}$}

\end{deluxetable}

\clearpage
\twocolumngrid

\acknowledgments{This publication has made use of the Carnegie Astrometric Program parallax reduction software as well as the data products from the Two Micron All-Sky Survey, which is a joint project of the University of Massachusetts and the Infrared Processing and Analysis Center/California Institute of Technology, funded by the National Aeronautics and Space Administration and the National Science Foundation. This research has alsp made use of the NASA/ IPAC Infrared Science Archive, which is operated by the Jet Propulsion Laboratory, California Institute of Technology, under contract with the National Aeronautics and Space Administration. Furthermore, this publication makes use of data products from the Wide-field Infrared Survey Explorer (WISE), which is a joint project of the University of California, Los Angeles, and the Jet Propulsion Laboratory/California Institute of Technology, and NEOWISE, which is a project of the Jet Propulsion Laboratory/California Institute of Technology. WISE and NEOWISE are funded by the US National Aeronautics and Space Administration. Australian access to the Magellan Telescopes was supported through the National Collaborative Research Infrastructure and Collaborative Research Infrastructure Strategies of the Australian Federal Government. CGT acknowledges the support in this research of Australian Research Council grants DP0774000 and DP130102695. JRT gratefully acknowledges support from NSF grants AST-0708810 and AST-1008217}

\bibliographystyle{apj.bst}
\bibliography{check}

\end{document}